\documentclass[%
 reprint,
 amsmath,amssymb,
 aps,
prd,
longbibliography,
]{revtex4-1}
\usepackage{graphicx}
\usepackage{epstopdf}
\usepackage{dcolumn}
\usepackage{bm}
\usepackage{color}
\usepackage{hyperref}

\hypersetup{
    colorlinks=true,
    linkcolor=blue,
    filecolor=magenta,
    urlcolor=cyan,
}

\usepackage[caption=false]{subfig}
\usepackage{xr}
\usepackage{cleveref}
\usepackage{siunitx}

\newcommand{\beq}{\begin{equation}}
\newcommand{\eeq}{\end{equation}}
\newcommand{\bea}{\begin{eqnarray}}
\newcommand{\eea}{\end{eqnarray}}

\begin{document}

\preprint{APS/123-QED}


\title{Laser Annealing  for Radiatively Broadened $\mathbf{MoSe_2}$ grown by Chemical Vapor Deposition}

\author{Christopher Rogers}
\email{cmrogers@stanford.edu}
\author{Dodd Gray}
\author{Nate Bogdanowicz}
\author{Hideo Mabuchi}
\email{hmabuchi@stanford.edu}
\affiliation{%
 Ginzton Laboratory, Stanford University, 348 Via Pueblo, Stanford, CA 94305
}%
%
%
%

\date{\today}


\begin{abstract}
We report on a laser annealing procedure which greatly improves the quality of suspended monolayers of chemical vapor deposition (CVD) grown $\mathrm{MoSe_2}$.
Annealing with a green laser locally heats the suspended flake to approximately 600 K, which both removes contaminants and reduces strain gradients.
At 4 K, we observe linewidths as narrow as 3.5 meV (1.6nm) full-width at half-max (FWHM) for both photoluminescence (PL) and reflection.
Large peak reflectances up to 47\% are also observed.
These values are comparable to those of the highest quality hexagonal boron nitride (hBN) encapsulated samples.
We demonstrate that this laser annealing process can yield highly spatially homogeneous samples, with the length scale of the homogeneity limited primarily by the size of the suspended area.
Annealed regions are very stable, exhibiting negligible deterioration over 24 hours at cryogenic temperatures.
The annealing method is also very repeatable, with substantial improvements of sample quality on every spot ($>40$) tested.

\end{abstract}

\pacs{Valid PACS appear here}
\maketitle



\section{\label{sec:Introduction} Introduction}

Recently, there has been tremendous interest in layered transition metal dichalcogenides (TMDCs).
This was sparked by the discovery that the canonical TMDC $\mathrm{MoS_2}$ becomes a direct bandgap semiconductor when isolated in monolayer form \cite{AtomicallyThinMoS2, EmergingPhotoluminescence}.
Many of the TMDCs have since been shown to exhibit a strong excitonic feature, showing strong and spectrally narrow PL and reflection.
Since then, many other TMDCs have been isolated in monolayer form, exhibiting interesting coherence, spin-valley, and strain properties \cite{ExcitonValleyCoherence, CoupledSpinValleyPhysics, ControlOfValleyPolarization, BandgapEngineeringOfStrainedMoS2}.
TMDCs have  become a popular testbed for many-body electron physics \cite{ObservationOfBiexcitonsInMonolayerWSe2, FermiPolaronPolaritons} and have attracted attention in the quantum optics community due to their potential applicability to exotic  exciton-polariton condensates and quantum  dot arrays deterministically patterned with electrostatic gating \cite{ElectricalControlofChargedCarriersInAtomicallyThin} or engineered strain \cite{LargeScaleQuantumEmitterArrays}.

Mechanical exfoliation \cite{AtomicallyThinMoS2, EmergingPhotoluminescence} is one of the most popular methods to produce monolayer flakes, and while it is possible to produce very high quality flakes in this manner, the samples tend to be small ($\sim$\SI{10}{\micro\meter} for TMDCs).
Flakes produced by mechanical exfoliation are also typically somewhat n-doped, so that electrostatic control is necessary to make the semiconductor neutral \cite{FermiPolaronPolaritons, TightlyBoundTrions}.

The highest quality flakes are produced by mechanical exfoliation and subsequent encapsulation in hBN \cite{ExcitonicLinewidthApproachingHomogenousLimit, LargeExcitonicReflectivity, RealizationOfAnElectricallyTunableMirror}.
Typically, encapsulation leads to even smaller flake sizes and is a time-consuming process.
These samples are often annealed in air or inert gas either during or after heterostructure fabrication to reduce contaminants between layers \cite{ExcitonicLinewidthApproachingHomogenousLimit}.
Microscopic spatial inhomogeneity of the exciton resonance energy is ubiquitous  in exfoliated and encapsulated  TMDC samples.
Even the highest quality samples demonstrated to date (as measured by linewidth and peak reflectance) exhibit uncontrolled linewidth-scale spatial variation of the exciton  resonance  over $<3$ \SI{}{\micro\meter} \cite{LargeExcitonicReflectivity}, and $10$ meV variation over $\sim 15$ \SI{}{\micro\meter} \cite{ExcitonicLinewidthApproachingHomogenousLimit}.
This severely limits the rate of scientific progress in understanding  TMDC exciton  physics and presents a barrier that must be overcome for practical scaling of TMDC-based optical and electronic devices.
The inhomogeneity is due to a combination of strain gradients\cite{OpticalImagingOfStrain}, material defects \cite{ExploringAtomicDefectsInMoS2, UltrafastDynamicsofDefectAssistedRecombiniation}, electrostatic substrate doping effects \cite{ScanningTunnelingMicroscopyOfGraphene} and surface contamination.

Monolayers can also be directly grown on substrates, typically by CVD.
This leads to much larger flakes, up to \SI{100}{\micro\meter} single crystal and continuous wafer-scale polycrystalline \cite{ScalableSynthesisofUniformMonolayerMoS2, SynthesisOfLargeAreaMoS2}.
While many interesting effects and devices have been demonstrated using CVD grown TMDCs, these monolayers are much lower quality \cite{StrongPLEnhancementOfMoS2Defect} by at least several important metrics (most notably PL and reflection linewidth at low temperature) when compared to mechanically exfoliated flakes.
This is despite the fact that defect densities in CVD and exfoliated monolayers are comparable \cite{ExploringAtomicDefectsInMoS2}.
The lower quality of CVD monolayers is likely due to a combination of adsorbed contaminants that are difficult to remove, and strain gradients from the growth/transfer processes \cite{AnnealingFreeCleanGrapheneTransfer, HighLuminescenceEfficiencyInMoS2}.

\section{\label{sec:Results} Results}
\begin{figure*}
    \centering
    \subfloat[ \label{fig:schematic}]
    {\includegraphics[width=0.6\textwidth]
    {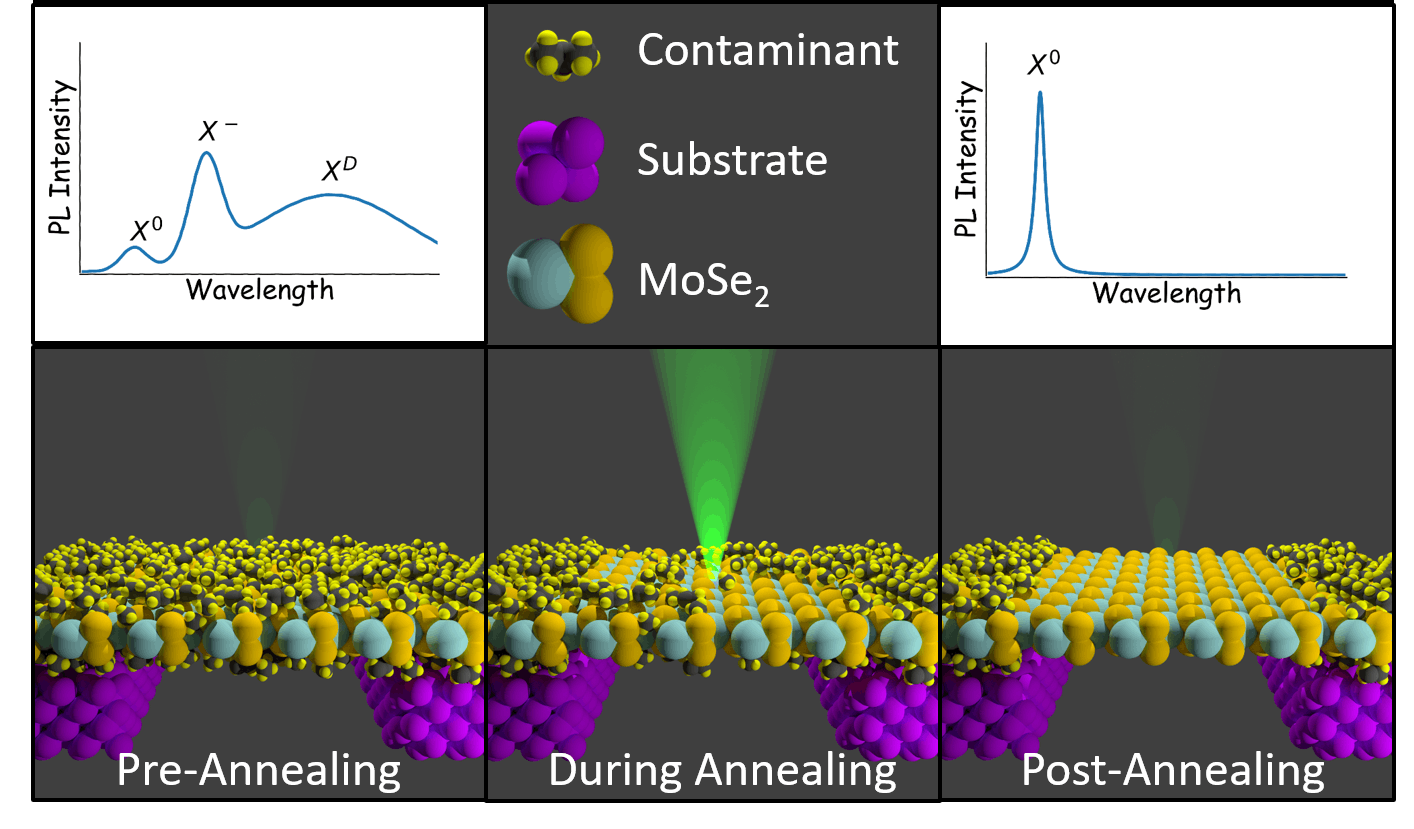}}
    \subfloat[ \label{fig:SampleImage}]
    {\includegraphics[width=0.2\textwidth]
    {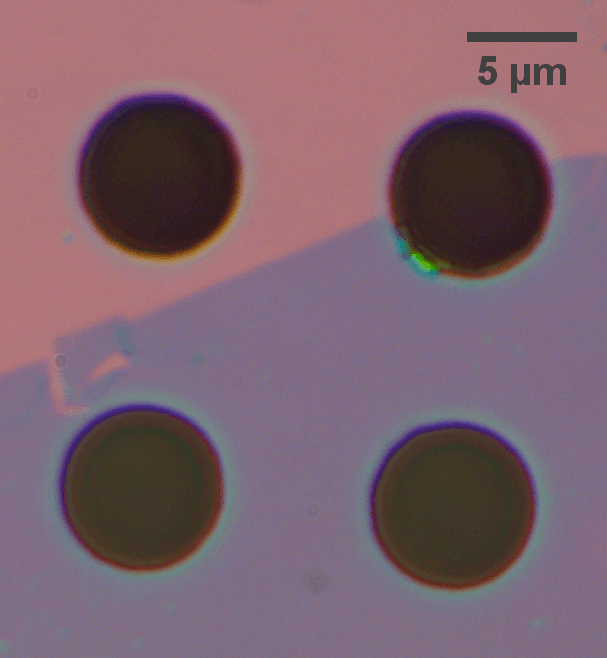}}

    \subfloat[ \label{fig:NarrowFWHM:PL}]
    {\includegraphics[width=0.22\textwidth]
    {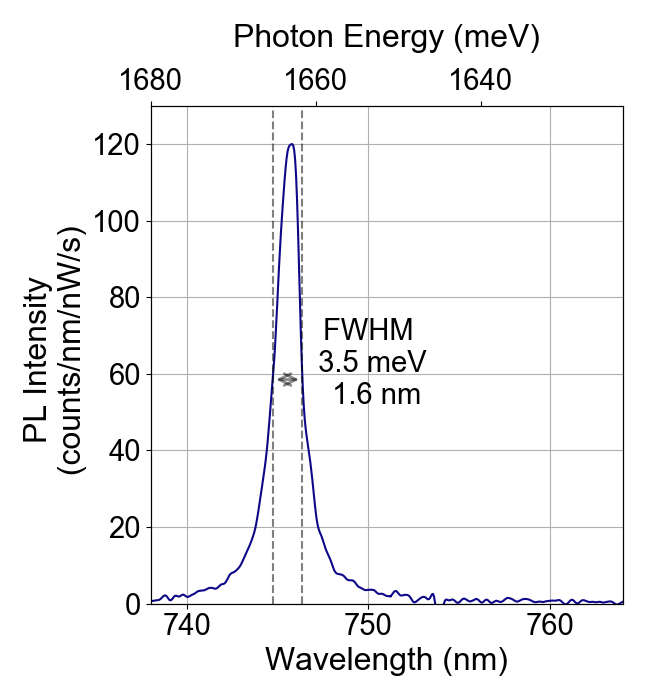}}
    \subfloat[ \label{fig:NarrowFWHM:Reflection}]
    {\includegraphics[width=0.22\textwidth]
    {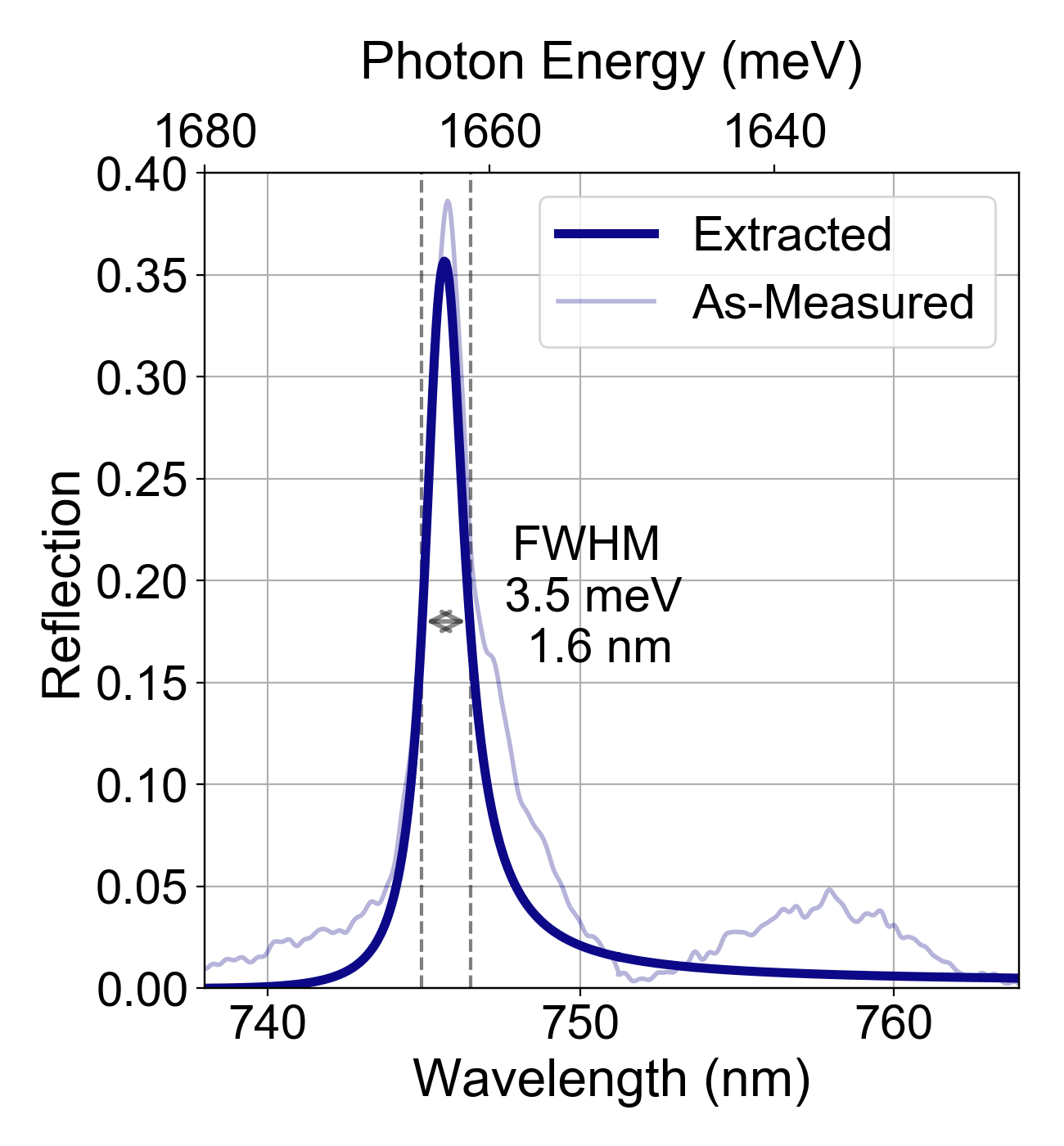}}
    \subfloat[ \label{fig:BeforeAfter:PL}]
    {\includegraphics[width=0.22\textwidth]
    {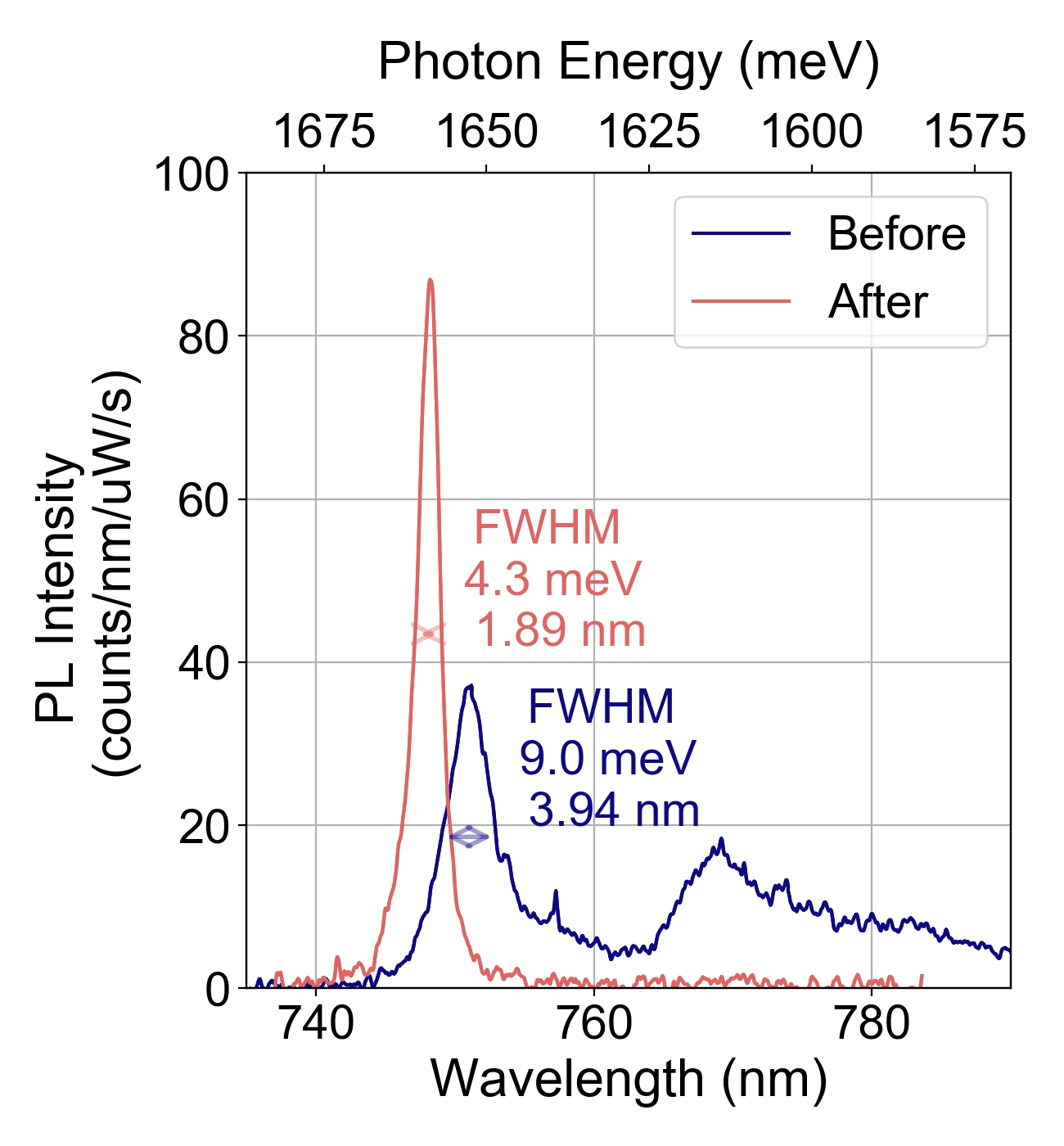}}

    \subfloat[ \label{fig:BeforeAfter:Reflection}]
    {\includegraphics[width=0.22\textwidth]
    {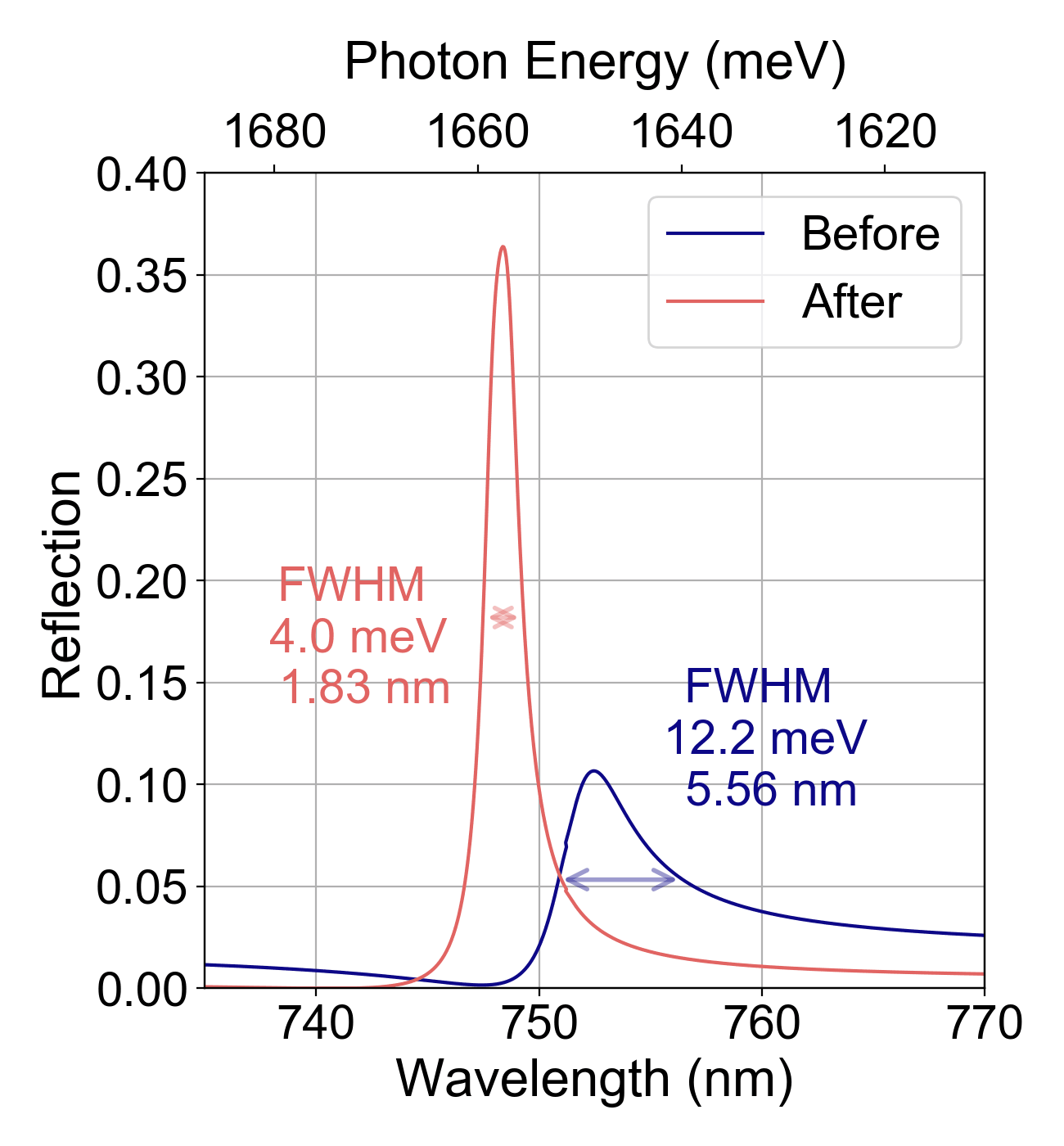}}
    \subfloat[ \label{fig:BeforeAfter:LargeReflection}]
    {\includegraphics[width=0.22\textwidth]
    {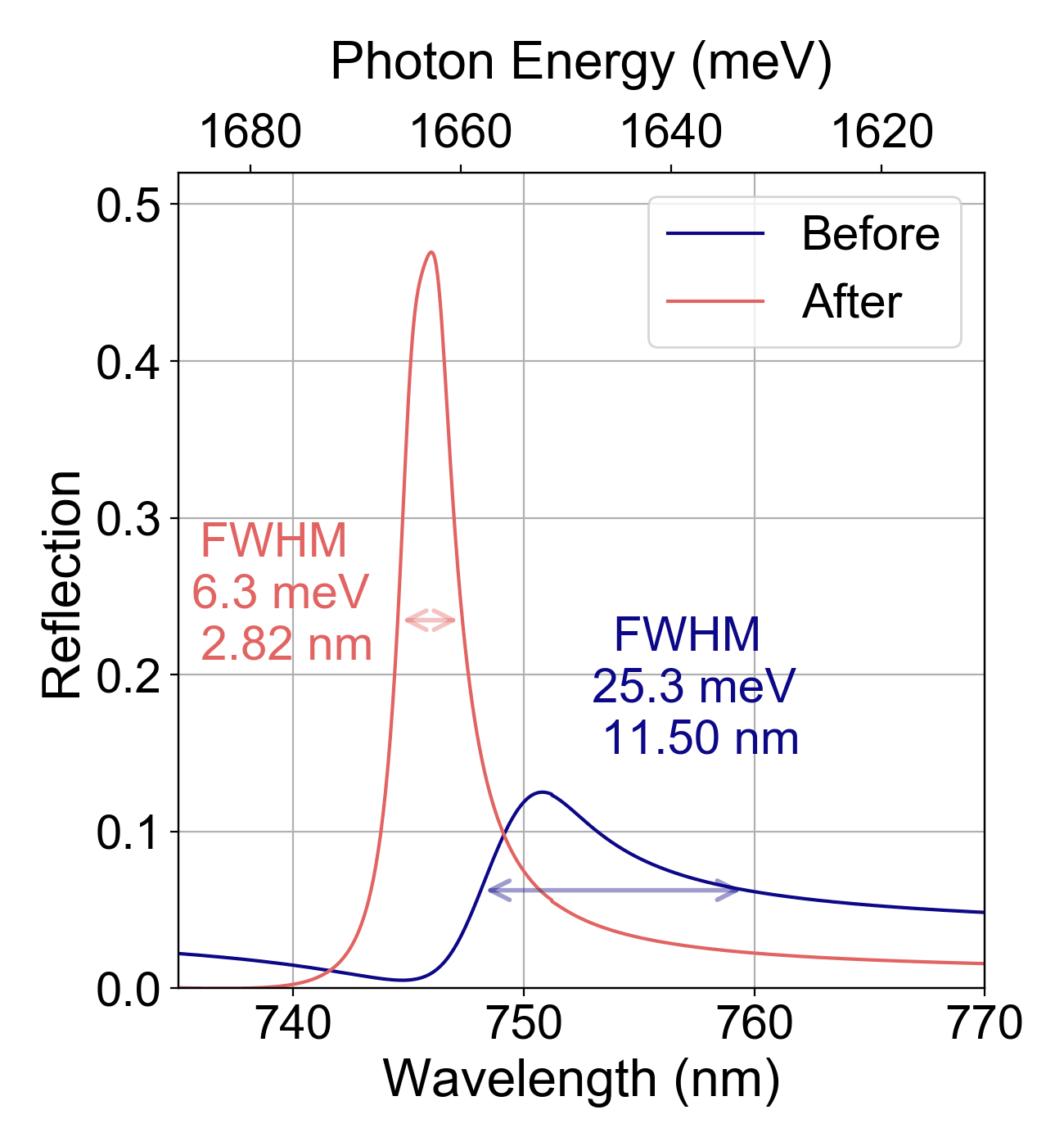}}
    \subfloat[ \label{fig:BeforeAfter:Reflection_24h}]
    {\includegraphics[width=0.22\textwidth]
    {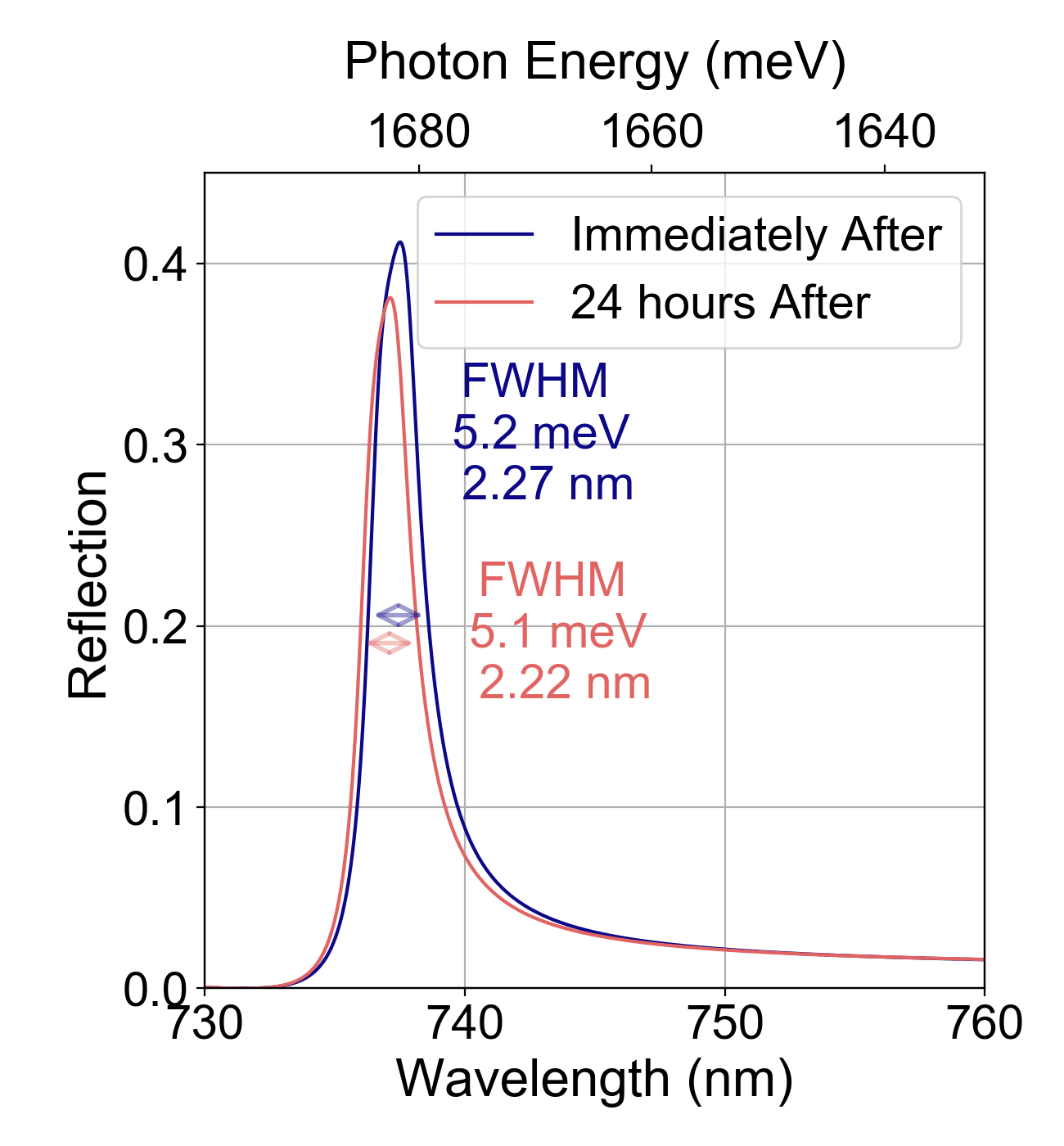}}
    \caption{Sample improvement from annealing.
    A schematic (a) of the annealing process showing a suspended flake before, during, and after annealing.
    Note that atoms sizes are not to scale with the laser spot and sample dimensions.
    The inset spectra illustrate typical PL before and after annealing.
    Neutral exciton emission is labelled $X^0$, negatively charged trion emission is labelled $X^-$ and the broad peak typically associated with defect emission is labelled $X^D$.
    The optical microscope image (b) of a typical sample with four dark holes visible in the substrate, the lower two of which have suspended monolayer $\mathrm{MoSe_2}$.
    The darker portion of the substrate has supported monolayer $\mathrm{MoSe_2}$.
    PL (c) and reflection (d) from the same position of suspended monolayer CVD-grown $\mathrm{MoSe_2}$ annealed for 1 minute with 9 mW laser power and a \SI{10}{\micro\meter} spot size.
    Note that in (d) the reflection from the suspended film alone has been extracted from the raw reflection using a fitting procedure described in Sec. \ref{subsec:Methods:ReflectionAnalysis} to remove interference fringes from reflections off the bottom of hole in the silicon substrate.
    PL (e) and reflection (f) from a second position of monolayer suspended CVD-grown $\mathrm{MoSe_2}$ annealed for 5 minutes with \SI{500}{\micro\watt} laser power with a 700 nm spot size.
    Reflection (g) before and after from a third positioin, which had the highest reflection after annealing.
    Reflection (h) immediately after annealing and 24 hours after annealing, from a fourth position.
    Measurement and annealing was at 4K, and $2\cdot 10^{-7}$ Torr pressure.
    The corresponding unfitted reflection spectra for (f), (g) and (h) are shown in Fig. \ref{sfig:NarrowFWHM}.
    \label{fig:NarrowFWHM}}
\end{figure*}
We report on a laser annealing procedure for improving the quality of monolayers of the TMDC $\mathrm{MoSe_2}$ to levels comparable with hBN-encapsulated samples.
The process can be tuned to yield highly spatially homogeneous samples.
It is also repeatable, in the sense that annealing significantly improved sample quality on every spot tested.
The annealing process involves heating suspended monolayers of $\mathrm{MoSe_2}$ to between 500 and 600 K in high vacuum using an above-bandgap green laser.
Note that this is done in a cryostat, so that the nominal substrate temperature is 4 K.
Unless otherwise mentioned, all measurements were performed at 4 K and $2\cdot 10^{-7}$ Torr.
We hypothesize that this heating effectively removes contaminants and extrinsic dopants adsorbed on the monolayer.
The heating in a radially symmetric manner also relaxes the strain pinning at the edge of the suspended layer and the spatial strain gradient is dramatically reduced.
A schematic of the annealing process is shown in Fig. \ref{fig:schematic}.

Figs. \ref{fig:NarrowFWHM:PL},\ref{fig:NarrowFWHM:Reflection} show some of the narrowest PL and reflection from the annealed monolayers.
The PL reflection full-width at half-max (FWHM) are both 3.5 meV.
This is comparable to the range of 2.4-4.9 meV, for PL from high quality encapsulated $\mathrm{MoSe_2}$ samples from \cite{ExcitonicLinewidthApproachingHomogenousLimit}.
It is worth noting that the dielectric environment for our suspended samples is much different than that for encapsulated samples.
Excitons in encapsulated $\mathrm{MoSe_2}$ see a larger effective dielectric constant and thus have a smaller binding energy \cite{DielectricImpactOnExcitonBindingEnergies}.
This higher effective dielectric constant in turn leads to a longer radiative lifetime and smaller exciton radius \cite{ExcitonRadiativeLifetimeInTMDCs}.
Thus, the intrinsic radiative lifetime in suspended layers is smaller than that for encapsulated layers.
Therefore, even in a system free of defects, contaminants and strain, at zero temperature (where the exciton is purely radiatively broadened), the linewidth of encapsulated samples would be smaller than that of suspended samples.

\begin{figure*}
    \centering
    \subfloat[ \label{fig:RasterAnneal:ExtractedFWHM}]
    {\includegraphics[width=0.175\textwidth]
    {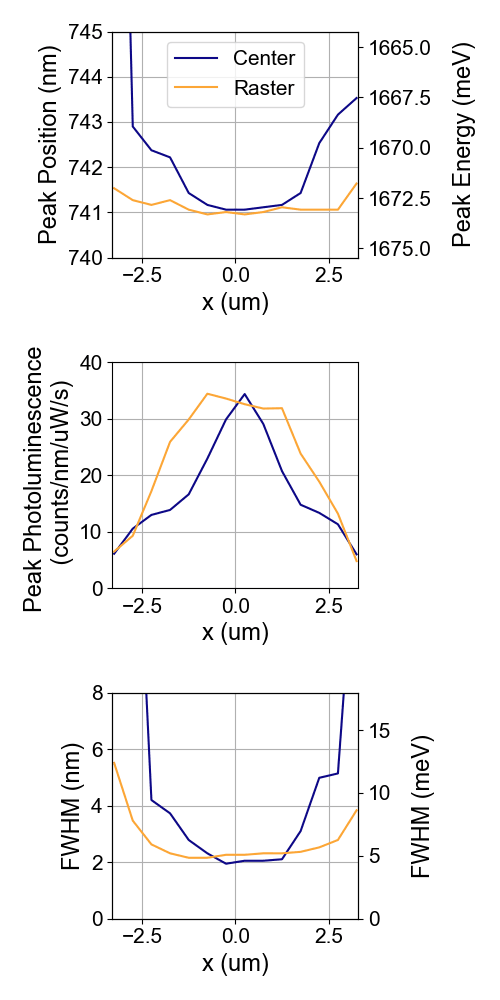}}
    \subfloat[ \label{fig:RasterAnneal:CenterAnneal}]
    {\includegraphics[width=0.35\textwidth]
    {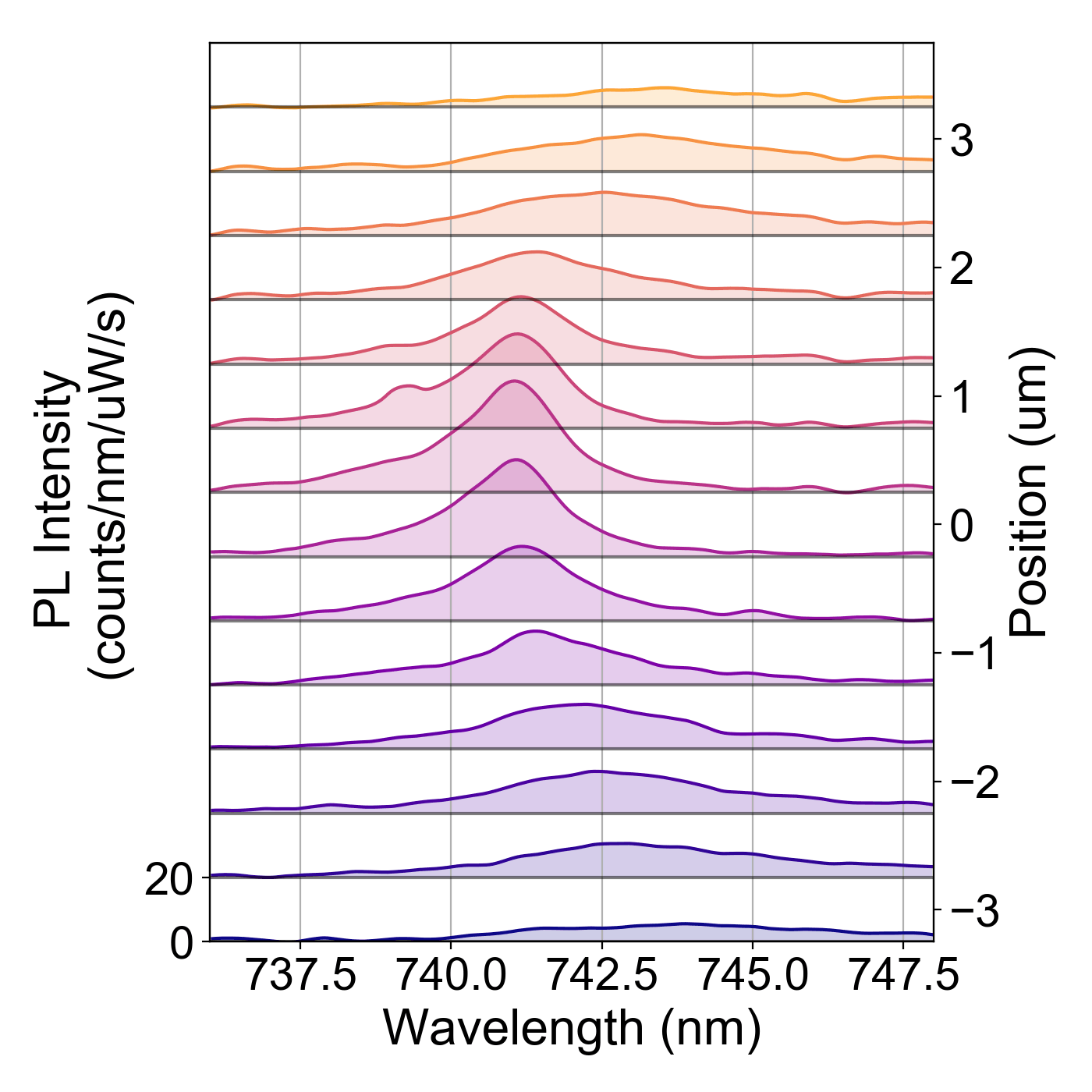}}
    \subfloat[ \label{fig:RasterAnneal:ReCenterAnneal}]
    {\includegraphics[width=0.35\textwidth]
    {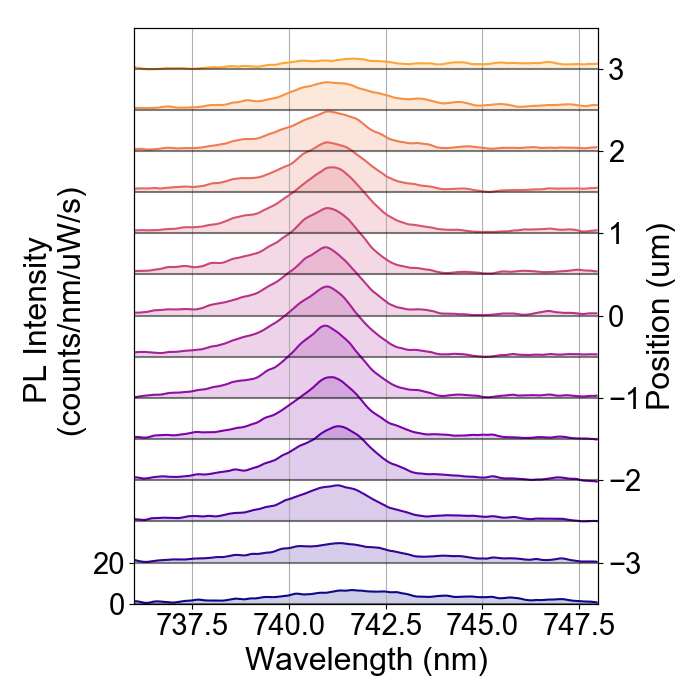}}
    \caption{PL homogeneity improvement from raster scanning the annealing laser.
    FWHM, peak amplitude and peak position (a) for the exciton PL from a line across a hole first annealed in the center, then annealed by raster scanning a spot over the entire hole.
    Each spot was annealed at \SI{400}{\micro\watt} for 1 s with a 700 nm spot size.
    PL along the same line for an initial anneal at the center (b), and for a raster-scanned anneal (c).
    \label{fig:RasterAnneal}}
\end{figure*}

Typical PL and reflection spectra both before and after annealing are shown in Fig \ref{fig:BeforeAfter:PL}.
After annealing, the PL shows a drastic reduction in linewidth.
The trion PL $X^-$ is also reduced to negligible intensity, indicating that the annealed sample is neutral.
We attribute these effects primarily to removal of contaminants from the surface of the sample.
It has been shown that water and oxygen are among the primary adsorbents that can cause PL modulation through charge transfer \cite{ModulationOfLightEmissionByPhysisorption}, and thus it seems likely that the annealing process reduces doping by removing these and other adsorbed molecules.
Adsorbed molecules could lead to atomic-scale variations in the dielectric and electrostatic environment seen by the exciton, manifesting as inhomogeneous broadening of the $X^0$ and $X^-$ frequency.
The intensity of the $X^D$ emission (often associated with defect emission \cite{ModulationOfLightEmissionByPhysisorption}) is also reduced below the noise floor of our measurements.
This suggests that $X^D$ is also due to the interaction of excitons with molecules adsorbed on the $\mathrm{MoSe_2}$ monolayer.

The spectrum with the highest peak reflection of from an annealed spot is shown in \ref{fig:BeforeAfter:LargeReflection}.
Assuming a simple Lorentzian model for the reflection including dephasing, radiative decay, and non-radiative decay terms, a peak reflection of 47\% indicates that the exciton is radiatively broadened.
The maximum reflection $R_\mathrm{max}$ of the exciton feature is related to the radiative broadening $\Gamma$ and the total linewidth broadening $\gamma_\mathrm{tot}$ by $R_\mathrm{max} = \frac{\Gamma^2}{\gamma_\mathrm{tot}^2}$ \cite{LargeExcitonicReflectivity, RealizationOfAnElectricallyTunableMirror}.
This leads to a ratio between radiative and all other broadening of $\frac{\Gamma}{\gamma_\mathrm{tot} - \Gamma} = \frac{\sqrt{R_\mathrm{max}}}{1-\sqrt{R_\mathrm{max}}}$.
The maximum reflection of 47\% observed here thus corresponds to a ratio of 2.2 between radiative broadening and all other broadening and a radiative broadening of 4.3 meV.

Here we note that after annealing the sample properties were very stable.
There were no appreciable changes over 24 hours (the longest time over which we measured), when the samples were kept at high vacuum ($2\cdot 10^{-7}$) and the cryostat was kept at its base temperature of 4 K.
An example of the reflection immediately after annealing and 24 hours after annealing is shown in Fig \ref{fig:BeforeAfter:Reflection_24h}.
The reflection changes very little over 24 hours.
The small changes that are present may be due to the fact that there is some unavoidable drift in the system, which leads to change in both the coupling efficiency of the collection and the location of the spot on the sample.
We expect that the samples would also be stable under high vacuum at room temperature, although we were not able to test this in our system, which relies on cryo-pumping to achieve high vaccum.
The flake quality degrades somewhat after warming and leaving at room temperature and 1 Torr for 24 hours, and then cooling back down.
For the data in Fig. \ref{sfig:BeforeAfter}, the PL linewidth went from 5.6 meV after annealing to 7.0 meV after the warm-up cycle.
Note that since the holes in the substrate are closed, any contaminants removed from the bottom of the suspended film by the annealing are trapped, and may re-adsorb when the entire sample is heated to room temperature.
However, these degraded flakes are higher quality than before annealing, and the degradation can be fully reversed by re-annealing.
In the case of Fig. \ref{sfig:BeforeAfter}, the linewidth is improved to 5.3 meV after re-annealing.
See Fig. \ref{sfig:BeforeAfter} for details.
\begin{figure*}
    \centering
    \subfloat[Before anneal \label{fig:MultiHole:PL_Before}]
    {\includegraphics[width=0.45\textwidth]
    {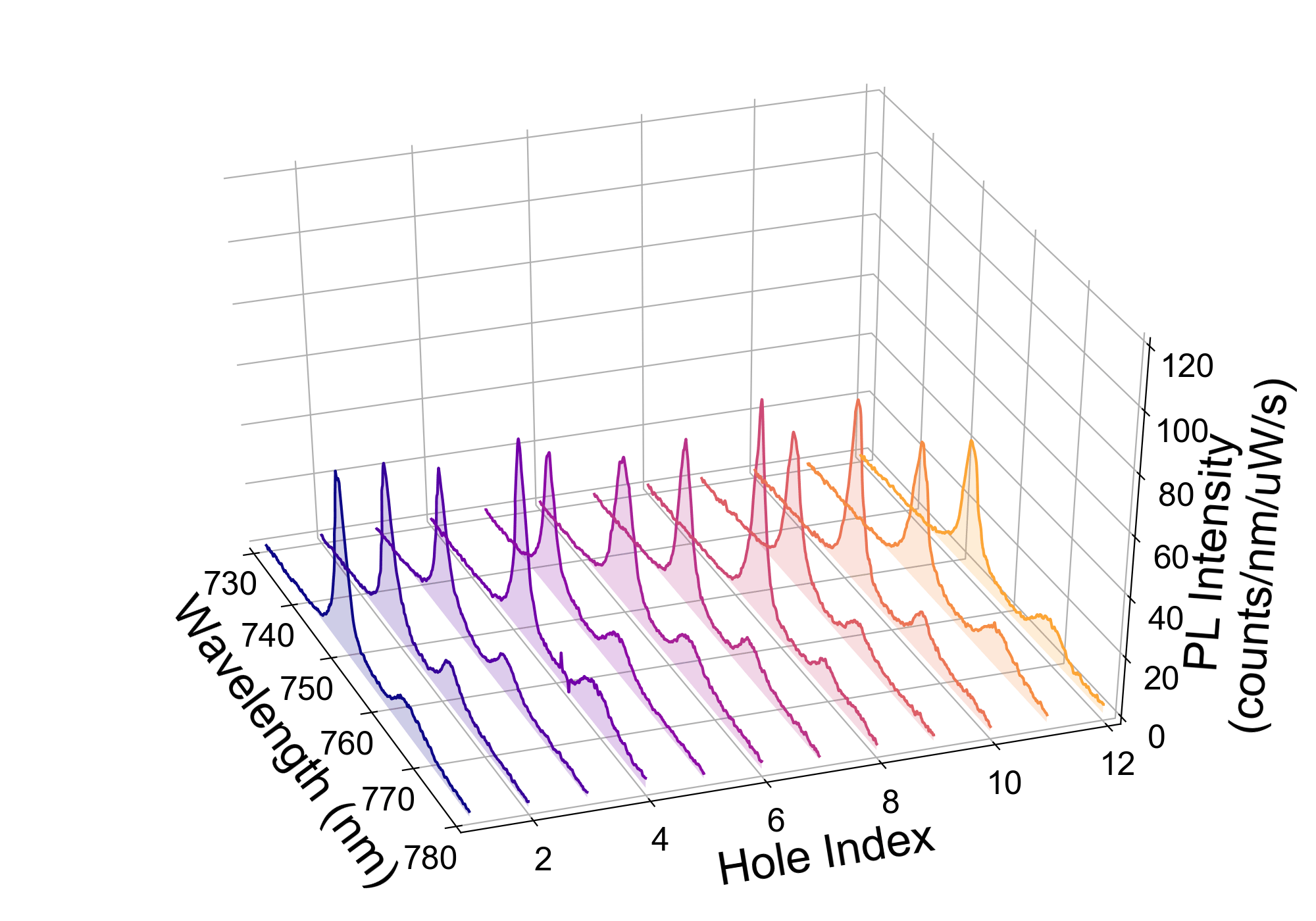}}
    \subfloat[Before anneal \label{fig:MultiHole:ReflectionBefore}]
    {\includegraphics[width=0.45\textwidth]
    {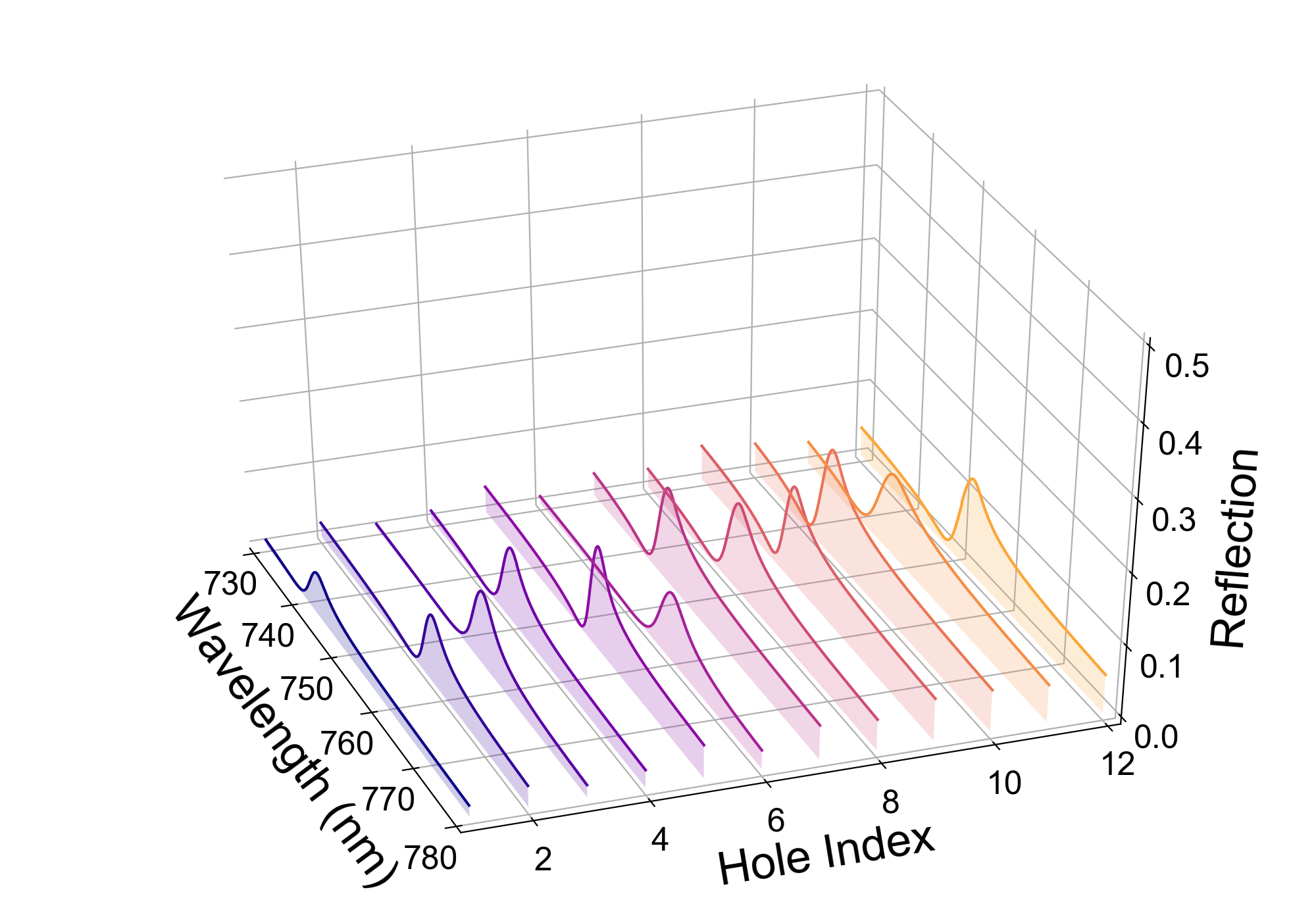}}

    \subfloat[After anneal \label{fig:MultiHole:PL_After}]
    {\includegraphics[width=0.45\textwidth]
    {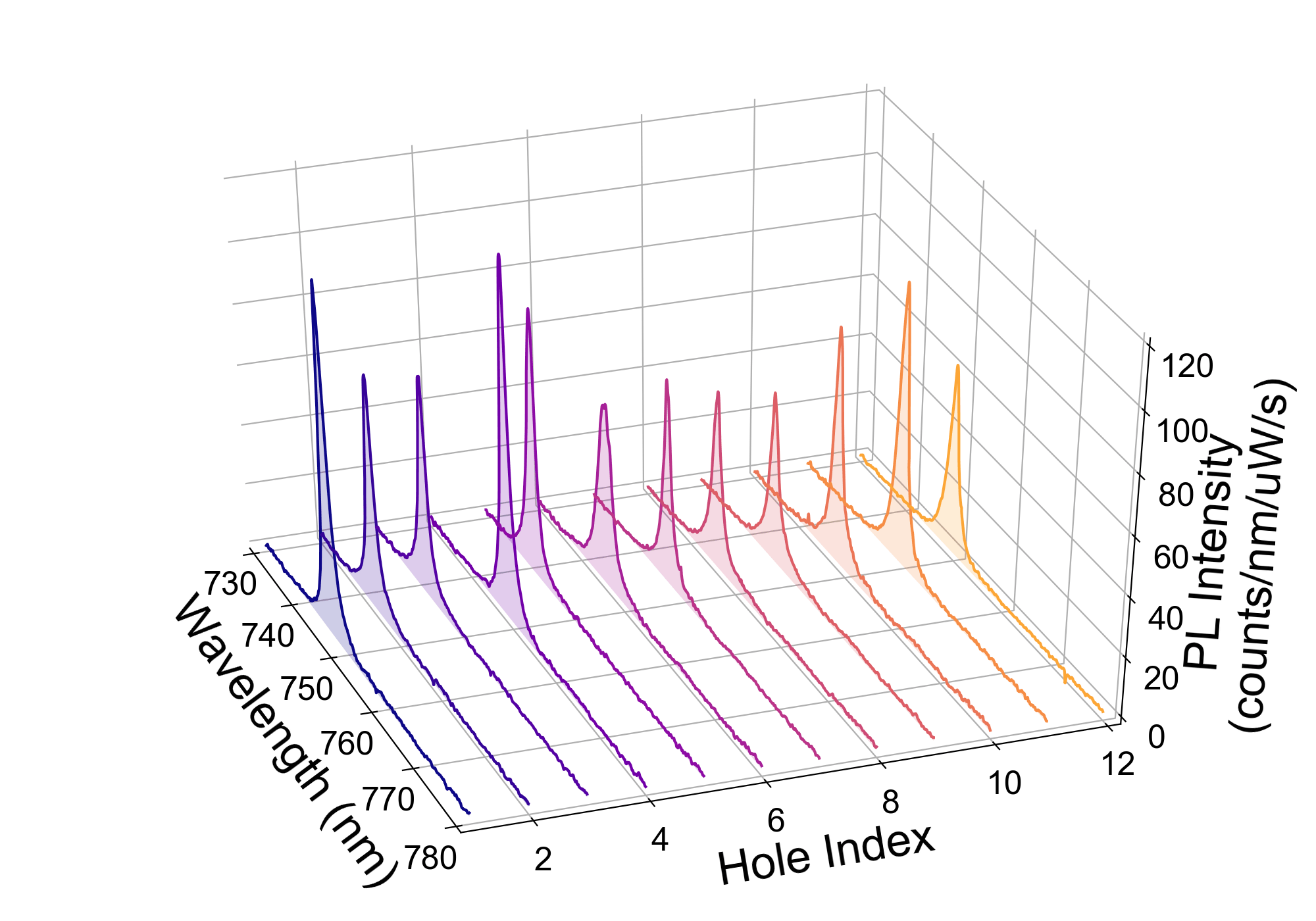}}
    \subfloat[After anneal \label{fig:MultiHole:ReflectionAfter}]
    {\includegraphics[width=0.45\textwidth]
    {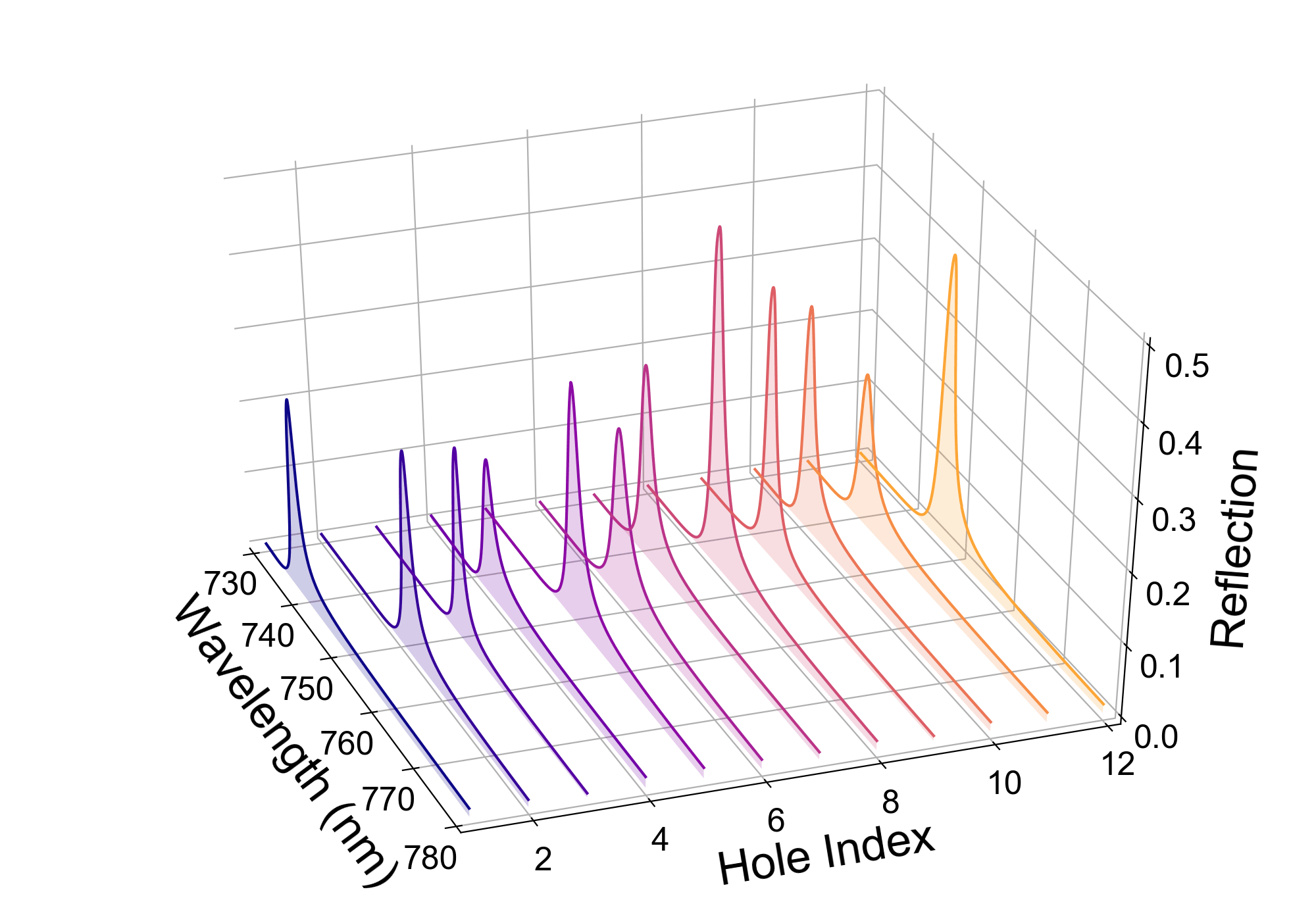}}
    \caption{Repeatability of the annealing process.
    PL and reflection before and after for 12 separate holes.
    The PL before annealing (a) and after annealing (c).
    The reflection before annealing (b) and after annealing (d).
    Note that the PL and reflection are from different sets of 12 holes.
    The holes were annealed in the center using a 700 nm diameter laser spot.
    Power was increased incrementally, and the best spectra was chosen.
    There was no pre- or post-selection of holes.
    The reflection spectra are fit as explained in Sec. \ref{subsec:Methods:ReflectionAnalysis}, and the corresponding raw spectra are shown in Fig. \ref{sfig:MultiHole}.
    \label{fig:MultiHole}}
\end{figure*}

It is possible to improve the spatial homogeneity of the annealed samples by raster scanning the annealing laser over the entire suspended area.
This more fully removes contaminants from the entire suspended area, and can also have implications for the spatial strain gradient present in the flake.
PL after annealing in the center with a laser spot much smaller than the hole (no raster scan) is shown in \ref{fig:RasterAnneal:CenterAnneal}.
The spectra were taken in a line across a hole with suspended $\mathrm{MoSe_2}$.
The extracted peak position, FWHM and amplitude are also shown.
It is apparent that both the line position and the quality of the flake, as measured by PL intensity and linewidth, vary substantially across the flake.
The quality is much higher in the middle (where the annealing laser was centered) than at the edges.
The line position varies parabolically about the center, possibly due to some combination of larger strain gradient and more contaminants near the edges of the suspended area.

PL after raster scanning the annealing beam over the entire suspended area is shown in Fig. \ref{fig:RasterAnneal:ReCenterAnneal}.
The spectra were taken across the same as above.
In this case, the line position is almost constant over most of the suspended area, changing by less than 1 meV (less than half the linewidth) over \SI{6}{\micro\meter}.
The PL intensity is also much more uniform, with the spatial FWHM being about \SI{4}{\micro\meter}, as opposed to \SI{2.5}{\micro\meter} before.
The length scale of the spatial homogeneity in this case appears to be limited by the size of the hole.
This is in stark contrast to exfoliated supported flakes, where the line center can shift by more than 10 meV over only a few microns.
Even very high quality encapsulated samples can show linewidth-scale variation over $<3$ \SI{}{\micro\meter} \cite{LargeExcitonicReflectivity}, and $10$ meV variation over $\sim 15$ \SI{}{\micro\meter} \cite{ExcitonicLinewidthApproachingHomogenousLimit}.
We note that raster annealing with position-dependent annealing power could further improve homogeneity near the edges of the suspended.
For the constant-power anneal performed here, the annealing temperature is higher when the beam is near the center because the center is more thermally isolated from the substrate.
\begin{figure*}
    \centering
    \subfloat[ \label{fig:AnnealPower:PL_Anneal}]
    {\includegraphics[width=0.37\textwidth]
    {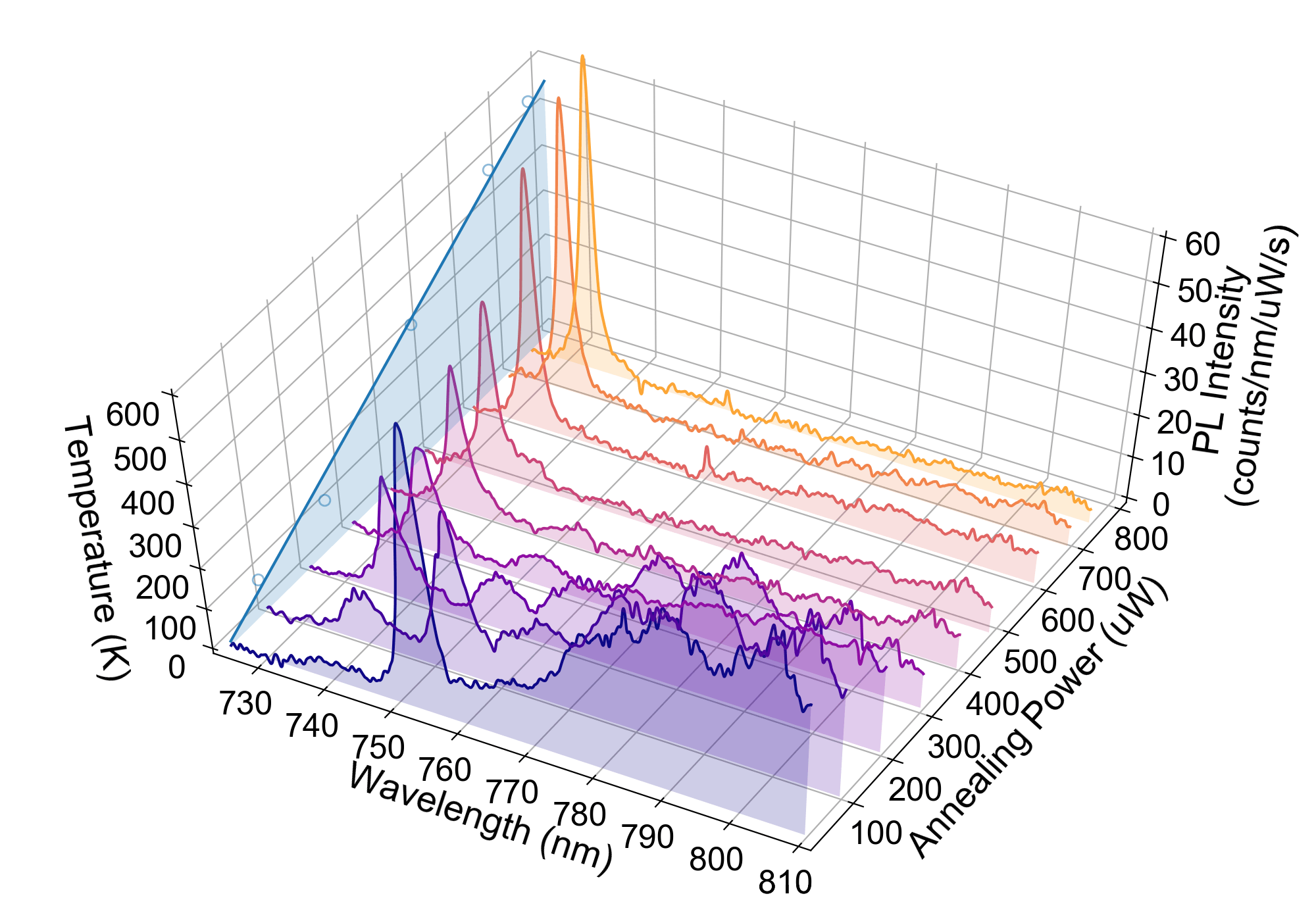}}
    \subfloat[ \label{fig:AnnealPower:ReflectionAnneal}]
    {\includegraphics[width=0.37\textwidth]
    {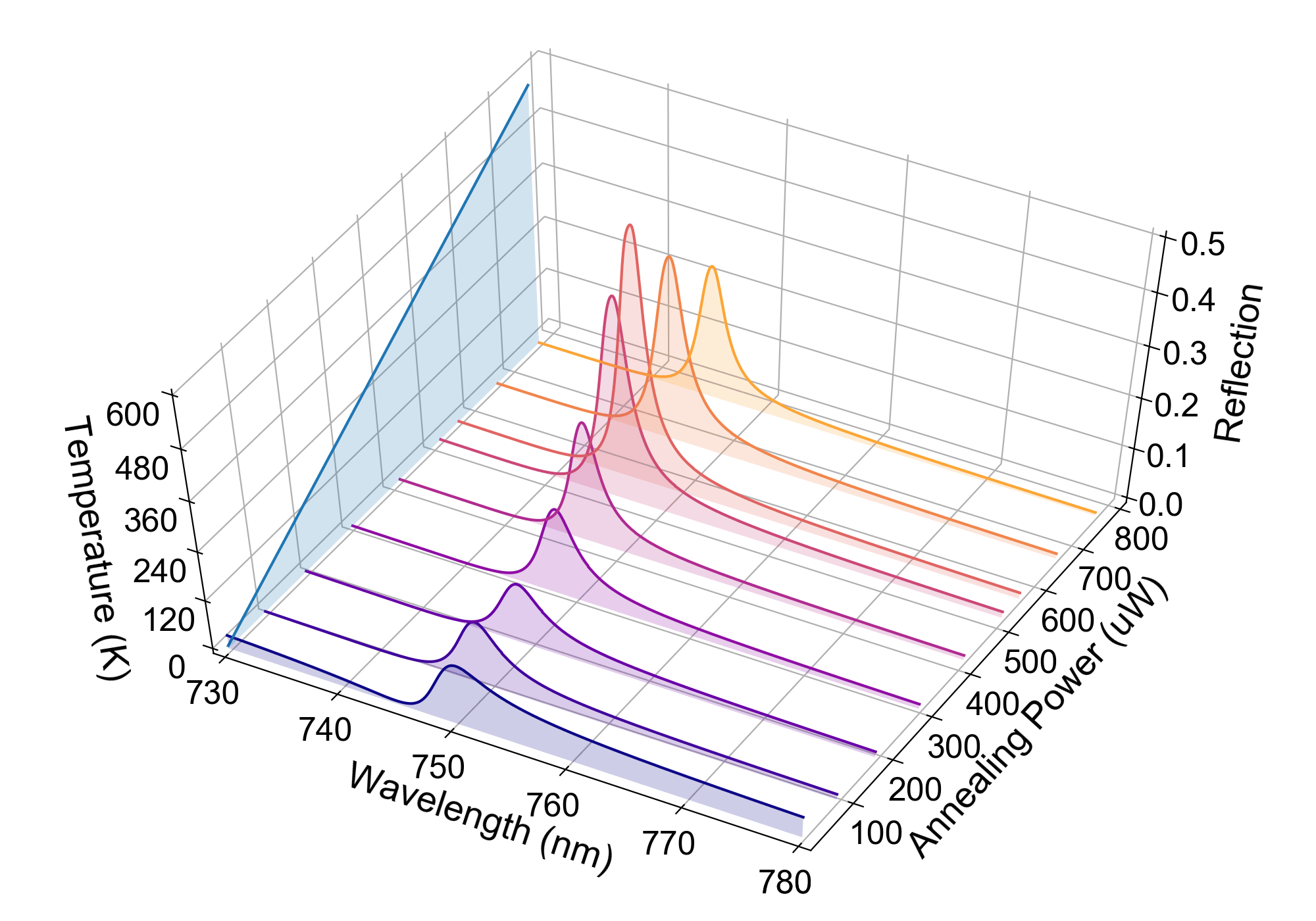}}

    \subfloat[ \label{fig:AnnealPower:TemperatureFWHM}]
    {\includegraphics[width=0.34\textwidth]
    {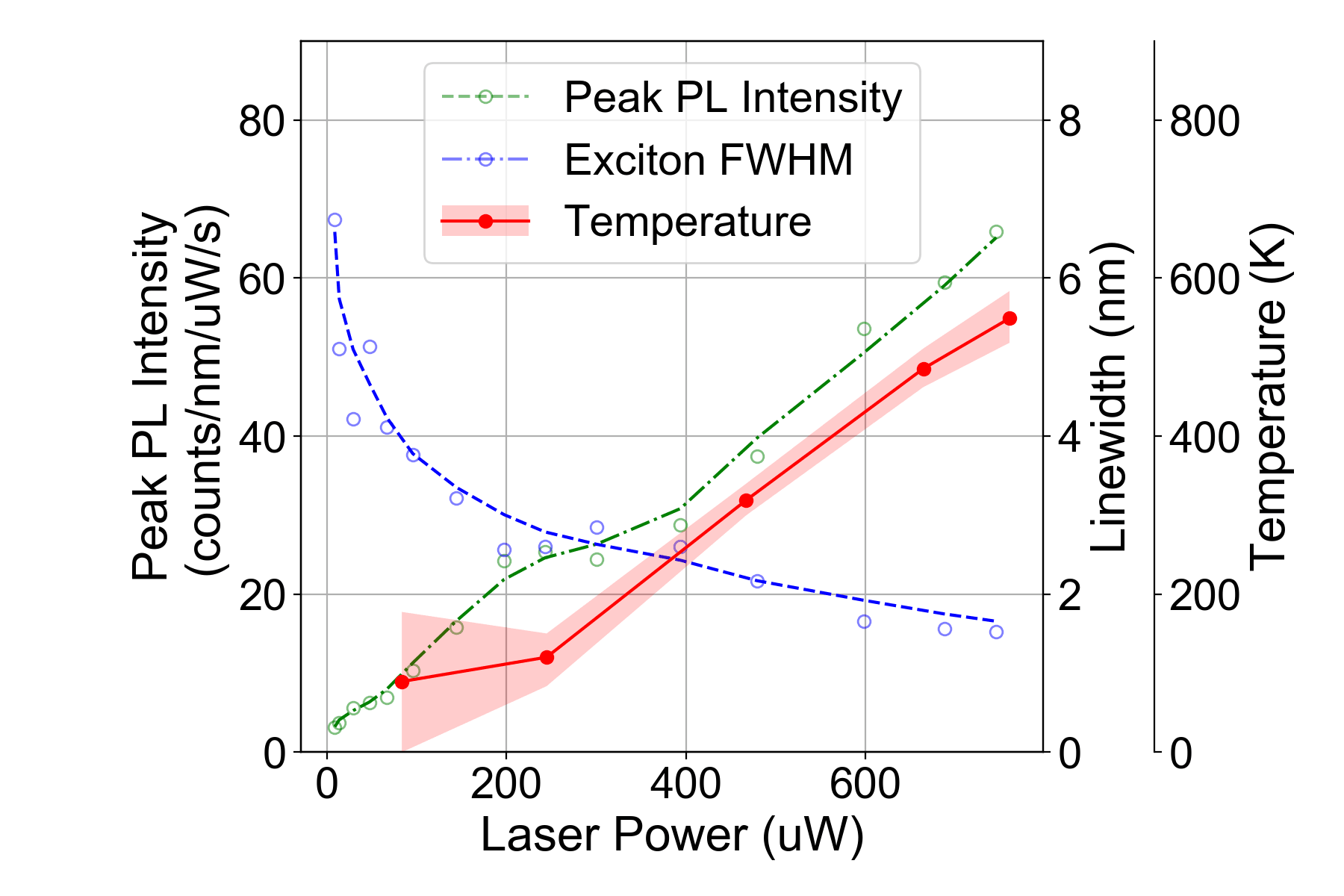}}
    \subfloat[ \label{fig:AnnealSiO2:SiO2_PL}]
    {\includegraphics[width=0.37\textwidth]
    {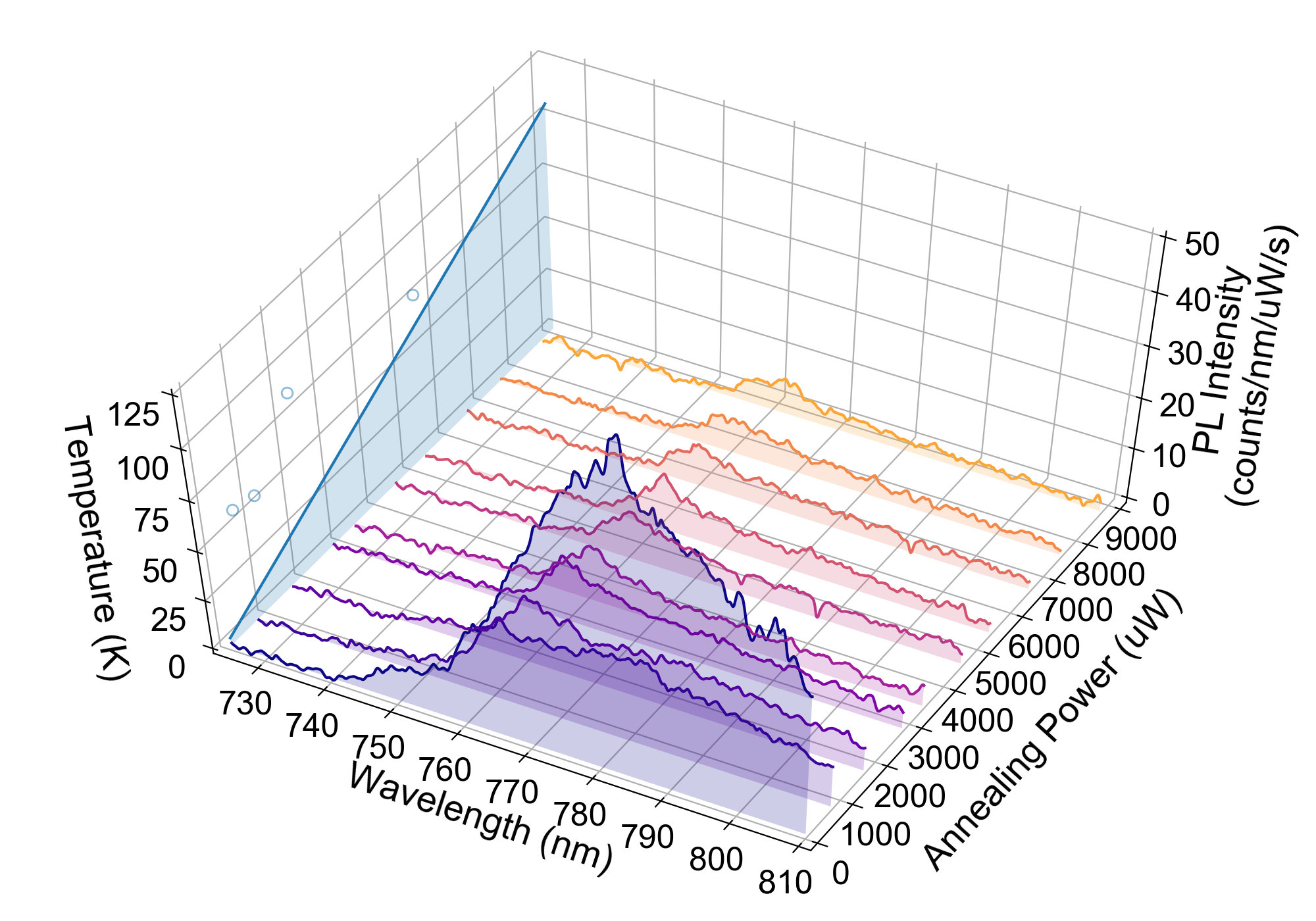}}
    \subfloat[ \label{fig:AnnealSiO2:Temperature}]
    {\includegraphics[width=0.25\textwidth]
    {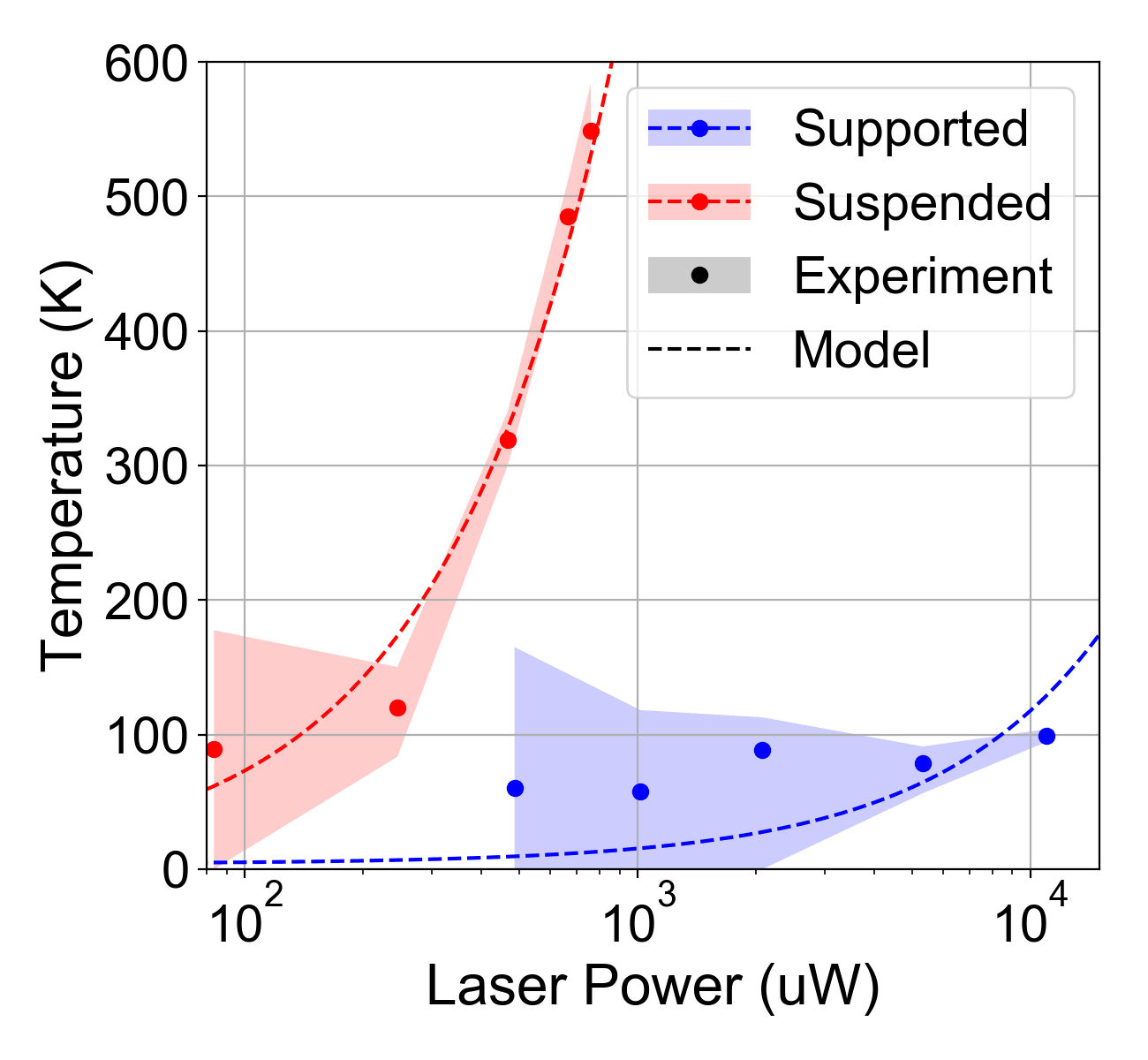}}
    \caption{Annealing power dependence of sample quality.
    The PL (a) and reflection (b) after annealing a suspended $\mathrm{MoSe_2}$ monolayer for successively increasing laser annealing power.
    Note that the PL and reflection are from different holes.
    Peak amplitude and width of the exciton PL (c), alongside the measured annealing temperature.
    The peak and width are extracted from the data in (a), and the lines shown are filtered versions of the data (circles) meant as a guide to the eye.
    PL after annealing a supported monolayer (d) on 280 nm $\mathrm{SiO_2}$ on Si for successively increasing powers.
    Measured and modeled temperature during annealing (e) for both supported and suspended samples.
    The shaded area represents the uncertainty in the temperature measurement.
    The raw reflection spectra corresponding to the fitted reflection spectra are shown in \ref{sfig:AnnealPower:ReflectionAnneal}.
    \label{fig:AnnealPower}}
\end{figure*}

When raster scanning the annealing beam, it is important that the final annealing position of the beam be at or near the center of the hole.
This is likely due to strain effects.
When the sample is cold, we hypothesize that the flake is essentially pinned on the substrate.
However, when supported areas of the flake (especially those near the edge of the hole) are heated during the annealing, the flake can more readily slide across the substrate, changing the strain distribution.
When annealing in the center, the areas of the flake at the edge of the hole are equally mobile.
This ensures that the strain gradient in the suspended flake is minimized.
If the final anneal spot is not centered, the flake could be more mobile on one side of the hole than on another, creating a non-uniform tensile force pulling from the edge of the hole.
This would result in a strain gradient in the suspended monolayer.
This strain gradient effect based upon the location of the final anneal spot is very repeatable --- that is, after a full raster anneal one can alternately anneal near the edge of the sample and then in the middle, and the line position homogeneity changes accordingly.
See Fig. \ref{sfig:RasterAnneal} for related plots of the PL homogeneity when the final anneal spot is not in the center, and for the repeatability of this effect.

Annealing significantly improved sample quality on every spot investigated ($>40$).
However, there was variability in line position, linewidth and peak reflection/PL intensity.
Fig. \ref{fig:MultiHole} shows reflection spectra both before and after annealing at 12 spots, as well as  PL before and after for 12 different spots.
Note that the holes were not pre-or post-selected in any way, other than to choose holes that did in fact have monolayer suspended $\mathrm{MoSe_2}$.
The data collected at each hole was post-selected in that each hole was annealed with successively higher laser power and the qualitatively best spectrum from each hole was selected.
There is some variation in the optimal annealing power from hole to hole, which is discussed further below.

Reflection shows a marked change; in some cases the peak reflection increases by as much as  $5\times$ and in all cases improves significantly.
Before annealing, there is a quite large reflectance from the suspended film, in some cases above 5\%.
For such a broadband reflection, this is much larger than one would expect from $\mathrm{MoSe_2}$ alone \cite{OpticalDielectricFunctionOfmTMDCs}.
We attribute this to a relatively thick layer of hydrocarbon contaminants on the suspended flake.
The much lower broadband reflection after annealing indicates the removal of this layer.
X-ray-photoelectron spectroscopy (XPS) measurements indicate that there is a hydrocarbon layer as thick as 5 nm on the $\mathrm{MoSe_2}$.
See Fig. \ref{sfig:XPS} for more discussion, and XPS data.

In PL, we see that both the $X^-$ and $X^D$ emission are reduced to below the noise floor for all 12 spots.
The PL peak intensity increases for almost all spots, and the linewidth decreases by a factor of two for most spots.
There is also a consistent blue shift of 10-15 meV after annealing.
This is most likely due to a reduction in tensile strain \cite{BandgapEngineeringOfStrainedMoS2}, but could also be related to removing certain contaminants and dopants from the surface, possibly changing the effective dielectric environment seen by the exciton.

Samples exfoliated from bulk were also annealed, and behaved similarly to the CVD samples presented here.
See Fig. \ref{sfig:BeforeAfterSiN} for more details.

The evolution of the annealed PL and reflection is shown as a function of annealing power in Fig. \ref{fig:AnnealPower}.
The $X^-$ and $X^D$ emission disappear at relatively low powers, and the red shift of the exciton primarily occurs at comparable or even lower powers.
However, the PL linewidth and peak intensity improve to their best values only at significantly higher powers.
The reflection linewidth and peak follow similar trends.
The fact that the trion and defect emission are suppressed at relatively low annealing powers could indicate that the doping and defect emission are due to a species of adsorbate that binds less tightly to the layer.
However, it is also possible that the linewidth and peak intensity only improve at higher powers simply because they are more sensitive to small amounts of contaminants.
\begin{figure*}
    \centering
    \subfloat[ \label{fig:AnnealTime:PL}]
    {\includegraphics[width=0.45\textwidth]
    {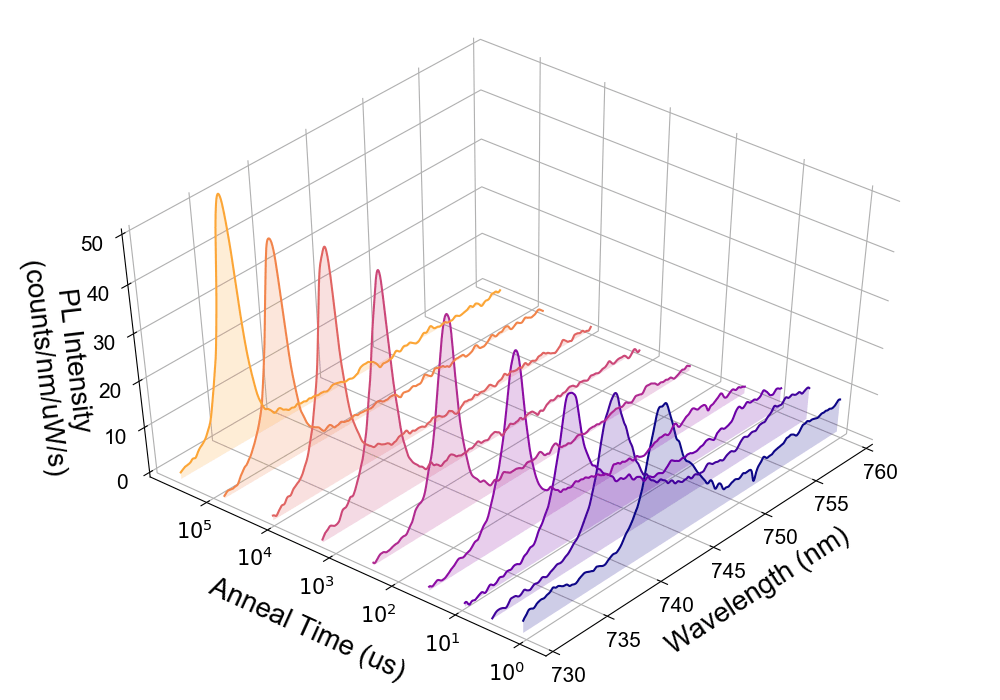}}
    \subfloat[ \label{fig:AnnealTime:Reflection}]
    {\includegraphics[width=0.45\textwidth]
    {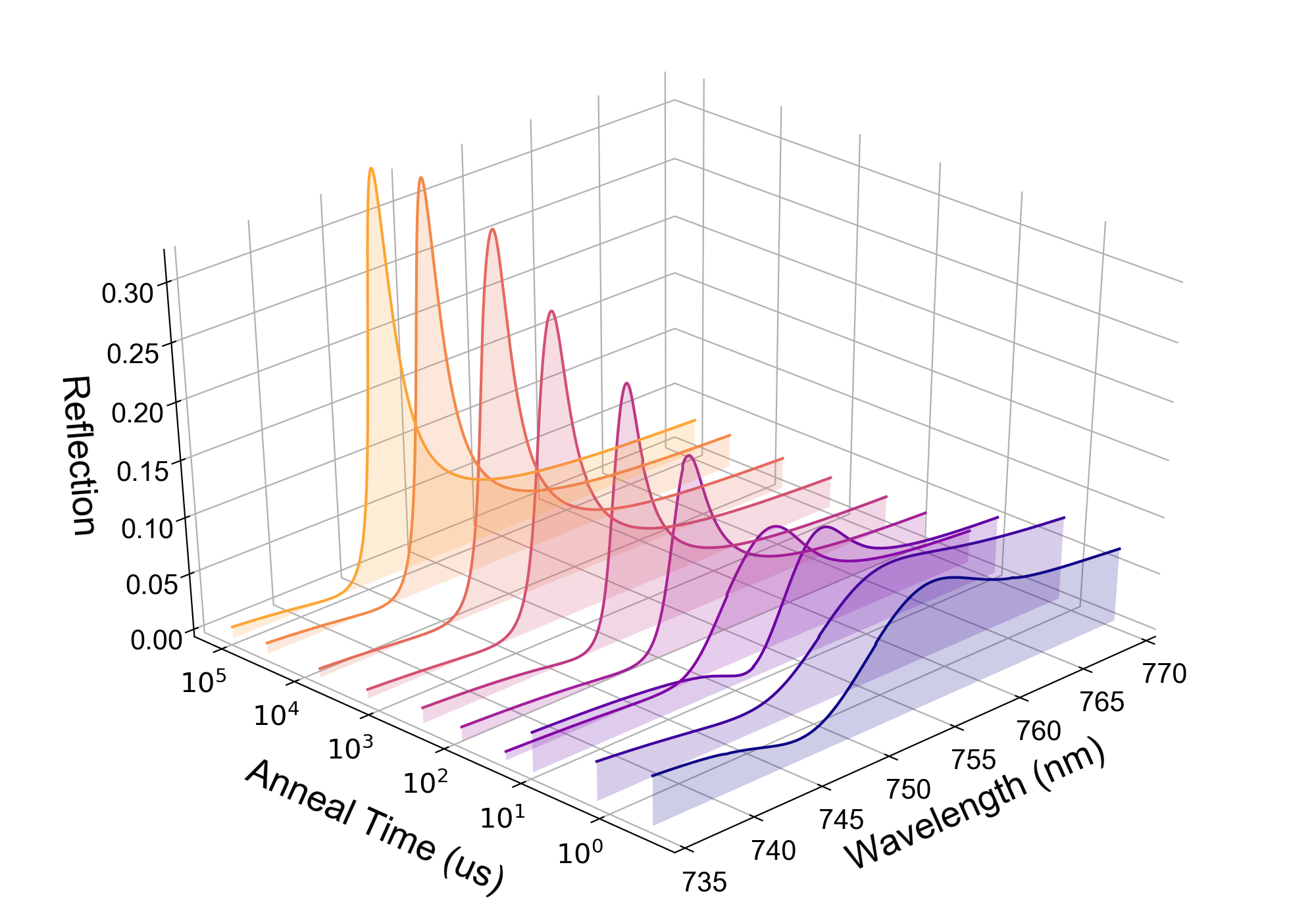}}

    \subfloat[ \label{fig:AnnealTime:PL_FWHM}]
    {\includegraphics[width=0.35\textwidth]
    {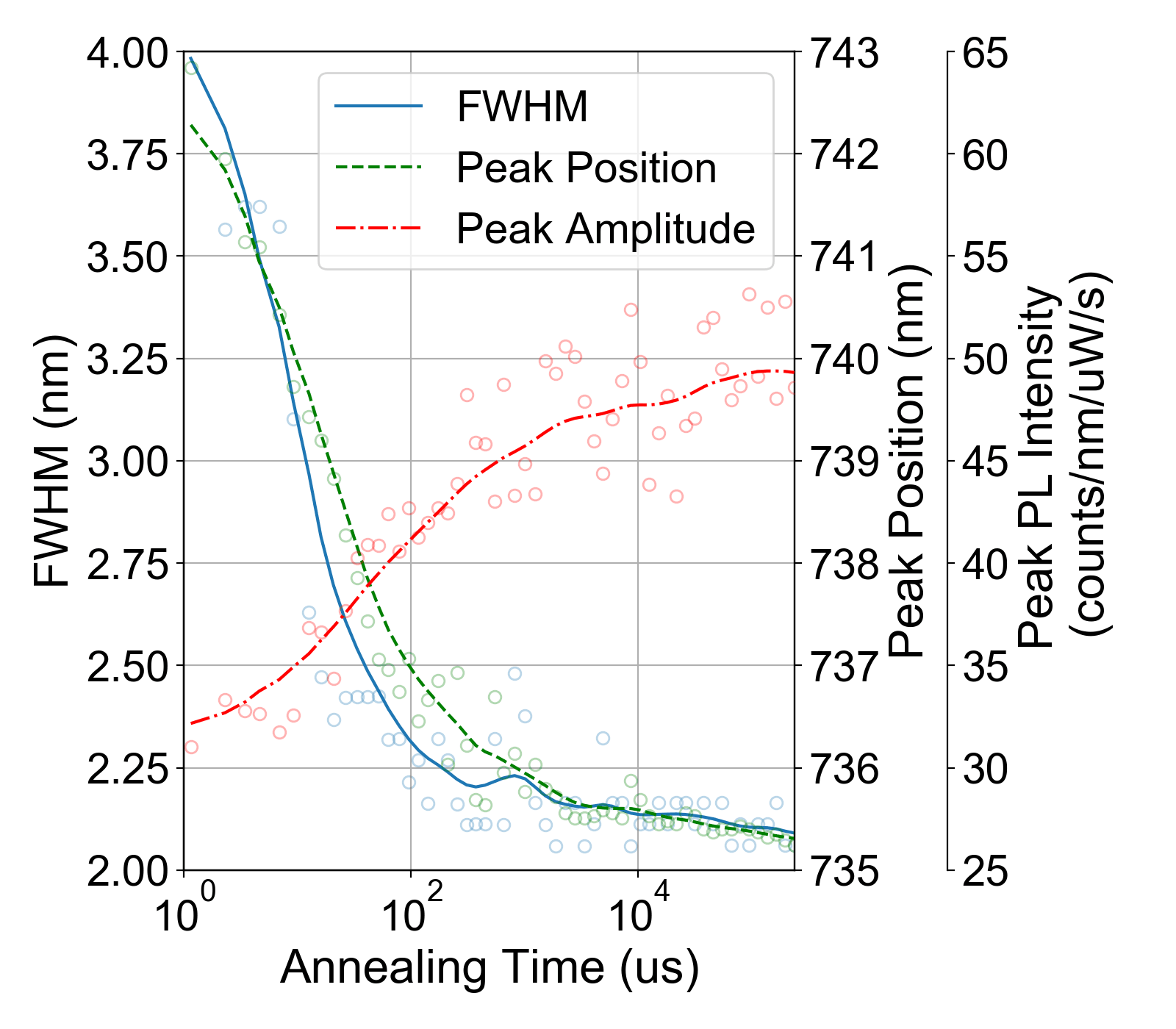}}
    \subfloat[ \label{fig:AnnealTime:ReflectionFWHM}]
    {\includegraphics[width=0.35\textwidth]
    {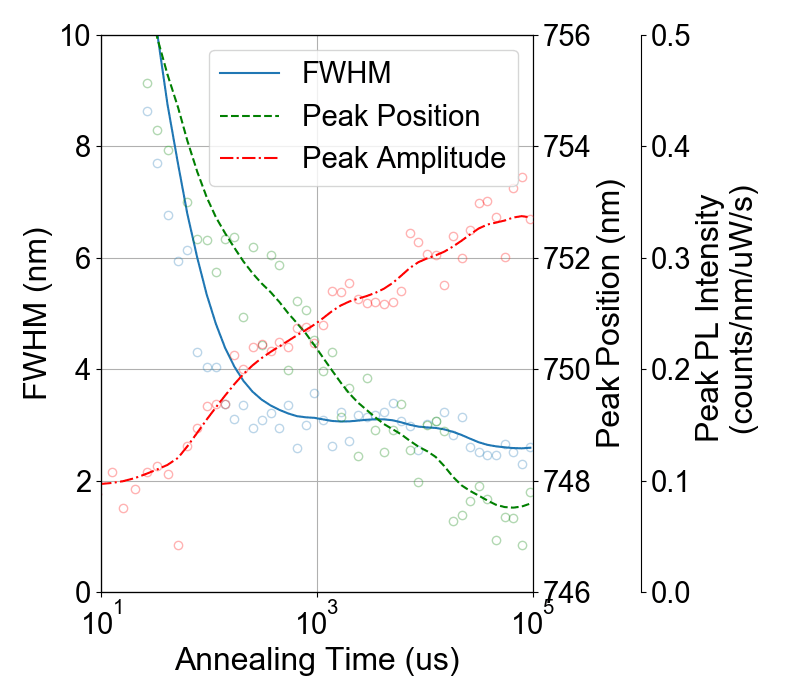}}
    \caption{Temporal evolution of the annealing process.
    Behavior as a function of total annealing time of the PL (a) and reflection (b) for a CVD-grown suspended $\mathrm{MoSe_2}$ monolayer.
    The flake was annealed in \SI{1}{\micro\second} increments, with a laser power of \SI{500}{\micro\watt} at 532 nm.
    Spectra were taken at selected intervals.
    The behavior of the extracted peak position, amplitude and FWHM for PL (c) and for reflection (d).
    The lines in (c) and (d) are filtered versions of the data, shown by the dots, and serve as a guide to the eye.
    Corresponding plots of the unfitted reflection spectra for (b) and (d) are shown in Fig. \ref{sfig:AnnealTime}
    \label{fig:AnnealTime}}
\end{figure*}

When annealing with a nominally 700 nm diameter spot, the optimal annealing power (as measured by the power at which the PL/reflection linewidths are narrowest) varied from \SI{400}{\micro\watt} to \SI{800}{\micro\watt} for the CVD grown samples.
Some, but not all, of this variation is likely due to small changes in the focusing conditions between different anneals.
As the power was increased, some suspended flakes eventually began to degrade in quality.
Others would snap and break before showing any signs of degradation.
The PL in Fig. \ref{fig:AnnealPower:PL_Anneal} is an example of a film that snapped before degrading, while the reflection in Fig. \ref{fig:AnnealPower:ReflectionAnneal} is an example of a film that degraded before snapping.
We posit that the main mode of failure is tearing of the flake due to tensile forces --- those flakes with a higher concentration of defects as well as those under more tensile stress from their boundary conditions would break at lower temperatures and lower annealing powers.

The annealing temperature of the suspended $\mathrm{MoSe_2}$ measured using Raman thermometry \cite{RamanEffectInCrystals} is shown in \ref{fig:AnnealPower:TemperatureFWHM}.
The maximum temperature for this anneal was about 550 K, and as expected, the temperature varies linearly with excitation power.
The PL intensity almost linearly tracks the excitation power.

For comparison, the PL after different-strength anneals and the associated annealing temperatures is shown for a supported flake in Fig. \ref{fig:AnnealSiO2:SiO2_PL}.
For the same annealing power, the supported flake does not reach as high temperature during annealing as the suspended flake because of heat conduction into the substrate.
The PL of the supported $\mathrm{MoSe_2}$ nowhere shows any strong, narrow peak that could definitively be identified as exciton or trion emission.
Rather, before annealing it has an extremely broad emission which slowly decreases and somewhat narrows as the annealing power is increased.
While at the highest annealing powers there is an extremely faint peak which might in some way be connected to an exciton, the annealing is clearly not effective for supported samples.
This could be due to contaminants trapped between the substrate and the $\mathrm{MoSe_2}$ that cannot be removed by heating.
There may also be roughness and strain pinning due to the substrate which is not affected by the annealing.

The PL and reflection as a function of annealing time are shown in Fig. \ref{fig:AnnealTime}.
The annealing power is kept fixed, and the sample is annealed with \SI{1}{\micro\second} pulses.
Since the in-plane speed of sound in monolayer $\mathrm{MoSe_2}$ should be similar to the $\sim 600$ m/s speed of $\mathrm{MoS_2}$ \cite{LatticeVibrationalModesOfMoS2}, the time to reach thermal equilibrium for a hole with radius $\sim$ \SI{4}{\micro\meter} is on the order of 7 ns, which is much shorter than the pulse length.
This is experimentally supported by the observation that after fully annealing using many pulses, annealing using a CW beam of the same power does not change the PL/reflection of the sample.
We can see that the PL improves quite quickly over the first ms, and afterwards changes much more slowly.
The blue shift and reduction in broad defect PL both happen even more quickly, in about 10 and 50 \SI{}{\micro\second}, respectively.
This is complementary to the fact that the blue shift of the exciton and trion/defect emission suppression occur at low powers in Fig. \ref{fig:AnnealPower:PL_Anneal}.
On the other hand, the reflection continues to improve even after 10 ms of annealing.
This indicates that the peak reflectance is more sensitive and thus a better metric of sample quality than PL.
This is supported by \cite{LargeExcitonicReflectivity, RealizationOfAnElectricallyTunableMirror}, in which the peak reflectance is shown to be a measure of the ratio of radiative to non-radiative broadening.

\section{\label{sec:Conclusion} Conclusions}
The results presented here strongly indicate that the intrinsic quality of CVD-grown $\mathrm{MoSe_2}$, and by extension other TMDCs including $\mathrm{MoS_2}$, $\mathrm{WS_2}$, and $\mathrm{WSe_2}$ is comparably high to that of their mechanically-exfoliated-from-bulk counterparts.
It seems necessary to completely remove contaminants that come from the growth and transfer processes, as well as from the ambient atmosphere.
We have shown here that sample quality can be repeatably and dramatically increased, in some cases reaching PL linewidths and peak reflectances close to those of the highest quality encapsulated samples.
The homogeneity of the samples appears to be limited by the size of the suspended area, offering a path forward for larger-area homogeneity than yet achieved in exfoliated samples.

We believe that this annealing procedure will enable further studies of exciton-polaritons, cavity QED with excitons, fundamental exciton physics and other many-body physics in TMDCs by providing a repeatable path to high-quality spatially-homogeneous TMDCs (and possibly all 2D materials).
This annealing procedure will greatly aid efforts to create high quality arrays of quantum dots for quantum information processing, as well as networks of optical cavities coupled to TMDCs.
The repeatability of the annealing as well as the large area and flexibility afforded by CVD-grown samples provides an avenue for scaling to multiple devices.
Finally, the annealing procedure may also prove useful for electronic applications through an improvement in carrier mobility.

\section{\label{sec:Methods} Methods}
\subsection{\label{subsec:Methods:Samples} Sample Preparation}
The suspended CVD samples were purchased from 2DLayer.
Their process involves growing monolayer $\mathrm{MoSe_2}$ by CVD, and then transferring to 280 nm SiO$_2$ on Si substrates with patterned holes.
The holes were approximately \SI{20}{\micro\meter} deep and nominally \SI{8}{\micro\meter} in diameter.
The data in all figures except Fig. \ref{sfig:BeforeAfterSiN} was taken using these samples.

A suspended sample was also prepared by starting with bulk $\mathrm{MoSe_2}$ from 2D Semiconductor.
The flake was exfoliated on to polydimethylsiloxane (PDMS) \cite{DeterministicTransfer}, and then directly transferred on to a holey silicon nitride grid from SPI Supplies (2 micron diameter holes in a 50 nm thick silicon nitride membrane).
Data from this sample is shown in Fig. \ref{sfig:BeforeAfterSiN}.

\subsection{\label{subsec:Methods:Setup} Experimental Setup}
All PL and reflection measurements were performed in a Montana Instruments Cryostation at a nominal base temperature of 4 K.
The pressure was typically between $1\cdot 10^{-7}$ and $2\cdot 10^{-7}$ Torr.
The cryostat is equipped with an XYZ piezo stage stack from Attocube, on top of which the samples are mounted.
A custom microscope assembly looks into the cryostat using a $20\times$ Mitutoyu long-working-distance near-infra-red (NIR) objective with 0.7 mm of glass correction and a numerical aperture (NA) of 0.4.
A removable beamsplitter couples light to a wide-field imaging path.
See Fig. \ref{sfig:schematic}.
There is a 0.5 mm thick glass window on the cryostat, and a 0.2 mm thick glass inner window on the cryostat heat shield, both of which are anti-reflection coated.
An XY galvanometer pair is mounted directly behind the objective, which allows scanning of the spot over about \SI{50}{\micro\meter}$\times$\SI{50}{\micro\meter}.
The entire microscope is also mounted on motorized XY translation stages which allow for translation of several mm, as well as a manual Z-stage which is used for focusing the microscope on the sample.
Measurements were automated using the python instrument control package Instrumental, available on Github at \url{https://github.com/mabuchilab/Instrumental}.

\subsection{\label{subsec:Methods:PL} PL Measurements}
PL was measured by exciting using narrow linewidth ($<1$ cm$^{-1}$ FWHM) 532 nm continuous wave (CW) beam obtained by frequency doubling a 1064 nm pump laser in periodically polled LiNbO$_3$ (PPLN).
For the pump, a 10 mW 1064 nm seed from an NKT Koheras Adjustik laser was amplified using a Nufern nuamp amplifier.
The 532 nm excitation laser was coupled onto the microscope by a single mode optical fiber and a reflective collimator.
It is then coupled into the light path using a volume Bragg grating, which essentially functioned as a dichroic mirror (reflecting 532 nm while transmitting other wavelengths).
This separates the PL from the excitation.
Again, see Fig. \ref{sfig:schematic}.
The excitation is then coupled into another single mode fiber, and the spectrum is measured using a home-built spectrometer.
The spectrometer has an 1800 line/mm grating on a computer-controlled rotation stage, and spectra are measured on a Princeton Instruments PIXIS 2048 camera.
The nominal resolution of the spectrometer is about 1 cm$^{-1}$.
The excitation power is measured before each spectrum by a power meter on a computer-controlled flip mount in the microscope setup.
Typical excitation powers were about \SI{3}{\micro\watt} at the sample.
We did not observe significant changes in line shape when measuring at lower excitation powers.

\subsection{\label{subsec:Methods:Annealing} Annealing}
The annealing procedure was performed using the same 532 nm laser used for PL, except at much higher powers.
The annealing was effective over a large range of spot sizes (we tested from 700 nm to \SI{10}{\micro\meter}).
As expected, the larger spots required correspondinly more laser power for optimal annealing.
We typically annealed using a 700 nm spot for several seconds, although from Fig. \ref{fig:AnnealTime} it can be seen that the annealing itself occurs in the first few ms.
For a beam size of 700 nm the optimal annealing power the optimal annealing power was typically between \SI{400}{\micro\watt} to \SI{800}{\micro\watt}.

\subsection{\label{subsec:Methods:Reflection} Reflection Measurements}
Reflection was measured using a Thorlabs SLS201 broadband stabilized light source.
Again, the incident beam was coupled on to the microscope through a single mode fiber.
The power at the sample in the range of 700-800 nm was about 5 nW.
The reflected beam was separated from the incident beam by a non-polarizing beamsplitter, and was coupled into a single mode fiber.
Again, see Fig. \ref{sfig:schematic}.
The spectrum was measured using the same spectrometer as for PL.
Each reflection spectrum was calibrated by referencing it to the reflection of the bare substrate.
The absolute reflection of the bare substrate was measured by in turn referencing to a silver mirror.

\subsection{\label{subsec:Methods:ReflectionAnalysis} Reflection Analysis}
When measuring reflection spectra from near the center of a suspended monolayer, the small back reflection (between 1\% and 3\% in most cases) from the bottom of the hole caused etalon interference fringes to appear in the measured spectrum.
Even though the reflection was small, the etalon interference fringes were much larger, up to 10\%.
These fringes make the raw spectra difficult to interpret.
Note that this was not an issue for the sample suspended on silicon nitride, since the holes in this case were through holes.

We fit the reflection spectra to a multilayer Fresnel model, which included a reflection from the bottom of the hole to account for the etalon.
The dielectric constant of the film was the sum of a real background permittivity and an inhomogenously broadened Lorentzian oscillator to account for interaction with the exciton.
Reflection from the trion is ignored, since it is assumed to contribute negligibly to the reflection.
This is accurate for all but the highest carrier doping levels \cite{FermiPolaronPolaritons}.
The background permittivity is meant to account for both the background permittivity of the $\mathrm{MoSe_2}$ and any (possibly thick) layer of contaminants.
The dielectric permittivity for a Lorentzian oscillator is:
\begin{equation}
\epsilon_r =  \left[ 1 + \frac{\omega_p^2}{(\omega_0 - \omega^2) + i \gamma \omega } \right]
\label{eq:lorentz}
\end{equation}
where $\epsilon_r$ is the relative permittivity, $\omega$ is the optical frequency, $\omega_0$ is the center frequency, $\gamma$ represents homogenous broadening, and $\omega_p$ is related to the strength of the oscillator.
The permittivity of the exciton is obtained by convolving this Lorentzian with a Gaussian to account for inhomogeneous broadening.

The etalon length (hole depth), etalon reflection, oscillator center frequency, oscillator strength, oscillator homogenous broadening width, oscillator inhomogeneous broadening width, and film background permittivity were all fitted parameters.
Once this model was fit to the raw data, the reflection from the suspended film alone could be extracted from the fitting parameters.
Examples of the fits to raw data along with the raw data and extracted spectra are shown in Fig. \ref{sfig:FitExample}.
The raw spectra for each of the figures in the main text that shows fitted reflection data is also shown in the SI.

\subsection{\label{subsec:Methods:ThermalModel} Thermal Model}
The peak temperature during annealing is modeled using a steady-state radial heat equation with the absorption of the laser in the $\mathrm{MoSe_2}$ treated as a point heat source.
The annealing temperature is calculated as the temperature at one beam radius away from the point source.
In the suspended case we set the temperature boundary condition at the edge of the hole to be the nominal base temperature of the cryostat, 4 K.
With $u(r)$ as the radial temperature distribution, $q(r)$ as the heat flux density and $k$ the thermal conductivity, the steady state heat equation is:
$$  $$
\begin{equation}
-k \nabla^2 u(r) = q(r)
\label{eq:heatequation}
\end{equation}
In the suspended 2D case with the boundary conditions above, the temperature distribution is:
\begin{equation}
\label{eq:2Dheat}
u(r) = \frac{T_1 - T_0}{\ln\left(\frac{r_1}{r_0}\right)} \ln\left(\frac{r}{r_0}\right) + T_0
\end{equation}
where $T_1$ is the temperature at radius $r_1$ (the beam radius), and $T_0$ is the temperature at $r_0$ (the edge of the hole).
For a given input heat flux $Q$, the annealing temperature is then:
\begin{equation}
T_1 = \frac{Q \ln\left(\frac{r_0}{r_1}\right) }{2 \pi k t_\textrm{flake}} + T_0
\label{eq:2DTemperature}
\end{equation}
where $t_\mathrm{flake}$ is the thickness of the $\mathrm{MoSe_2}$.
We use a thermal conductivity of 40 W/m$\cdot$K for the $\mathrm{MoSe_2}$, which is consistent with both theory and experiment \cite{ThermalConductivityOfMoSe2, LateralThermalConductivityOfMoSe2}.

The supported case follows a similar analysis, except in this case we use infinite boundary conditions (the temperature goes to 4 K at large radius), and we assume that the thermal conductivity of the substrate dominates that of the flake.
In this case, the annealing temperature is:
\begin{equation}
T_1 = \frac{Q}{2\pi k r_1} + T_0
\label{eq:3DTemperature}
\end{equation}
We use a thermal conductivity of 2 W/m$\cdot$K for the substrate, which is intermediate between that of $\mathrm{SiO_2}$ and Si (but closer to that of $\mathrm{SiO_2}$) in the relevant temperature range \cite{ThermalConductivityOfSilicon, TemperatureDependenceOfThermalConductivityForSiO2}.
The laser spot diameter was $r_1 = 350$ nm, and the hole radius was $r_0=4$ \SI{}{\micro\meter}.
Note that both models above ignore any temperature dependence of the thermal conductivity.

\subsection{\label{subsec:Methods:Raman} Raman Thermometry}
Raman measurements were performed using the same 532 nm laser and the same beam path as for the PL measurements, except that an extra volume Bragg grating was used on the microscope to further knock out the excitation laser.
This was necessary to prevent fiber Raman from overwhelming the Raman scattering from the $\mathrm{MoSe_2}$, since the microscope was coupled to the spectrometer using a single mode fiber.
A different home-built spectrometer was used for Raman measurements.
This spectrometer has a 1200 line/mm blazed grating, and spectra are measured using a PIXIS 256 camera.
The nominal resolution is about 1.5 cm$^{-1}$.
There are two more volume Bragg gratings within the spectrometer, to further knock out the 532 nm excitation.

We used the $\mathrm{MoSe_2}$ raman peak at ~242 cm$^{-1}$ to perform Raman thermometry during the annealing process.
Since the annealing process was in most cases at relatively low laser powers, the scattering signal was very weak and typical spectra would involve up to 10 exposures of 20 minutes each.
The ratio between Stokes and anti-Stokes scattering is:
\begin{equation}
R = \frac{I_\text{Stokes}}{I_\text{Anti-Stokes}} = \left[ \frac{\nu_0 - \Delta\nu}{\nu_0 + \Delta\nu} \right]^4 \exp\left[\frac{h \Delta \nu}{k_BT}\right]
\label{eq:stokes_ratio}
\end{equation}
where $I_\text{Stokes}$ and $I_\text{Anti-Stokes}$ are respectively the Stokes and anti-Stokes intensities, $\nu_0$ is the laser frequency, $\Delta\nu$ is the Raman transition frequency, $T$ is the temperature, $h$ is Planck's constant and $k_B$ is the Boltzmann constant.
The temperature is then:
\begin{equation}
T = \frac{h \Delta \nu}{k_B  \ln \left[ R \left[ \frac{\nu_0 + \Delta\nu}{\nu_0 - \Delta\nu} \right]^4 \right] }
\label{eq:temperature}
\end{equation}
Some examples of the Raman spectra used in calculating the annealing temperature are shown in Fig. \ref{sfig:raman}.

\subsection{\label{subsec:Methods:XPS} XPS}
We took XPS data from a supported area, assuming that the contaminants are qualitatively similar on suspended and supported areas.
Data was taken on a PHI Versaprobe 3, on a monolayer of $\sim$ \SI{100}{\micro\meter} dimensions using a \SI{100}{\micro\meter} X-ray spot.
This instrument uses 1486 eV X-rays and is sensitive to the top $\sim$ \SI{10}{\nano\meter} of the sample.
We used the machine calibration to extract atomic abundances using MultiPak software.
A bulk sample of $\mathrm{MoSe_2}$ from HQ Graphene was used as a further calibration for the relative abundance of Mo and Se.
The raw XPS data for a supported flake is shown in Fig. \ref{sfig:XPS:Spectrum}.

\begin{acknowledgments}
This work was funded in part by the National Science Foundation (NSF) award PHY-1648807, and also by a seed grant from the Precourt Institute for Energy at Stanford University.
Part of this work was performed at the Stanford Nano Shared Facilities (SNSF), supported by the NSF under award ECCS-1542152.
CR, DG, and NB were supported in part by Stanford Graduate Fellowships.
CR was also supported in part by a Natural Sciences and Engineering Research Council of Canada doctoral postgraduate scholarship.
CR thanks Charles Hitzman for help performing and interpreting the XPS data.
CR also thanks Ozgur Burak Aslan for discussions of strain effects in TMDC materials.
\end{acknowledgments}

\section{\label{sec:Contributions} Author Contributions}
CR first noted the annealing effect, conceived the experiments, performed the experiments and performed the data analysis.
HM suggested raster scanning the anneal, and annealing with a large diameter beam.
CR and DG built the reflection/PL/raman microscope setup, as well as the custom spectrometers.
DG built the frequency doubling setup.
CR and NB automated the measurements.
All authors contributed to the manuscript.


%
%

\bibliography{LaserAnnealing}

\providecommand{\noopsort}[1]{}\providecommand{\singleletter}[1]{#1}%
\begin{thebibliography}{34}%
\makeatletter
\providecommand \@ifxundefined [1]{%
 \@ifx{#1\undefined}
}%
\providecommand \@ifnum [1]{%
 \ifnum #1\expandafter \@firstoftwo
 \else \expandafter \@secondoftwo
 \fi
}%
\providecommand \@ifx [1]{%
 \ifx #1\expandafter \@firstoftwo
 \else \expandafter \@secondoftwo
 \fi
}%
\providecommand \natexlab [1]{#1}%
\providecommand \enquote  [1]{``#1''}%
\providecommand \bibnamefont  [1]{#1}%
\providecommand \bibfnamefont [1]{#1}%
\providecommand \citenamefont [1]{#1}%
\providecommand \href@noop [0]{\@secondoftwo}%
\providecommand \href [0]{\begingroup \@sanitize@url \@href}%
\providecommand \@href[1]{\@@startlink{#1}\@@href}%
\providecommand \@@href[1]{\endgroup#1\@@endlink}%
\providecommand \@sanitize@url [0]{\catcode `\\12\catcode `\$12\catcode
  `\&12\catcode `\#12\catcode `\^12\catcode `\_12\catcode `\%12\relax}%
\providecommand \@@startlink[1]{}%
\providecommand \@@endlink[0]{}%
\providecommand \url  [0]{\begingroup\@sanitize@url \@url }%
\providecommand \@url [1]{\endgroup\@href {#1}{\urlprefix }}%
\providecommand \urlprefix  [0]{URL }%
\providecommand \Eprint [0]{\href }%
\providecommand \doibase [0]{http://dx.doi.org/}%
\providecommand \selectlanguage [0]{\@gobble}%
\providecommand \bibinfo  [0]{\@secondoftwo}%
\providecommand \bibfield  [0]{\@secondoftwo}%
\providecommand \translation [1]{[#1]}%
\providecommand \BibitemOpen [0]{}%
\providecommand \bibitemStop [0]{}%
\providecommand \bibitemNoStop [0]{.\EOS\space}%
\providecommand \EOS [0]{\spacefactor3000\relax}%
\providecommand \BibitemShut  [1]{\csname bibitem#1\endcsname}%
\let\auto@bib@innerbib\@empty
\bibitem [{\citenamefont {Mak}\ \emph {et~al.}(2010)\citenamefont {Mak},
  \citenamefont {Lee}, \citenamefont {Hone}, \citenamefont {Shan},\ and\
  \citenamefont {Heinz}}]{AtomicallyThinMoS2}%
  \BibitemOpen
  \bibfield  {author} {\bibinfo {author} {\bibfnamefont {Kin~Fai}\ \bibnamefont
  {Mak}}, \bibinfo {author} {\bibfnamefont {Changgu}\ \bibnamefont {Lee}},
  \bibinfo {author} {\bibfnamefont {James}\ \bibnamefont {Hone}}, \bibinfo
  {author} {\bibfnamefont {Jie}\ \bibnamefont {Shan}}, \ and\ \bibinfo {author}
  {\bibfnamefont {Tony~F.}\ \bibnamefont {Heinz}},\ }\bibfield  {title}
  {\enquote {\bibinfo {title} {Atomically thin ${\mathrm{{mos}}}_{2}$: A new
  direct-gap semiconductor},}\ }\href {\doibase 10.1103/PhysRevLett.105.136805}
  {\bibfield  {journal} {\bibinfo  {journal} {Phys. Rev. Lett.}\ }\textbf
  {\bibinfo {volume} {105}},\ \bibinfo {pages} {136805} (\bibinfo {year}
  {2010})}\BibitemShut {NoStop}%
\bibitem [{\citenamefont {Splendiani}\ \emph {et~al.}(2010)\citenamefont
  {Splendiani}, \citenamefont {Sun}, \citenamefont {Zhang}, \citenamefont {Li},
  \citenamefont {Kim}, \citenamefont {Chim}, \citenamefont {Galli},\ and\
  \citenamefont {Wang}}]{EmergingPhotoluminescence}%
  \BibitemOpen
  \bibfield  {author} {\bibinfo {author} {\bibfnamefont {Andrea}\ \bibnamefont
  {Splendiani}}, \bibinfo {author} {\bibfnamefont {Liang}\ \bibnamefont {Sun}},
  \bibinfo {author} {\bibfnamefont {Yuanbo}\ \bibnamefont {Zhang}}, \bibinfo
  {author} {\bibfnamefont {Tianshu}\ \bibnamefont {Li}}, \bibinfo {author}
  {\bibfnamefont {Jonghwan}\ \bibnamefont {Kim}}, \bibinfo {author}
  {\bibfnamefont {Chi-Yung}\ \bibnamefont {Chim}}, \bibinfo {author}
  {\bibfnamefont {Giulia}\ \bibnamefont {Galli}}, \ and\ \bibinfo {author}
  {\bibfnamefont {Feng}\ \bibnamefont {Wang}},\ }\bibfield  {title} {\enquote
  {\bibinfo {title} {Emerging photoluminescence in monolayer mos2},}\ }\href
  {\doibase 10.1021/nl903868w} {\bibfield  {journal} {\bibinfo  {journal} {Nano
  Letters}\ }\textbf {\bibinfo {volume} {10}},\ \bibinfo {pages} {1271--1275}
  (\bibinfo {year} {2010})}\BibitemShut {NoStop}%
\bibitem [{\citenamefont {Jones}\ \emph {et~al.}(2013)\citenamefont {Jones},
  \citenamefont {Yu}, \citenamefont {Ghimire}, \citenamefont {Wu},
  \citenamefont {Aivazian}, \citenamefont {Ross}, \citenamefont {Zhao},
  \citenamefont {Yan}, \citenamefont {Mandrus}, \citenamefont {Xiao},
  \citenamefont {Yao},\ and\ \citenamefont {Xu}}]{ExcitonValleyCoherence}%
  \BibitemOpen
  \bibfield  {author} {\bibinfo {author} {\bibfnamefont {Aaron~M.}\
  \bibnamefont {Jones}}, \bibinfo {author} {\bibfnamefont {Hongyi}\
  \bibnamefont {Yu}}, \bibinfo {author} {\bibfnamefont {Nirmal~J.}\
  \bibnamefont {Ghimire}}, \bibinfo {author} {\bibfnamefont {Sanfeng}\
  \bibnamefont {Wu}}, \bibinfo {author} {\bibfnamefont {Grant}\ \bibnamefont
  {Aivazian}}, \bibinfo {author} {\bibfnamefont {Jason~S.}\ \bibnamefont
  {Ross}}, \bibinfo {author} {\bibfnamefont {Bo}~\bibnamefont {Zhao}}, \bibinfo
  {author} {\bibfnamefont {Jiaqiang}\ \bibnamefont {Yan}}, \bibinfo {author}
  {\bibfnamefont {David~G.}\ \bibnamefont {Mandrus}}, \bibinfo {author}
  {\bibfnamefont {Di}~\bibnamefont {Xiao}}, \bibinfo {author} {\bibfnamefont
  {Wang}\ \bibnamefont {Yao}}, \ and\ \bibinfo {author} {\bibfnamefont
  {Xiaodong}\ \bibnamefont {Xu}},\ }\bibfield  {title} {\enquote {\bibinfo
  {title} {Optical generation of excitonic valley coherence in monolayer
  wse2},}\ }\href {http://dx.doi.org/10.1038/nnano.2013.151} {\bibfield
  {journal} {\bibinfo  {journal} {Nature Nanotechnology}\ }\textbf {\bibinfo
  {volume} {8}},\ \bibinfo {pages} {634 EP --} (\bibinfo {year}
  {2013})}\BibitemShut {NoStop}%
\bibitem [{\citenamefont {Xiao}\ \emph {et~al.}(2012)\citenamefont {Xiao},
  \citenamefont {Liu}, \citenamefont {Feng}, \citenamefont {Xu},\ and\
  \citenamefont {Yao}}]{CoupledSpinValleyPhysics}%
  \BibitemOpen
  \bibfield  {author} {\bibinfo {author} {\bibfnamefont {Di}~\bibnamefont
  {Xiao}}, \bibinfo {author} {\bibfnamefont {Gui-Bin}\ \bibnamefont {Liu}},
  \bibinfo {author} {\bibfnamefont {Wanxiang}\ \bibnamefont {Feng}}, \bibinfo
  {author} {\bibfnamefont {Xiaodong}\ \bibnamefont {Xu}}, \ and\ \bibinfo
  {author} {\bibfnamefont {Wang}\ \bibnamefont {Yao}},\ }\bibfield  {title}
  {\enquote {\bibinfo {title} {Coupled spin and valley physics in monolayers of
  ${\mathrm{mos}}_{2}$ and other group-vi dichalcogenides},}\ }\href {\doibase
  10.1103/PhysRevLett.108.196802} {\bibfield  {journal} {\bibinfo  {journal}
  {Phys. Rev. Lett.}\ }\textbf {\bibinfo {volume} {108}},\ \bibinfo {pages}
  {196802} (\bibinfo {year} {2012})}\BibitemShut {NoStop}%
\bibitem [{\citenamefont {Mak}\ \emph {et~al.}(2012{\natexlab{a}})\citenamefont
  {Mak}, \citenamefont {He}, \citenamefont {Shan},\ and\ \citenamefont
  {Heinz}}]{ControlOfValleyPolarization}%
  \BibitemOpen
  \bibfield  {author} {\bibinfo {author} {\bibfnamefont {Kin~Fai}\ \bibnamefont
  {Mak}}, \bibinfo {author} {\bibfnamefont {Keliang}\ \bibnamefont {He}},
  \bibinfo {author} {\bibfnamefont {Jie}\ \bibnamefont {Shan}}, \ and\ \bibinfo
  {author} {\bibfnamefont {Tony~F.}\ \bibnamefont {Heinz}},\ }\bibfield
  {title} {\enquote {\bibinfo {title} {Control of valley polarization in
  monolayer mos2 by optical helicity},}\ }\href
  {http://dx.doi.org/10.1038/nnano.2012.96} {\bibfield  {journal} {\bibinfo
  {journal} {Nature Nanotechnology}\ }\textbf {\bibinfo {volume} {7}},\
  \bibinfo {pages} {494 EP --} (\bibinfo {year}
  {2012}{\natexlab{a}})}\BibitemShut {NoStop}%
\bibitem [{\citenamefont {Conley}\ \emph {et~al.}(2013)\citenamefont {Conley},
  \citenamefont {Wang}, \citenamefont {Ziegler}, \citenamefont {Haglund},
  \citenamefont {Pantelides},\ and\ \citenamefont
  {Bolotin}}]{BandgapEngineeringOfStrainedMoS2}%
  \BibitemOpen
  \bibfield  {author} {\bibinfo {author} {\bibfnamefont {Hiram~J.}\
  \bibnamefont {Conley}}, \bibinfo {author} {\bibfnamefont {Bin}\ \bibnamefont
  {Wang}}, \bibinfo {author} {\bibfnamefont {Jed~I.}\ \bibnamefont {Ziegler}},
  \bibinfo {author} {\bibfnamefont {Richard~F.}\ \bibnamefont {Haglund}},
  \bibinfo {author} {\bibfnamefont {Sokrates~T.}\ \bibnamefont {Pantelides}}, \
  and\ \bibinfo {author} {\bibfnamefont {Kirill~I.}\ \bibnamefont {Bolotin}},\
  }\bibfield  {title} {\enquote {\bibinfo {title} {Bandgap engineering of
  strained monolayer and bilayer mos2},}\ }\href {\doibase 10.1021/nl4014748}
  {\bibfield  {journal} {\bibinfo  {journal} {Nano Letters}\ }\textbf {\bibinfo
  {volume} {13}},\ \bibinfo {pages} {3626--3630} (\bibinfo {year}
  {2013})}\BibitemShut {NoStop}%
\bibitem [{\citenamefont {You}\ \emph {et~al.}(2015)\citenamefont {You},
  \citenamefont {Zhang}, \citenamefont {Berkelbach}, \citenamefont {Hybertsen},
  \citenamefont {Reichman},\ and\ \citenamefont
  {Heinz}}]{ObservationOfBiexcitonsInMonolayerWSe2}%
  \BibitemOpen
  \bibfield  {author} {\bibinfo {author} {\bibfnamefont {Yumeng}\ \bibnamefont
  {You}}, \bibinfo {author} {\bibfnamefont {Xiao-Xiao}\ \bibnamefont {Zhang}},
  \bibinfo {author} {\bibfnamefont {Timothy~C.}\ \bibnamefont {Berkelbach}},
  \bibinfo {author} {\bibfnamefont {Mark~S.}\ \bibnamefont {Hybertsen}},
  \bibinfo {author} {\bibfnamefont {David~R.}\ \bibnamefont {Reichman}}, \ and\
  \bibinfo {author} {\bibfnamefont {Tony~F.}\ \bibnamefont {Heinz}},\
  }\bibfield  {title} {\enquote {\bibinfo {title} {Observation of biexcitons in
  monolayer wse2},}\ }\href {http://dx.doi.org/10.1038/nphys3324} {\bibfield
  {journal} {\bibinfo  {journal} {Nature Physics}\ }\textbf {\bibinfo {volume}
  {11}},\ \bibinfo {pages} {477 EP --} (\bibinfo {year} {2015})}\BibitemShut
  {NoStop}%
\bibitem [{\citenamefont {Sidler}\ \emph {et~al.}(2016)\citenamefont {Sidler},
  \citenamefont {Back}, \citenamefont {Cotlet}, \citenamefont {Srivastava},
  \citenamefont {Fink}, \citenamefont {Kroner}, \citenamefont {Demler},\ and\
  \citenamefont {Imamoglu}}]{FermiPolaronPolaritons}%
  \BibitemOpen
  \bibfield  {author} {\bibinfo {author} {\bibfnamefont {Meinrad}\ \bibnamefont
  {Sidler}}, \bibinfo {author} {\bibfnamefont {Patrick}\ \bibnamefont {Back}},
  \bibinfo {author} {\bibfnamefont {Ovidiu}\ \bibnamefont {Cotlet}}, \bibinfo
  {author} {\bibfnamefont {Ajit}\ \bibnamefont {Srivastava}}, \bibinfo {author}
  {\bibfnamefont {Thomas}\ \bibnamefont {Fink}}, \bibinfo {author}
  {\bibfnamefont {Martin}\ \bibnamefont {Kroner}}, \bibinfo {author}
  {\bibfnamefont {Eugene}\ \bibnamefont {Demler}}, \ and\ \bibinfo {author}
  {\bibfnamefont {Atac}\ \bibnamefont {Imamoglu}},\ }\bibfield  {title}
  {\enquote {\bibinfo {title} {Fermi polaron-polaritons in charge-tunable
  atomically thin semiconductors},}\ }\href
  {http://dx.doi.org/10.1038/nphys3949} {\bibfield  {journal} {\bibinfo
  {journal} {Nature Physics}\ }\textbf {\bibinfo {volume} {13}},\ \bibinfo
  {pages} {255 EP --} (\bibinfo {year} {2016})}\BibitemShut {NoStop}%
\bibitem [{\citenamefont {Wang}\ \emph {et~al.}(2018)\citenamefont {Wang},
  \citenamefont {De~Greve}, \citenamefont {Jauregui}, \citenamefont {Sushko},
  \citenamefont {High}, \citenamefont {Zhou}, \citenamefont {Scuri},
  \citenamefont {Taniguchi}, \citenamefont {Watanabe}, \citenamefont {Lukin},
  \citenamefont {Park},\ and\ \citenamefont
  {Kim}}]{ElectricalControlofChargedCarriersInAtomicallyThin}%
  \BibitemOpen
  \bibfield  {author} {\bibinfo {author} {\bibfnamefont {Ke}~\bibnamefont
  {Wang}}, \bibinfo {author} {\bibfnamefont {Kristiaan}\ \bibnamefont
  {De~Greve}}, \bibinfo {author} {\bibfnamefont {Luis~A.}\ \bibnamefont
  {Jauregui}}, \bibinfo {author} {\bibfnamefont {Andrey}\ \bibnamefont
  {Sushko}}, \bibinfo {author} {\bibfnamefont {Alexander}\ \bibnamefont
  {High}}, \bibinfo {author} {\bibfnamefont {You}\ \bibnamefont {Zhou}},
  \bibinfo {author} {\bibfnamefont {Giovanni}\ \bibnamefont {Scuri}}, \bibinfo
  {author} {\bibfnamefont {Takashi}\ \bibnamefont {Taniguchi}}, \bibinfo
  {author} {\bibfnamefont {Kenji}\ \bibnamefont {Watanabe}}, \bibinfo {author}
  {\bibfnamefont {Mikhail~D.}\ \bibnamefont {Lukin}}, \bibinfo {author}
  {\bibfnamefont {Hongkun}\ \bibnamefont {Park}}, \ and\ \bibinfo {author}
  {\bibfnamefont {Philip}\ \bibnamefont {Kim}},\ }\bibfield  {title} {\enquote
  {\bibinfo {title} {Electrical control of charged carriers and excitons in
  atomically thin materials},}\ }\href {\doibase 10.1038/s41565-017-0030-x}
  {\bibfield  {journal} {\bibinfo  {journal} {Nature Nanotechnology}\ }\textbf
  {\bibinfo {volume} {13}},\ \bibinfo {pages} {128--132} (\bibinfo {year}
  {2018})}\BibitemShut {NoStop}%
\bibitem [{\citenamefont {Palacios-Berraquero}\ \emph
  {et~al.}(2017)\citenamefont {Palacios-Berraquero}, \citenamefont {Kara},
  \citenamefont {Montblanch}, \citenamefont {Barbone}, \citenamefont
  {Latawiec}, \citenamefont {Yoon}, \citenamefont {Ott}, \citenamefont
  {Loncar}, \citenamefont {Ferrari},\ and\ \citenamefont
  {Atat{\"u}re}}]{LargeScaleQuantumEmitterArrays}%
  \BibitemOpen
  \bibfield  {author} {\bibinfo {author} {\bibfnamefont {Carmen}\ \bibnamefont
  {Palacios-Berraquero}}, \bibinfo {author} {\bibfnamefont {Dhiren~M.}\
  \bibnamefont {Kara}}, \bibinfo {author} {\bibfnamefont {Alejandro R.~P}\
  \bibnamefont {Montblanch}}, \bibinfo {author} {\bibfnamefont {Matteo}\
  \bibnamefont {Barbone}}, \bibinfo {author} {\bibfnamefont {Pawel}\
  \bibnamefont {Latawiec}}, \bibinfo {author} {\bibfnamefont {Duhee}\
  \bibnamefont {Yoon}}, \bibinfo {author} {\bibfnamefont {Anna~K.}\
  \bibnamefont {Ott}}, \bibinfo {author} {\bibfnamefont {Marko}\ \bibnamefont
  {Loncar}}, \bibinfo {author} {\bibfnamefont {Andrea~C.}\ \bibnamefont
  {Ferrari}}, \ and\ \bibinfo {author} {\bibfnamefont {Mete}\ \bibnamefont
  {Atat{\"u}re}},\ }\bibfield  {title} {\enquote {\bibinfo {title} {Large-scale
  quantum-emitter arrays in atomically thin semiconductors},}\ }\href
  {http://dx.doi.org/10.1038/ncomms15093} {\bibfield  {journal} {\bibinfo
  {journal} {Nature Communications}\ }\textbf {\bibinfo {volume} {8}},\
  \bibinfo {pages} {15093 EP --} (\bibinfo {year} {2017})},\ \bibinfo {note}
  {article}\BibitemShut {NoStop}%
\bibitem [{\citenamefont {Mak}\ \emph {et~al.}(2012{\natexlab{b}})\citenamefont
  {Mak}, \citenamefont {He}, \citenamefont {Lee}, \citenamefont {Lee},
  \citenamefont {Hone}, \citenamefont {Heinz},\ and\ \citenamefont
  {Shan}}]{TightlyBoundTrions}%
  \BibitemOpen
  \bibfield  {author} {\bibinfo {author} {\bibfnamefont {Kin~Fai}\ \bibnamefont
  {Mak}}, \bibinfo {author} {\bibfnamefont {Keliang}\ \bibnamefont {He}},
  \bibinfo {author} {\bibfnamefont {Changgu}\ \bibnamefont {Lee}}, \bibinfo
  {author} {\bibfnamefont {Gwan~Hyoung}\ \bibnamefont {Lee}}, \bibinfo {author}
  {\bibfnamefont {James}\ \bibnamefont {Hone}}, \bibinfo {author}
  {\bibfnamefont {Tony~F.}\ \bibnamefont {Heinz}}, \ and\ \bibinfo {author}
  {\bibfnamefont {Jie}\ \bibnamefont {Shan}},\ }\bibfield  {title} {\enquote
  {\bibinfo {title} {Tightly bound trions in monolayer mos2},}\ }\href
  {http://dx.doi.org/10.1038/nmat3505} {\bibfield  {journal} {\bibinfo
  {journal} {Nature Materials}\ }\textbf {\bibinfo {volume} {12}},\ \bibinfo
  {pages} {207 EP --} (\bibinfo {year} {2012}{\natexlab{b}})}\BibitemShut
  {NoStop}%
\bibitem [{\citenamefont {Cadiz}\ \emph {et~al.}(2017)\citenamefont {Cadiz},
  \citenamefont {Courtade}, \citenamefont {Robert}, \citenamefont {Wang},
  \citenamefont {Shen}, \citenamefont {Cai}, \citenamefont {Taniguchi},
  \citenamefont {Watanabe}, \citenamefont {Carrere}, \citenamefont {Lagarde},
  \citenamefont {Manca}, \citenamefont {Amand}, \citenamefont {Renucci},
  \citenamefont {Tongay}, \citenamefont {Marie},\ and\ \citenamefont
  {Urbaszek}}]{ExcitonicLinewidthApproachingHomogenousLimit}%
  \BibitemOpen
  \bibfield  {author} {\bibinfo {author} {\bibfnamefont {F.}~\bibnamefont
  {Cadiz}}, \bibinfo {author} {\bibfnamefont {E.}~\bibnamefont {Courtade}},
  \bibinfo {author} {\bibfnamefont {C.}~\bibnamefont {Robert}}, \bibinfo
  {author} {\bibfnamefont {G.}~\bibnamefont {Wang}}, \bibinfo {author}
  {\bibfnamefont {Y.}~\bibnamefont {Shen}}, \bibinfo {author} {\bibfnamefont
  {H.}~\bibnamefont {Cai}}, \bibinfo {author} {\bibfnamefont {T.}~\bibnamefont
  {Taniguchi}}, \bibinfo {author} {\bibfnamefont {K.}~\bibnamefont {Watanabe}},
  \bibinfo {author} {\bibfnamefont {H.}~\bibnamefont {Carrere}}, \bibinfo
  {author} {\bibfnamefont {D.}~\bibnamefont {Lagarde}}, \bibinfo {author}
  {\bibfnamefont {M.}~\bibnamefont {Manca}}, \bibinfo {author} {\bibfnamefont
  {T.}~\bibnamefont {Amand}}, \bibinfo {author} {\bibfnamefont
  {P.}~\bibnamefont {Renucci}}, \bibinfo {author} {\bibfnamefont
  {S.}~\bibnamefont {Tongay}}, \bibinfo {author} {\bibfnamefont
  {X.}~\bibnamefont {Marie}}, \ and\ \bibinfo {author} {\bibfnamefont
  {B.}~\bibnamefont {Urbaszek}},\ }\bibfield  {title} {\enquote {\bibinfo
  {title} {Excitonic linewidth approaching the homogeneous limit in
  ${\mathrm{mos}}_{2}$-based van der waals heterostructures},}\ }\href
  {\doibase 10.1103/PhysRevX.7.021026} {\bibfield  {journal} {\bibinfo
  {journal} {Phys. Rev. X}\ }\textbf {\bibinfo {volume} {7}},\ \bibinfo {pages}
  {021026} (\bibinfo {year} {2017})}\BibitemShut {NoStop}%
\bibitem [{\citenamefont {Scuri}\ \emph {et~al.}(2018)\citenamefont {Scuri},
  \citenamefont {Zhou}, \citenamefont {High}, \citenamefont {Wild},
  \citenamefont {Shu}, \citenamefont {De~Greve}, \citenamefont {Jauregui},
  \citenamefont {Taniguchi}, \citenamefont {Watanabe}, \citenamefont {Kim},
  \citenamefont {Lukin},\ and\ \citenamefont
  {Park}}]{LargeExcitonicReflectivity}%
  \BibitemOpen
  \bibfield  {author} {\bibinfo {author} {\bibfnamefont {Giovanni}\
  \bibnamefont {Scuri}}, \bibinfo {author} {\bibfnamefont {You}\ \bibnamefont
  {Zhou}}, \bibinfo {author} {\bibfnamefont {Alexander~A.}\ \bibnamefont
  {High}}, \bibinfo {author} {\bibfnamefont {Dominik~S.}\ \bibnamefont {Wild}},
  \bibinfo {author} {\bibfnamefont {Chi}\ \bibnamefont {Shu}}, \bibinfo
  {author} {\bibfnamefont {Kristiaan}\ \bibnamefont {De~Greve}}, \bibinfo
  {author} {\bibfnamefont {Luis~A.}\ \bibnamefont {Jauregui}}, \bibinfo
  {author} {\bibfnamefont {Takashi}\ \bibnamefont {Taniguchi}}, \bibinfo
  {author} {\bibfnamefont {Kenji}\ \bibnamefont {Watanabe}}, \bibinfo {author}
  {\bibfnamefont {Philip}\ \bibnamefont {Kim}}, \bibinfo {author}
  {\bibfnamefont {Mikhail~D.}\ \bibnamefont {Lukin}}, \ and\ \bibinfo {author}
  {\bibfnamefont {Hongkun}\ \bibnamefont {Park}},\ }\bibfield  {title}
  {\enquote {\bibinfo {title} {Large excitonic reflectivity of monolayer
  ${\mathrm{mose}}_{2}$ encapsulated in hexagonal boron nitride},}\ }\href
  {\doibase 10.1103/PhysRevLett.120.037402} {\bibfield  {journal} {\bibinfo
  {journal} {Phys. Rev. Lett.}\ }\textbf {\bibinfo {volume} {120}},\ \bibinfo
  {pages} {037402} (\bibinfo {year} {2018})}\BibitemShut {NoStop}%
\bibitem [{\citenamefont {Back}\ \emph {et~al.}(2018)\citenamefont {Back},
  \citenamefont {Zeytinoglu}, \citenamefont {Ijaz}, \citenamefont {Kroner},\
  and\ \citenamefont {Imamo\ifmmode~\breve{g}\else
  \u{g}\fi{}lu}}]{RealizationOfAnElectricallyTunableMirror}%
  \BibitemOpen
  \bibfield  {author} {\bibinfo {author} {\bibfnamefont {Patrick}\ \bibnamefont
  {Back}}, \bibinfo {author} {\bibfnamefont {Sina}\ \bibnamefont {Zeytinoglu}},
  \bibinfo {author} {\bibfnamefont {Aroosa}\ \bibnamefont {Ijaz}}, \bibinfo
  {author} {\bibfnamefont {Martin}\ \bibnamefont {Kroner}}, \ and\ \bibinfo
  {author} {\bibfnamefont {Atac}\ \bibnamefont {Imamo\ifmmode~\breve{g}\else
  \u{g}\fi{}lu}},\ }\bibfield  {title} {\enquote {\bibinfo {title} {Realization
  of an electrically tunable narrow-bandwidth atomically thin mirror using
  monolayer ${\mathrm{mose}}_{2}$},}\ }\href {\doibase
  10.1103/PhysRevLett.120.037401} {\bibfield  {journal} {\bibinfo  {journal}
  {Phys. Rev. Lett.}\ }\textbf {\bibinfo {volume} {120}},\ \bibinfo {pages}
  {037401} (\bibinfo {year} {2018})}\BibitemShut {NoStop}%
\bibitem [{\citenamefont {Mennel}\ \emph {et~al.}(2018)\citenamefont {Mennel},
  \citenamefont {Furchi}, \citenamefont {Wachter}, \citenamefont {Paur},
  \citenamefont {Polyushkin},\ and\ \citenamefont
  {Mueller}}]{OpticalImagingOfStrain}%
  \BibitemOpen
  \bibfield  {author} {\bibinfo {author} {\bibfnamefont {Lukas}\ \bibnamefont
  {Mennel}}, \bibinfo {author} {\bibfnamefont {Marco~M.}\ \bibnamefont
  {Furchi}}, \bibinfo {author} {\bibfnamefont {Stefan}\ \bibnamefont
  {Wachter}}, \bibinfo {author} {\bibfnamefont {Matthias}\ \bibnamefont
  {Paur}}, \bibinfo {author} {\bibfnamefont {Dmitry~K.}\ \bibnamefont
  {Polyushkin}}, \ and\ \bibinfo {author} {\bibfnamefont {Thomas}\ \bibnamefont
  {Mueller}},\ }\bibfield  {title} {\enquote {\bibinfo {title} {Optical imaging
  of strain in two-dimensional crystals},}\ }\href {\doibase
  10.1038/s41467-018-02830-y} {\bibfield  {journal} {\bibinfo  {journal}
  {Nature Communications}\ }\textbf {\bibinfo {volume} {9}},\ \bibinfo {pages}
  {516} (\bibinfo {year} {2018})}\BibitemShut {NoStop}%
\bibitem [{\citenamefont {Hong}\ \emph {et~al.}(2015)\citenamefont {Hong},
  \citenamefont {Hu}, \citenamefont {Probert}, \citenamefont {Li},
  \citenamefont {Lv}, \citenamefont {Yang}, \citenamefont {Gu}, \citenamefont
  {Mao}, \citenamefont {Feng}, \citenamefont {Xie}, \citenamefont {Zhang},
  \citenamefont {Wu}, \citenamefont {Zhang}, \citenamefont {Jin}, \citenamefont
  {Ji}, \citenamefont {Zhang}, \citenamefont {Yuan},\ and\ \citenamefont
  {Zhang}}]{ExploringAtomicDefectsInMoS2}%
  \BibitemOpen
  \bibfield  {author} {\bibinfo {author} {\bibfnamefont {Jinhua}\ \bibnamefont
  {Hong}}, \bibinfo {author} {\bibfnamefont {Zhixin}\ \bibnamefont {Hu}},
  \bibinfo {author} {\bibfnamefont {Matt}\ \bibnamefont {Probert}}, \bibinfo
  {author} {\bibfnamefont {Kun}\ \bibnamefont {Li}}, \bibinfo {author}
  {\bibfnamefont {Danhui}\ \bibnamefont {Lv}}, \bibinfo {author} {\bibfnamefont
  {Xinan}\ \bibnamefont {Yang}}, \bibinfo {author} {\bibfnamefont {Lin}\
  \bibnamefont {Gu}}, \bibinfo {author} {\bibfnamefont {Nannan}\ \bibnamefont
  {Mao}}, \bibinfo {author} {\bibfnamefont {Qingliang}\ \bibnamefont {Feng}},
  \bibinfo {author} {\bibfnamefont {Liming}\ \bibnamefont {Xie}}, \bibinfo
  {author} {\bibfnamefont {Jin}\ \bibnamefont {Zhang}}, \bibinfo {author}
  {\bibfnamefont {Dianzhong}\ \bibnamefont {Wu}}, \bibinfo {author}
  {\bibfnamefont {Zhiyong}\ \bibnamefont {Zhang}}, \bibinfo {author}
  {\bibfnamefont {Chuanhong}\ \bibnamefont {Jin}}, \bibinfo {author}
  {\bibfnamefont {Wei}\ \bibnamefont {Ji}}, \bibinfo {author} {\bibfnamefont
  {Xixiang}\ \bibnamefont {Zhang}}, \bibinfo {author} {\bibfnamefont {Jun}\
  \bibnamefont {Yuan}}, \ and\ \bibinfo {author} {\bibfnamefont
  {Ze}~\bibnamefont {Zhang}},\ }\bibfield  {title} {\enquote {\bibinfo {title}
  {Exploring atomic defects in molybdenum disulphide monolayers},}\ }\href
  {http://dx.doi.org/10.1038/ncomms7293} {\bibfield  {journal} {\bibinfo
  {journal} {Nature Communications}\ }\textbf {\bibinfo {volume} {6}},\
  \bibinfo {pages} {6293 EP --} (\bibinfo {year} {2015})},\ \bibinfo {note}
  {article}\BibitemShut {NoStop}%
\bibitem [{\citenamefont {Wang}\ \emph {et~al.}(2015)\citenamefont {Wang},
  \citenamefont {Zhang},\ and\ \citenamefont
  {Rana}}]{UltrafastDynamicsofDefectAssistedRecombiniation}%
  \BibitemOpen
  \bibfield  {author} {\bibinfo {author} {\bibfnamefont {Haining}\ \bibnamefont
  {Wang}}, \bibinfo {author} {\bibfnamefont {Changjian}\ \bibnamefont {Zhang}},
  \ and\ \bibinfo {author} {\bibfnamefont {Farhan}\ \bibnamefont {Rana}},\
  }\bibfield  {title} {\enquote {\bibinfo {title} {Ultrafast dynamics of
  defect-assisted electron-hole recombination in monolayer mos2},}\ }\href
  {\doibase 10.1021/nl503636c} {\bibfield  {journal} {\bibinfo  {journal} {Nano
  Letters}\ }\textbf {\bibinfo {volume} {15}},\ \bibinfo {pages} {339--345}
  (\bibinfo {year} {2015})}\BibitemShut {NoStop}%
\bibitem [{\citenamefont {Xue}\ \emph {et~al.}(2011)\citenamefont {Xue},
  \citenamefont {Sanchez-Yamagishi}, \citenamefont {Bulmash}, \citenamefont
  {Jacquod}, \citenamefont {Deshpande}, \citenamefont {Watanabe}, \citenamefont
  {Taniguchi}, \citenamefont {Jarillo-Herrero},\ and\ \citenamefont
  {LeRoy}}]{ScanningTunnelingMicroscopyOfGraphene}%
  \BibitemOpen
  \bibfield  {author} {\bibinfo {author} {\bibfnamefont {Jiamin}\ \bibnamefont
  {Xue}}, \bibinfo {author} {\bibfnamefont {Javier}\ \bibnamefont
  {Sanchez-Yamagishi}}, \bibinfo {author} {\bibfnamefont {Danny}\ \bibnamefont
  {Bulmash}}, \bibinfo {author} {\bibfnamefont {Philippe}\ \bibnamefont
  {Jacquod}}, \bibinfo {author} {\bibfnamefont {Aparna}\ \bibnamefont
  {Deshpande}}, \bibinfo {author} {\bibfnamefont {K.}~\bibnamefont {Watanabe}},
  \bibinfo {author} {\bibfnamefont {T.}~\bibnamefont {Taniguchi}}, \bibinfo
  {author} {\bibfnamefont {Pablo}\ \bibnamefont {Jarillo-Herrero}}, \ and\
  \bibinfo {author} {\bibfnamefont {Brian~J.}\ \bibnamefont {LeRoy}},\
  }\bibfield  {title} {\enquote {\bibinfo {title} {Scanning tunnelling
  microscopy and spectroscopy of ultra-flat graphene on hexagonal boron
  nitride},}\ }\href {http://dx.doi.org/10.1038/nmat2968} {\bibfield  {journal}
  {\bibinfo  {journal} {Nature Materials}\ }\textbf {\bibinfo {volume} {10}},\
  \bibinfo {pages} {282 EP --} (\bibinfo {year} {2011})}\BibitemShut {NoStop}%
\bibitem [{\citenamefont {Yu}\ \emph {et~al.}(2013)\citenamefont {Yu},
  \citenamefont {Li}, \citenamefont {Liu}, \citenamefont {Su}, \citenamefont
  {Zhang},\ and\ \citenamefont
  {Cao}}]{ScalableSynthesisofUniformMonolayerMoS2}%
  \BibitemOpen
  \bibfield  {author} {\bibinfo {author} {\bibfnamefont {Yifei}\ \bibnamefont
  {Yu}}, \bibinfo {author} {\bibfnamefont {Chun}\ \bibnamefont {Li}}, \bibinfo
  {author} {\bibfnamefont {Yi}~\bibnamefont {Liu}}, \bibinfo {author}
  {\bibfnamefont {Liqin}\ \bibnamefont {Su}}, \bibinfo {author} {\bibfnamefont
  {Yong}\ \bibnamefont {Zhang}}, \ and\ \bibinfo {author} {\bibfnamefont
  {Linyou}\ \bibnamefont {Cao}},\ }\bibfield  {title} {\enquote {\bibinfo
  {title} {Controlled scalable synthesis of uniform, high-quality monolayer and
  few-layer mos2 films},}\ }\href {http://dx.doi.org/10.1038/srep01866}
  {\bibfield  {journal} {\bibinfo  {journal} {Scientific Reports}\ }\textbf
  {\bibinfo {volume} {3}},\ \bibinfo {pages} {1866 EP --} (\bibinfo {year}
  {2013})},\ \bibinfo {note} {article}\BibitemShut {NoStop}%
\bibitem [{\citenamefont {Lee}\ \emph {et~al.}()\citenamefont {Lee},
  \citenamefont {Zhang}, \citenamefont {Zhang}, \citenamefont {Chang},
  \citenamefont {Lin}, \citenamefont {Chang}, \citenamefont {Yu}, \citenamefont
  {Wang}, \citenamefont {Chang}, \citenamefont {Li},\ and\ \citenamefont
  {Lin}}]{SynthesisOfLargeAreaMoS2}%
  \BibitemOpen
  \bibfield  {author} {\bibinfo {author} {\bibfnamefont {Yi‐Hsien}\
  \bibnamefont {Lee}}, \bibinfo {author} {\bibfnamefont {Xin‐Quan}\
  \bibnamefont {Zhang}}, \bibinfo {author} {\bibfnamefont {Wenjing}\
  \bibnamefont {Zhang}}, \bibinfo {author} {\bibfnamefont {Mu‐Tung}\
  \bibnamefont {Chang}}, \bibinfo {author} {\bibfnamefont {Cheng‐Te}\
  \bibnamefont {Lin}}, \bibinfo {author} {\bibfnamefont {Kai‐Di}\
  \bibnamefont {Chang}}, \bibinfo {author} {\bibfnamefont {Ya‐Chu}\
  \bibnamefont {Yu}}, \bibinfo {author} {\bibfnamefont {Jacob~Tse‐Wei}\
  \bibnamefont {Wang}}, \bibinfo {author} {\bibfnamefont {Chia‐Seng}\
  \bibnamefont {Chang}}, \bibinfo {author} {\bibfnamefont {Lain‐Jong}\
  \bibnamefont {Li}}, \ and\ \bibinfo {author} {\bibfnamefont {Tsung‐Wu}\
  \bibnamefont {Lin}},\ }\bibfield  {title} {\enquote {\bibinfo {title}
  {Synthesis of large‐area mos2 atomic layers with chemical vapor
  deposition},}\ }\href {\doibase 10.1002/adma.201104798} {\bibfield  {journal}
  {\bibinfo  {journal} {Advanced Materials}\ }\textbf {\bibinfo {volume}
  {24}},\ \bibinfo {pages} {2320--2325}}\BibitemShut {NoStop}%
\bibitem [{\citenamefont {Nan}\ \emph {et~al.}(2014)\citenamefont {Nan},
  \citenamefont {Wang}, \citenamefont {Wang}, \citenamefont {Liang},
  \citenamefont {Lu}, \citenamefont {Chen}, \citenamefont {He}, \citenamefont
  {Tan}, \citenamefont {Miao}, \citenamefont {Wang}, \citenamefont {Wang},\
  and\ \citenamefont {Ni}}]{StrongPLEnhancementOfMoS2Defect}%
  \BibitemOpen
  \bibfield  {author} {\bibinfo {author} {\bibfnamefont {Haiyan}\ \bibnamefont
  {Nan}}, \bibinfo {author} {\bibfnamefont {Zilu}\ \bibnamefont {Wang}},
  \bibinfo {author} {\bibfnamefont {Wenhui}\ \bibnamefont {Wang}}, \bibinfo
  {author} {\bibfnamefont {Zheng}\ \bibnamefont {Liang}}, \bibinfo {author}
  {\bibfnamefont {Yan}\ \bibnamefont {Lu}}, \bibinfo {author} {\bibfnamefont
  {Qian}\ \bibnamefont {Chen}}, \bibinfo {author} {\bibfnamefont {Daowei}\
  \bibnamefont {He}}, \bibinfo {author} {\bibfnamefont {Pingheng}\ \bibnamefont
  {Tan}}, \bibinfo {author} {\bibfnamefont {Feng}\ \bibnamefont {Miao}},
  \bibinfo {author} {\bibfnamefont {Xinran}\ \bibnamefont {Wang}}, \bibinfo
  {author} {\bibfnamefont {Jinlan}\ \bibnamefont {Wang}}, \ and\ \bibinfo
  {author} {\bibfnamefont {Zhenhua}\ \bibnamefont {Ni}},\ }\bibfield  {title}
  {\enquote {\bibinfo {title} {Strong photoluminescence enhancement of mos2
  through defect engineering and oxygen bonding},}\ }\href {\doibase
  10.1021/nn500532f} {\bibfield  {journal} {\bibinfo  {journal} {ACS Nano}\
  }\textbf {\bibinfo {volume} {8}},\ \bibinfo {pages} {5738--5745} (\bibinfo
  {year} {2014})}\BibitemShut {NoStop}%
\bibitem [{\citenamefont {Wood}\ \emph {et~al.}(2015)\citenamefont {Wood},
  \citenamefont {Doidge}, \citenamefont {Carrion}, \citenamefont {Koepke},
  \citenamefont {Kaitz}, \citenamefont {Datye}, \citenamefont {Behnam},
  \citenamefont {Hewaparakrama}, \citenamefont {Aruin}, \citenamefont {Chen},
  \citenamefont {Dong}, \citenamefont {Haasch}, \citenamefont {Lyding},\ and\
  \citenamefont {Pop}}]{AnnealingFreeCleanGrapheneTransfer}%
  \BibitemOpen
  \bibfield  {author} {\bibinfo {author} {\bibfnamefont {Joshua~D}\
  \bibnamefont {Wood}}, \bibinfo {author} {\bibfnamefont {Gregory~P}\
  \bibnamefont {Doidge}}, \bibinfo {author} {\bibfnamefont {Enrique~A}\
  \bibnamefont {Carrion}}, \bibinfo {author} {\bibfnamefont {Justin~C}\
  \bibnamefont {Koepke}}, \bibinfo {author} {\bibfnamefont {Joshua~A}\
  \bibnamefont {Kaitz}}, \bibinfo {author} {\bibfnamefont {Isha}\ \bibnamefont
  {Datye}}, \bibinfo {author} {\bibfnamefont {Ashkan}\ \bibnamefont {Behnam}},
  \bibinfo {author} {\bibfnamefont {Jayan}\ \bibnamefont {Hewaparakrama}},
  \bibinfo {author} {\bibfnamefont {Basil}\ \bibnamefont {Aruin}}, \bibinfo
  {author} {\bibfnamefont {Yaofeng}\ \bibnamefont {Chen}}, \bibinfo {author}
  {\bibfnamefont {Hefei}\ \bibnamefont {Dong}}, \bibinfo {author}
  {\bibfnamefont {Richard~T}\ \bibnamefont {Haasch}}, \bibinfo {author}
  {\bibfnamefont {Joseph~W}\ \bibnamefont {Lyding}}, \ and\ \bibinfo {author}
  {\bibfnamefont {Eric}\ \bibnamefont {Pop}},\ }\bibfield  {title} {\enquote
  {\bibinfo {title} {Annealing free, clean graphene transfer using alternative
  polymer scaffolds},}\ }\href
  {http://stacks.iop.org/0957-4484/26/i=5/a=055302} {\bibfield  {journal}
  {\bibinfo  {journal} {Nanotechnology}\ }\textbf {\bibinfo {volume} {26}},\
  \bibinfo {pages} {055302} (\bibinfo {year} {2015})}\BibitemShut {NoStop}%
\bibitem [{\citenamefont {Amani}\ \emph {et~al.}(2016)\citenamefont {Amani},
  \citenamefont {Burke}, \citenamefont {Ji}, \citenamefont {Zhao},
  \citenamefont {Lien}, \citenamefont {Taheri}, \citenamefont {Ahn},
  \citenamefont {Kirya}, \citenamefont {Ager}, \citenamefont {Yablonovitch},
  \citenamefont {Kong}, \citenamefont {Dubey},\ and\ \citenamefont
  {Javey}}]{HighLuminescenceEfficiencyInMoS2}%
  \BibitemOpen
  \bibfield  {author} {\bibinfo {author} {\bibfnamefont {Matin}\ \bibnamefont
  {Amani}}, \bibinfo {author} {\bibfnamefont {Robert~A.}\ \bibnamefont
  {Burke}}, \bibinfo {author} {\bibfnamefont {Xiang}\ \bibnamefont {Ji}},
  \bibinfo {author} {\bibfnamefont {Peida}\ \bibnamefont {Zhao}}, \bibinfo
  {author} {\bibfnamefont {Der-Hsien}\ \bibnamefont {Lien}}, \bibinfo {author}
  {\bibfnamefont {Peyman}\ \bibnamefont {Taheri}}, \bibinfo {author}
  {\bibfnamefont {Geun~Ho}\ \bibnamefont {Ahn}}, \bibinfo {author}
  {\bibfnamefont {Daisuke}\ \bibnamefont {Kirya}}, \bibinfo {author}
  {\bibfnamefont {Joel~W.}\ \bibnamefont {Ager}}, \bibinfo {author}
  {\bibfnamefont {Eli}\ \bibnamefont {Yablonovitch}}, \bibinfo {author}
  {\bibfnamefont {Jing}\ \bibnamefont {Kong}}, \bibinfo {author} {\bibfnamefont
  {Madan}\ \bibnamefont {Dubey}}, \ and\ \bibinfo {author} {\bibfnamefont
  {Ali}\ \bibnamefont {Javey}},\ }\bibfield  {title} {\enquote {\bibinfo
  {title} {High luminescence efficiency in mos2 grown by chemical vapor
  deposition},}\ }\href {\doibase 10.1021/acsnano.6b03443} {\bibfield
  {journal} {\bibinfo  {journal} {ACS Nano}\ }\textbf {\bibinfo {volume}
  {10}},\ \bibinfo {pages} {6535--6541} (\bibinfo {year} {2016})}\BibitemShut
  {NoStop}%
\bibitem [{\citenamefont {Florian}\ \emph {et~al.}(2018)\citenamefont
  {Florian}, \citenamefont {Hartmann}, \citenamefont {Steinhoff}, \citenamefont
  {Klein}, \citenamefont {Holleitner}, \citenamefont {Finley}, \citenamefont
  {Wehling}, \citenamefont {Kaniber},\ and\ \citenamefont
  {Gies}}]{DielectricImpactOnExcitonBindingEnergies}%
  \BibitemOpen
  \bibfield  {author} {\bibinfo {author} {\bibfnamefont {Matthias}\
  \bibnamefont {Florian}}, \bibinfo {author} {\bibfnamefont {Malte}\
  \bibnamefont {Hartmann}}, \bibinfo {author} {\bibfnamefont {Alexander}\
  \bibnamefont {Steinhoff}}, \bibinfo {author} {\bibfnamefont {Julian}\
  \bibnamefont {Klein}}, \bibinfo {author} {\bibfnamefont {Alexander~W.}\
  \bibnamefont {Holleitner}}, \bibinfo {author} {\bibfnamefont {Jonathan~J.}\
  \bibnamefont {Finley}}, \bibinfo {author} {\bibfnamefont {Tim~O.}\
  \bibnamefont {Wehling}}, \bibinfo {author} {\bibfnamefont {Michael}\
  \bibnamefont {Kaniber}}, \ and\ \bibinfo {author} {\bibfnamefont
  {Christopher}\ \bibnamefont {Gies}},\ }\bibfield  {title} {\enquote {\bibinfo
  {title} {The dielectric impact of layer distances on exciton and trion
  binding energies in van der waals heterostructures},}\ }\href {\doibase
  10.1021/acs.nanolett.8b00840} {\bibfield  {journal} {\bibinfo  {journal}
  {Nano Letters}\ }\textbf {\bibinfo {volume} {18}},\ \bibinfo {pages}
  {2725--2732} (\bibinfo {year} {2018})}\BibitemShut {NoStop}%
\bibitem [{\citenamefont {Robert}\ \emph {et~al.}(2016)\citenamefont {Robert},
  \citenamefont {Lagarde}, \citenamefont {Cadiz}, \citenamefont {Wang},
  \citenamefont {Lassagne}, \citenamefont {Amand}, \citenamefont {Balocchi},
  \citenamefont {Renucci}, \citenamefont {Tongay}, \citenamefont {Urbaszek},\
  and\ \citenamefont {Marie}}]{ExcitonRadiativeLifetimeInTMDCs}%
  \BibitemOpen
  \bibfield  {author} {\bibinfo {author} {\bibfnamefont {C.}~\bibnamefont
  {Robert}}, \bibinfo {author} {\bibfnamefont {D.}~\bibnamefont {Lagarde}},
  \bibinfo {author} {\bibfnamefont {F.}~\bibnamefont {Cadiz}}, \bibinfo
  {author} {\bibfnamefont {G.}~\bibnamefont {Wang}}, \bibinfo {author}
  {\bibfnamefont {B.}~\bibnamefont {Lassagne}}, \bibinfo {author}
  {\bibfnamefont {T.}~\bibnamefont {Amand}}, \bibinfo {author} {\bibfnamefont
  {A.}~\bibnamefont {Balocchi}}, \bibinfo {author} {\bibfnamefont
  {P.}~\bibnamefont {Renucci}}, \bibinfo {author} {\bibfnamefont
  {S.}~\bibnamefont {Tongay}}, \bibinfo {author} {\bibfnamefont
  {B.}~\bibnamefont {Urbaszek}}, \ and\ \bibinfo {author} {\bibfnamefont
  {X.}~\bibnamefont {Marie}},\ }\bibfield  {title} {\enquote {\bibinfo {title}
  {Exciton radiative lifetime in transition metal dichalcogenide monolayers},}\
  }\href {\doibase 10.1103/PhysRevB.93.205423} {\bibfield  {journal} {\bibinfo
  {journal} {Phys. Rev. B}\ }\textbf {\bibinfo {volume} {93}},\ \bibinfo
  {pages} {205423} (\bibinfo {year} {2016})}\BibitemShut {NoStop}%
\bibitem [{\citenamefont {Tongay}\ \emph {et~al.}(2013)\citenamefont {Tongay},
  \citenamefont {Zhou}, \citenamefont {Ataca}, \citenamefont {Liu},
  \citenamefont {Kang}, \citenamefont {Matthews}, \citenamefont {You},
  \citenamefont {Li}, \citenamefont {Grossman},\ and\ \citenamefont
  {Wu}}]{ModulationOfLightEmissionByPhysisorption}%
  \BibitemOpen
  \bibfield  {author} {\bibinfo {author} {\bibfnamefont {Sefaattin}\
  \bibnamefont {Tongay}}, \bibinfo {author} {\bibfnamefont {Jian}\ \bibnamefont
  {Zhou}}, \bibinfo {author} {\bibfnamefont {Can}\ \bibnamefont {Ataca}},
  \bibinfo {author} {\bibfnamefont {Jonathan}\ \bibnamefont {Liu}}, \bibinfo
  {author} {\bibfnamefont {Jeong~Seuk}\ \bibnamefont {Kang}}, \bibinfo {author}
  {\bibfnamefont {Tyler~S.}\ \bibnamefont {Matthews}}, \bibinfo {author}
  {\bibfnamefont {Long}\ \bibnamefont {You}}, \bibinfo {author} {\bibfnamefont
  {Jingbo}\ \bibnamefont {Li}}, \bibinfo {author} {\bibfnamefont {Jeffrey~C.}\
  \bibnamefont {Grossman}}, \ and\ \bibinfo {author} {\bibfnamefont {Junqiao}\
  \bibnamefont {Wu}},\ }\bibfield  {title} {\enquote {\bibinfo {title}
  {Broad-range modulation of light emission in two-dimensional semiconductors
  by molecular physisorption gating},}\ }\href {\doibase 10.1021/nl4011172}
  {\bibfield  {journal} {\bibinfo  {journal} {Nano Letters}\ }\textbf {\bibinfo
  {volume} {13}},\ \bibinfo {pages} {2831--2836} (\bibinfo {year}
  {2013})}\BibitemShut {NoStop}%
\bibitem [{\citenamefont {Li}\ \emph {et~al.}(2014)\citenamefont {Li},
  \citenamefont {Chernikov}, \citenamefont {Zhang}, \citenamefont {Rigosi},
  \citenamefont {Hill}, \citenamefont {van~der Zande}, \citenamefont {Chenet},
  \citenamefont {Shih}, \citenamefont {Hone},\ and\ \citenamefont
  {Heinz}}]{OpticalDielectricFunctionOfmTMDCs}%
  \BibitemOpen
  \bibfield  {author} {\bibinfo {author} {\bibfnamefont {Yilei}\ \bibnamefont
  {Li}}, \bibinfo {author} {\bibfnamefont {Alexey}\ \bibnamefont {Chernikov}},
  \bibinfo {author} {\bibfnamefont {Xian}\ \bibnamefont {Zhang}}, \bibinfo
  {author} {\bibfnamefont {Albert}\ \bibnamefont {Rigosi}}, \bibinfo {author}
  {\bibfnamefont {Heather~M.}\ \bibnamefont {Hill}}, \bibinfo {author}
  {\bibfnamefont {Arend~M.}\ \bibnamefont {van~der Zande}}, \bibinfo {author}
  {\bibfnamefont {Daniel~A.}\ \bibnamefont {Chenet}}, \bibinfo {author}
  {\bibfnamefont {En-Min}\ \bibnamefont {Shih}}, \bibinfo {author}
  {\bibfnamefont {James}\ \bibnamefont {Hone}}, \ and\ \bibinfo {author}
  {\bibfnamefont {Tony~F.}\ \bibnamefont {Heinz}},\ }\bibfield  {title}
  {\enquote {\bibinfo {title} {Measurement of the optical dielectric function
  of monolayer transition-metal dichalcogenides: ${\mathrm{mos}}_{2}$,
  $\mathrm{Mo}\mathrm{S}{\mathrm{e}}_{2}$, ${\mathrm{ws}}_{2}$, and
  $\mathrm{WS}{\mathrm{e}}_{2}$},}\ }\href {\doibase
  10.1103/PhysRevB.90.205422} {\bibfield  {journal} {\bibinfo  {journal} {Phys.
  Rev. B}\ }\textbf {\bibinfo {volume} {90}},\ \bibinfo {pages} {205422}
  (\bibinfo {year} {2014})}\BibitemShut {NoStop}%
\bibitem [{\citenamefont {Loudon}(1964)}]{RamanEffectInCrystals}%
  \BibitemOpen
  \bibfield  {author} {\bibinfo {author} {\bibfnamefont {R.}~\bibnamefont
  {Loudon}},\ }\bibfield  {title} {\enquote {\bibinfo {title} {The raman effect
  in crystals},}\ }\href {\doibase 10.1080/00018736400101051} {\bibfield
  {journal} {\bibinfo  {journal} {Advances in Physics}\ }\textbf {\bibinfo
  {volume} {13}},\ \bibinfo {pages} {423--482} (\bibinfo {year}
  {1964})}\BibitemShut {NoStop}%
\bibitem [{\citenamefont {Cai}\ \emph {et~al.}(2014)\citenamefont {Cai},
  \citenamefont {Lan}, \citenamefont {Zhang},\ and\ \citenamefont
  {Zhang}}]{LatticeVibrationalModesOfMoS2}%
  \BibitemOpen
  \bibfield  {author} {\bibinfo {author} {\bibfnamefont {Yongqing}\
  \bibnamefont {Cai}}, \bibinfo {author} {\bibfnamefont {Jinghua}\ \bibnamefont
  {Lan}}, \bibinfo {author} {\bibfnamefont {Gang}\ \bibnamefont {Zhang}}, \
  and\ \bibinfo {author} {\bibfnamefont {Yong-Wei}\ \bibnamefont {Zhang}},\
  }\bibfield  {title} {\enquote {\bibinfo {title} {Lattice vibrational modes
  and phonon thermal conductivity of monolayer mos${}_{2}$},}\ }\href {\doibase
  10.1103/PhysRevB.89.035438} {\bibfield  {journal} {\bibinfo  {journal} {Phys.
  Rev. B}\ }\textbf {\bibinfo {volume} {89}},\ \bibinfo {pages} {035438}
  (\bibinfo {year} {2014})}\BibitemShut {NoStop}%
\bibitem [{\citenamefont {Castellanos-Gomez}\ \emph {et~al.}(2014)\citenamefont
  {Castellanos-Gomez}, \citenamefont {Buscema}, \citenamefont {Molenaar},
  \citenamefont {Singh}, \citenamefont {Janssen}, \citenamefont {van~der
  Zant},\ and\ \citenamefont {Steele}}]{DeterministicTransfer}%
  \BibitemOpen
  \bibfield  {author} {\bibinfo {author} {\bibfnamefont {Andres}\ \bibnamefont
  {Castellanos-Gomez}}, \bibinfo {author} {\bibfnamefont {Michele}\
  \bibnamefont {Buscema}}, \bibinfo {author} {\bibfnamefont {Rianda}\
  \bibnamefont {Molenaar}}, \bibinfo {author} {\bibfnamefont {Vibhor}\
  \bibnamefont {Singh}}, \bibinfo {author} {\bibfnamefont {Laurens}\
  \bibnamefont {Janssen}}, \bibinfo {author} {\bibfnamefont {Herre S~J}\
  \bibnamefont {van~der Zant}}, \ and\ \bibinfo {author} {\bibfnamefont
  {Gary~A}\ \bibnamefont {Steele}},\ }\bibfield  {title} {\enquote {\bibinfo
  {title} {Deterministic transfer of two-dimensional materials by all-dry
  viscoelastic stamping},}\ }\href
  {http://stacks.iop.org/2053-1583/1/i=1/a=011002} {\bibfield  {journal}
  {\bibinfo  {journal} {2D Materials}\ }\textbf {\bibinfo {volume} {1}},\
  \bibinfo {pages} {011002} (\bibinfo {year} {2014})}\BibitemShut {NoStop}%
\bibitem [{\citenamefont {Hong}\ \emph {et~al.}(2016)\citenamefont {Hong},
  \citenamefont {Zhang},\ and\ \citenamefont
  {Zeng}}]{ThermalConductivityOfMoSe2}%
  \BibitemOpen
  \bibfield  {author} {\bibinfo {author} {\bibfnamefont {Yang}\ \bibnamefont
  {Hong}}, \bibinfo {author} {\bibfnamefont {Jingchao}\ \bibnamefont {Zhang}},
  \ and\ \bibinfo {author} {\bibfnamefont {Xiao~Cheng}\ \bibnamefont {Zeng}},\
  }\bibfield  {title} {\enquote {\bibinfo {title} {Thermal conductivity of
  monolayer mose2 and mos2},}\ }\href {\doibase 10.1021/acs.jpcc.6b07262}
  {\bibfield  {journal} {\bibinfo  {journal} {The Journal of Physical Chemistry
  C}\ }\textbf {\bibinfo {volume} {120}},\ \bibinfo {pages} {26067--26075}
  (\bibinfo {year} {2016})}\BibitemShut {NoStop}%
\bibitem [{\citenamefont {Zhang}\ \emph {et~al.}(2015)\citenamefont {Zhang},
  \citenamefont {Sun}, \citenamefont {Li}, \citenamefont {Lee}, \citenamefont
  {Cui}, \citenamefont {Chenet}, \citenamefont {You}, \citenamefont {Heinz},\
  and\ \citenamefont {Hone}}]{LateralThermalConductivityOfMoSe2}%
  \BibitemOpen
  \bibfield  {author} {\bibinfo {author} {\bibfnamefont {Xian}\ \bibnamefont
  {Zhang}}, \bibinfo {author} {\bibfnamefont {Dezheng}\ \bibnamefont {Sun}},
  \bibinfo {author} {\bibfnamefont {Yilei}\ \bibnamefont {Li}}, \bibinfo
  {author} {\bibfnamefont {Gwan-Hyoung}\ \bibnamefont {Lee}}, \bibinfo {author}
  {\bibfnamefont {Xu}~\bibnamefont {Cui}}, \bibinfo {author} {\bibfnamefont
  {Daniel}\ \bibnamefont {Chenet}}, \bibinfo {author} {\bibfnamefont {Yumeng}\
  \bibnamefont {You}}, \bibinfo {author} {\bibfnamefont {Tony~F.}\ \bibnamefont
  {Heinz}}, \ and\ \bibinfo {author} {\bibfnamefont {James~C.}\ \bibnamefont
  {Hone}},\ }\bibfield  {title} {\enquote {\bibinfo {title} {Measurement of
  lateral and interfacial thermal conductivity of single- and bilayer mos2 and
  mose2 using refined optothermal raman technique},}\ }\href {\doibase
  10.1021/acsami.5b08580} {\bibfield  {journal} {\bibinfo  {journal} {ACS
  Applied Materials \& Interfaces}\ }\textbf {\bibinfo {volume} {7}},\ \bibinfo
  {pages} {25923--25929} (\bibinfo {year} {2015})}\BibitemShut {NoStop}%
\bibitem [{\citenamefont {Glassbrenner}\ and\ \citenamefont
  {Slack}(1964)}]{ThermalConductivityOfSilicon}%
  \BibitemOpen
  \bibfield  {author} {\bibinfo {author} {\bibfnamefont {C.~J.}\ \bibnamefont
  {Glassbrenner}}\ and\ \bibinfo {author} {\bibfnamefont {Glen~A.}\
  \bibnamefont {Slack}},\ }\bibfield  {title} {\enquote {\bibinfo {title}
  {Thermal conductivity of silicon and germanium from $3^\circ$k to the melting
  point},}\ }\href {\doibase 10.1103/PhysRev.134.A1058} {\bibfield  {journal}
  {\bibinfo  {journal} {Phys. Rev.}\ }\textbf {\bibinfo {volume} {134}},\
  \bibinfo {pages} {A1058--A1069} (\bibinfo {year} {1964})}\BibitemShut
  {NoStop}%
\bibitem [{\citenamefont {Yao}\ \emph {et~al.}(2008)\citenamefont {Yao},
  \citenamefont {Lai},\ and\ \citenamefont
  {Chien}}]{TemperatureDependenceOfThermalConductivityForSiO2}%
  \BibitemOpen
  \bibfield  {author} {\bibinfo {author} {\bibfnamefont {Da-Jeng}\ \bibnamefont
  {Yao}}, \bibinfo {author} {\bibfnamefont {Wei-Chih}\ \bibnamefont {Lai}}, \
  and\ \bibinfo {author} {\bibfnamefont {Heng-Chieh}\ \bibnamefont {Chien}},\
  }\bibfield  {title} {\enquote {\bibinfo {title} {Temperature dependence of
  thermal conductivity for silicon dioxide},}\ }\href {\doibase
  doi:10.1115/MNHT2008-52052} {\bibfield  {journal} {\bibinfo  {journal} {2008
  Proceedings of the ASME Micro/Nanoscale Heat Transfer International
  Conference, MNHT 2008}\ ,\ \bibinfo {pages} {435--439}} (\bibinfo {year}
  {2008})}\BibitemShut {NoStop}%
\end{thebibliography}%


\providecommand{\noopsort}[1]{}\providecommand{\singleletter}[1]{#1}%
\begin{thebibliography}{1}%
\makeatletter
\providecommand \@ifxundefined [1]{%
 \@ifx{#1\undefined}
}%
\providecommand \@ifnum [1]{%
 \ifnum #1\expandafter \@firstoftwo
 \else \expandafter \@secondoftwo
 \fi
}%
\providecommand \@ifx [1]{%
 \ifx #1\expandafter \@firstoftwo
 \else \expandafter \@secondoftwo
 \fi
}%
\providecommand \natexlab [1]{#1}%
\providecommand \enquote  [1]{``#1''}%
\providecommand \bibnamefont  [1]{#1}%
\providecommand \bibfnamefont [1]{#1}%
\providecommand \citenamefont [1]{#1}%
\providecommand \href@noop [0]{\@secondoftwo}%
\providecommand \href [0]{\begingroup \@sanitize@url \@href}%
\providecommand \@href[1]{\@@startlink{#1}\@@href}%
\providecommand \@@href[1]{\endgroup#1\@@endlink}%
\providecommand \@sanitize@url [0]{\catcode `\\12\catcode `\$12\catcode
  `\&12\catcode `\#12\catcode `\^12\catcode `\_12\catcode `\%12\relax}%
\providecommand \@@startlink[1]{}%
\providecommand \@@endlink[0]{}%
\providecommand \url  [0]{\begingroup\@sanitize@url \@url }%
\providecommand \@url [1]{\endgroup\@href {#1}{\urlprefix }}%
\providecommand \urlprefix  [0]{URL }%
\providecommand \Eprint [0]{\href }%
\providecommand \doibase [0]{http://dx.doi.org/}%
\providecommand \selectlanguage [0]{\@gobble}%
\providecommand \bibinfo  [0]{\@secondoftwo}%
\providecommand \bibfield  [0]{\@secondoftwo}%
\providecommand \translation [1]{[#1]}%
\providecommand \BibitemOpen [0]{}%
\providecommand \bibitemStop [0]{}%
\providecommand \bibitemNoStop [0]{.\EOS\space}%
\providecommand \EOS [0]{\spacefactor3000\relax}%
\providecommand \BibitemShut  [1]{\csname bibitem#1\endcsname}%
\let\auto@bib@innerbib\@empty
\bibitem [{\citenamefont {Soubelet}\ \emph {et~al.}(2016)\citenamefont
  {Soubelet}, \citenamefont {Bruchhausen}, \citenamefont {Fainstein},
  \citenamefont {Nogajewski},\ and\ \citenamefont
  {Faugeras}}]{ResonanceEffectsInRamanScatteringOfMoSe2}%
  \BibitemOpen
  \bibfield  {author} {\bibinfo {author} {\bibfnamefont {P.}~\bibnamefont
  {Soubelet}}, \bibinfo {author} {\bibfnamefont {A.~E.}\ \bibnamefont
  {Bruchhausen}}, \bibinfo {author} {\bibfnamefont {A.}~\bibnamefont
  {Fainstein}}, \bibinfo {author} {\bibfnamefont {K.}~\bibnamefont
  {Nogajewski}}, \ and\ \bibinfo {author} {\bibfnamefont {C.}~\bibnamefont
  {Faugeras}},\ }\bibfield  {title} {\enquote {\bibinfo {title} {Resonance
  effects in the raman scattering of monolayer and few-layer
  ${\mathrm{mose}}_{2}$},}\ }\href {\doibase 10.1103/PhysRevB.93.155407}
  {\bibfield  {journal} {\bibinfo  {journal} {Phys. Rev. B}\ }\textbf {\bibinfo
  {volume} {93}},\ \bibinfo {pages} {155407} (\bibinfo {year}
  {2016})}\BibitemShut {NoStop}%
\end{thebibliography}%

\end{document}




\title{Laser Annealing  for Radiatively Broadened $\mathbf{MoSe_2}$ grown by Chemical Vapor Deposition: \\ Supporting Information}

\author{Christopher Rogers}
\email{cmrogers@stanford.edu}
\author{Dodd Gray}
\author{Nate Bogdanowicz}
\author{Hideo Mabuchi}
\email{hmabuchi@stanford.edu}
\affiliation{%
 Ginzton Laboratory, Stanford University, 348 Via Pueblo, Stanford, CA 94305
}%
%
%
%

\date{\today}
\maketitle


\beginsupplement


\section{\label{sec:Supplement} Supplementary Info}

\begin{figure}
    \centering
    \subfloat[ \label{sfig:BeforeAfter:Reflection}]
    {\includegraphics[width=0.24\textwidth]
    {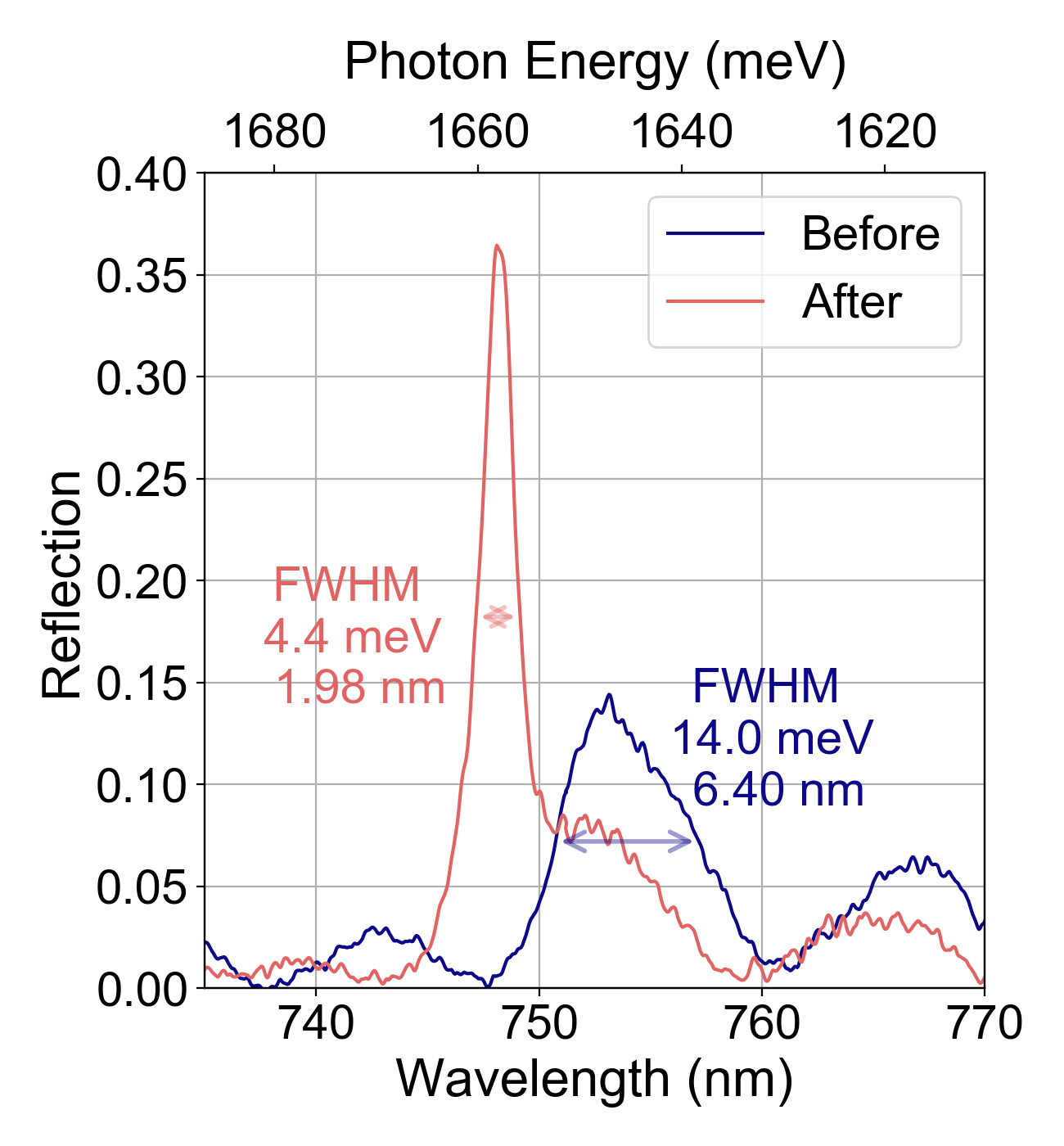}}
    \subfloat[ \label{sfig:BeforeAfter:LargeReflection}]
    {\includegraphics[width=0.24\textwidth]
    {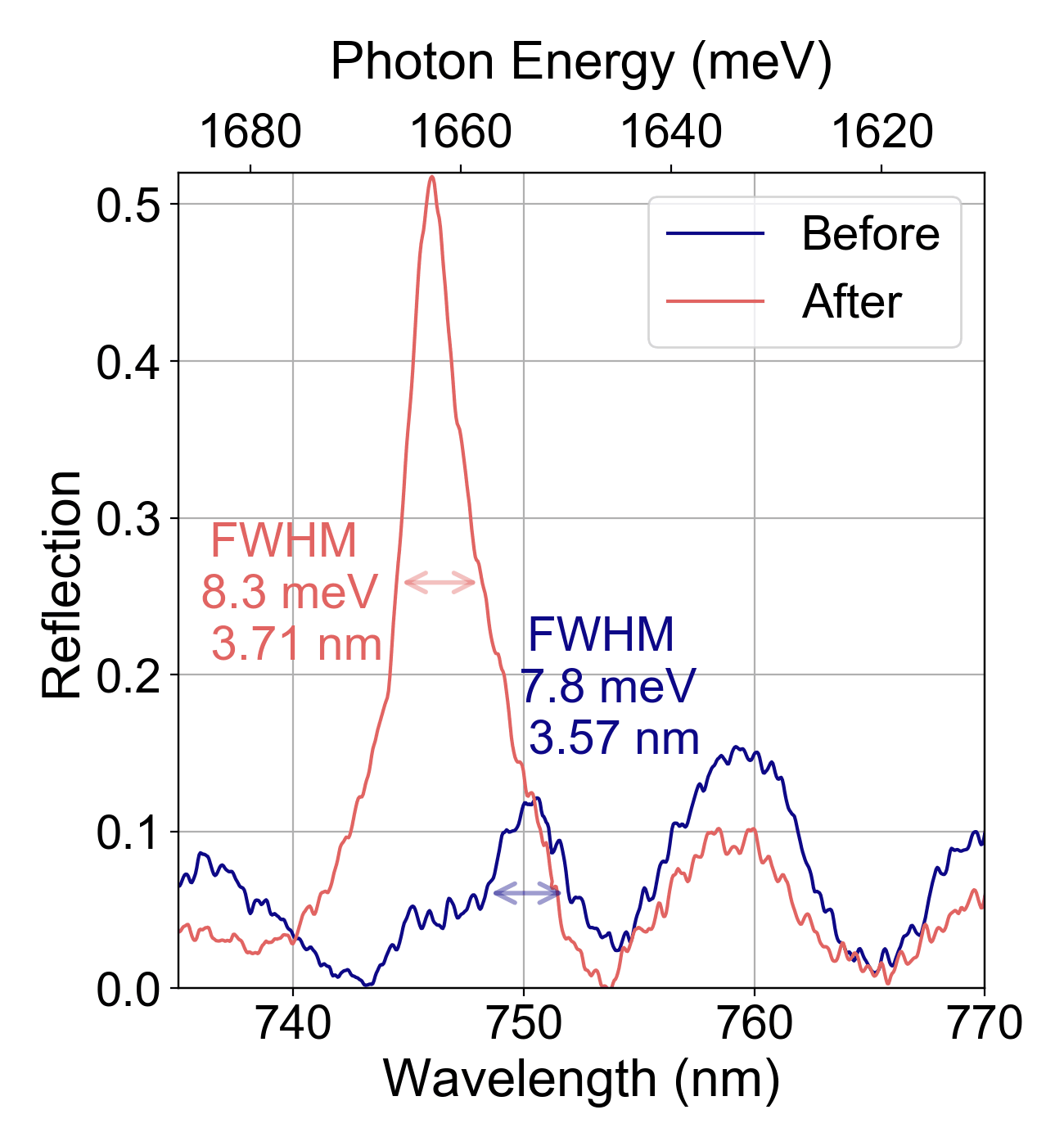}}
    \subfloat[ \label{sfig:BeforeAfter:Reflection_24h}]
    {\includegraphics[width=0.24\textwidth]
    {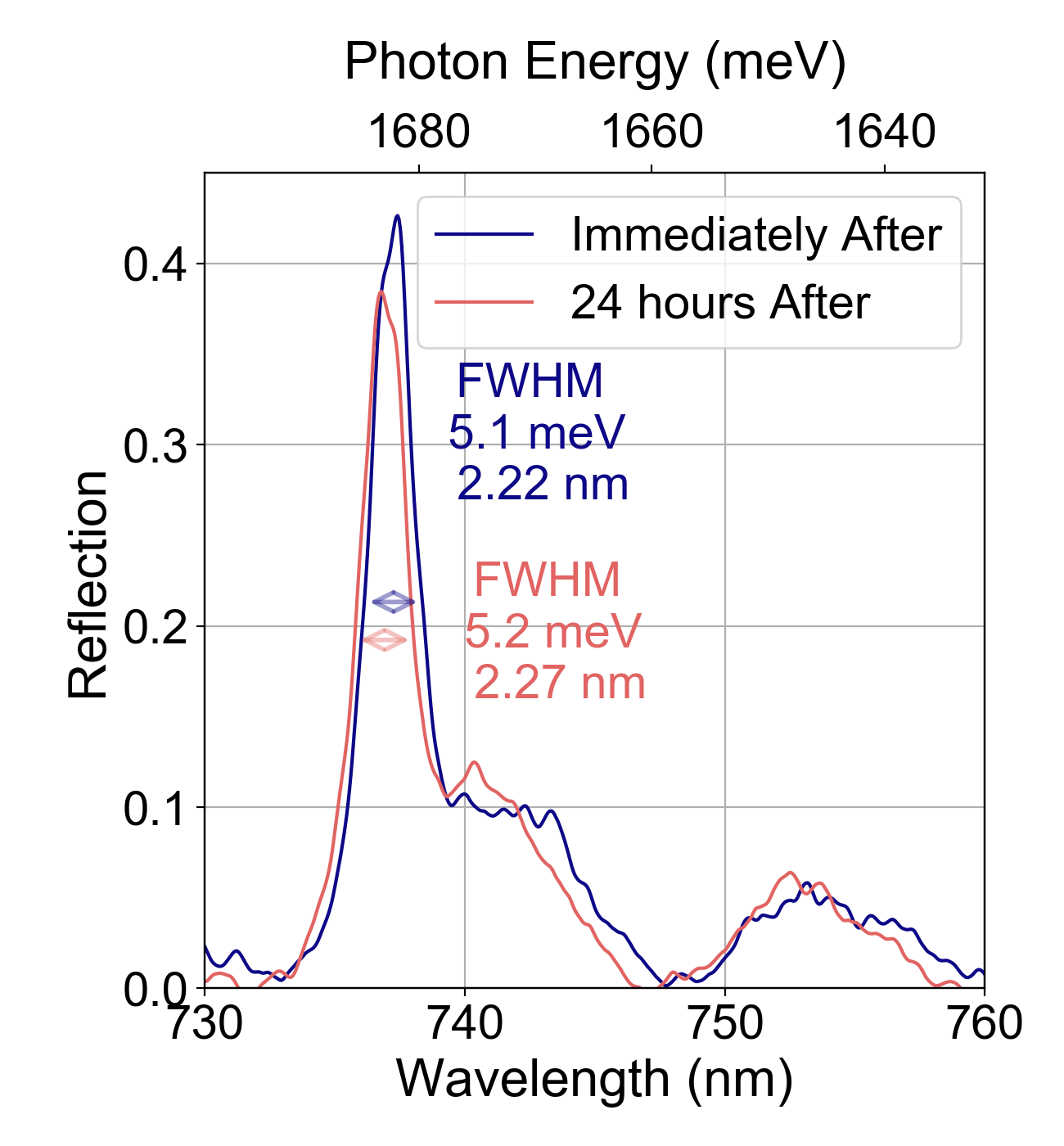}}
    \caption{Raw reflection corresponding to Fig. \ref{fig:BeforeAfter:Reflection} (b), corresponding to Fig. \ref{fig:BeforeAfter:Reflection_24h} (c), and corresponding to Fig. \ref{fig:BeforeAfter:LargeReflection} (d).
    \label{sfig:NarrowFWHM}}
\end{figure}

\begin{figure}
    \centering

    \subfloat[ \label{sfig:BeforeAfter:WarmupPL}]
    {\includegraphics[width=0.24\textwidth]
    {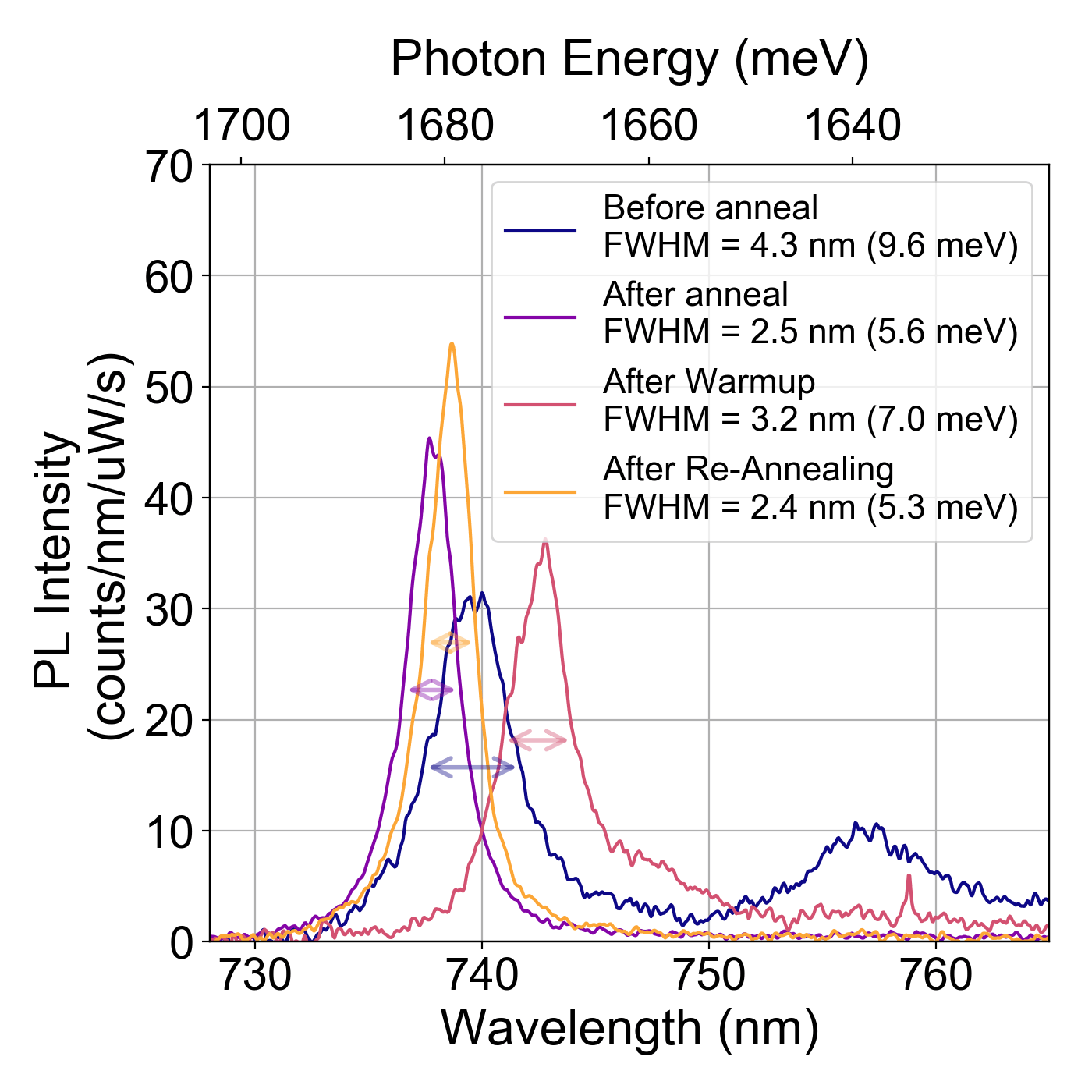}}
    \subfloat[ \label{sfig:BeforeAfter:WarmupReflection}]
    {\includegraphics[width=0.24\textwidth]
    {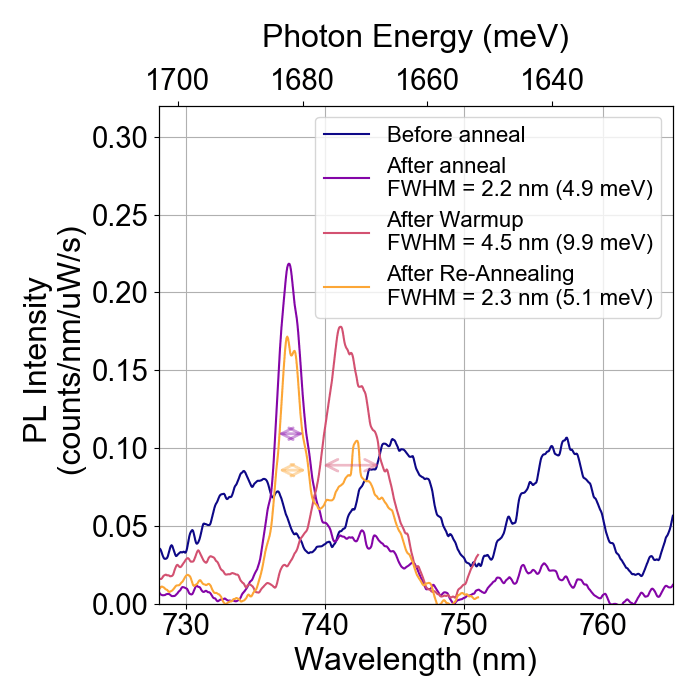}}
    \caption{The PL (a) and reflection (b) after warming up and then cooling the sample back down.
    The sample was first annealed at 4 K with \SI{500}{\micro\watt}.
    It was then brought to room temperature for 24 hours. During this time, the pressure was $\sim 1$ Torr.
    The sample was then cooled back down to 4 K, and subsequently re-annealed with \SI{500}{\micro\watt} laser power.
    \label{sfig:BeforeAfter}
    }
\end{figure}

\begin{figure*}
    \centering
    \subfloat[ Center Anneal \label{sfig:RasterAnneal:CenterAnneal1}]
    {\includegraphics[width=0.3\textwidth]
    {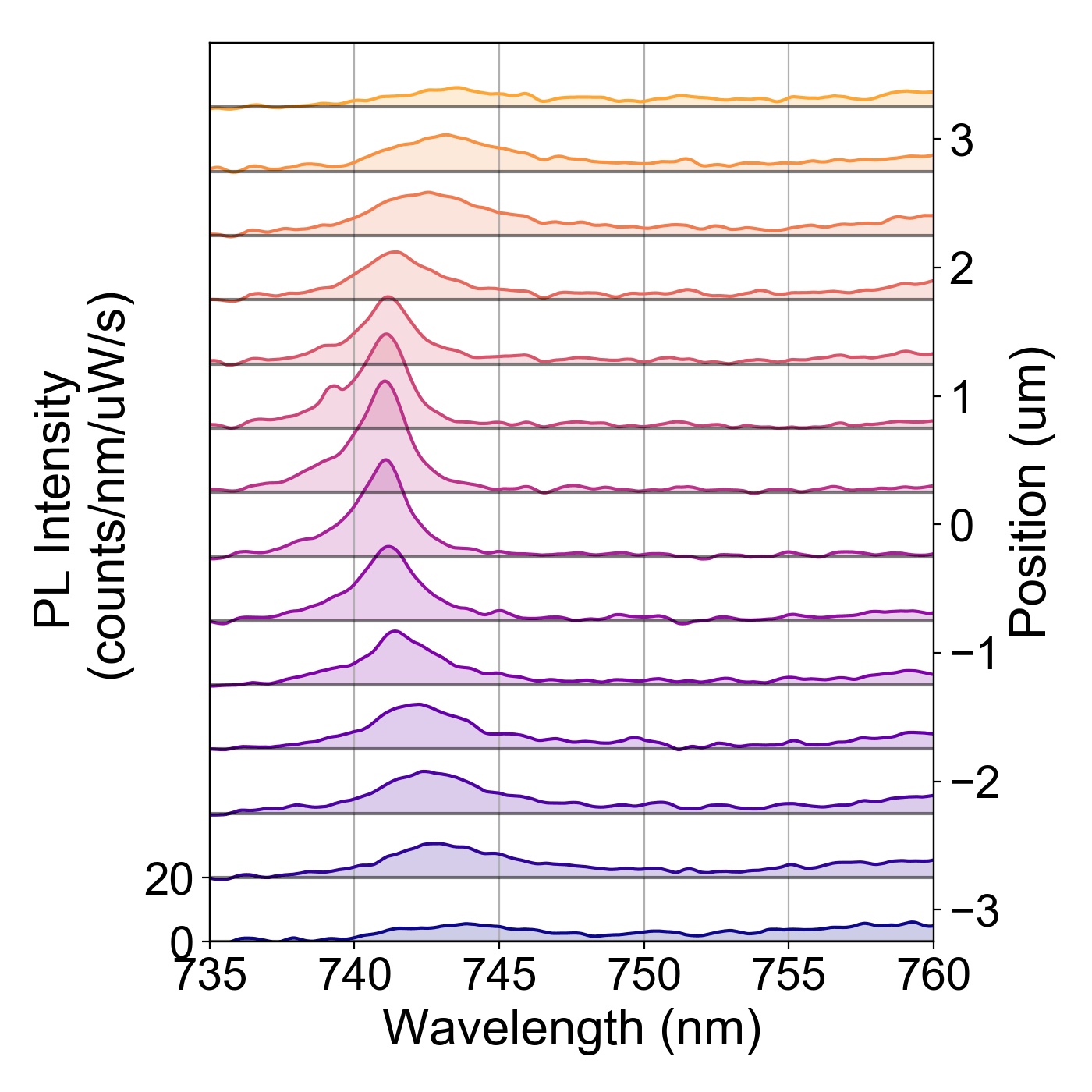}}
    \subfloat[Raster Scan Anneal \label{sfig:RasterAnneal:RasterAnneal2}]
    {\includegraphics[width=0.3\textwidth]
    {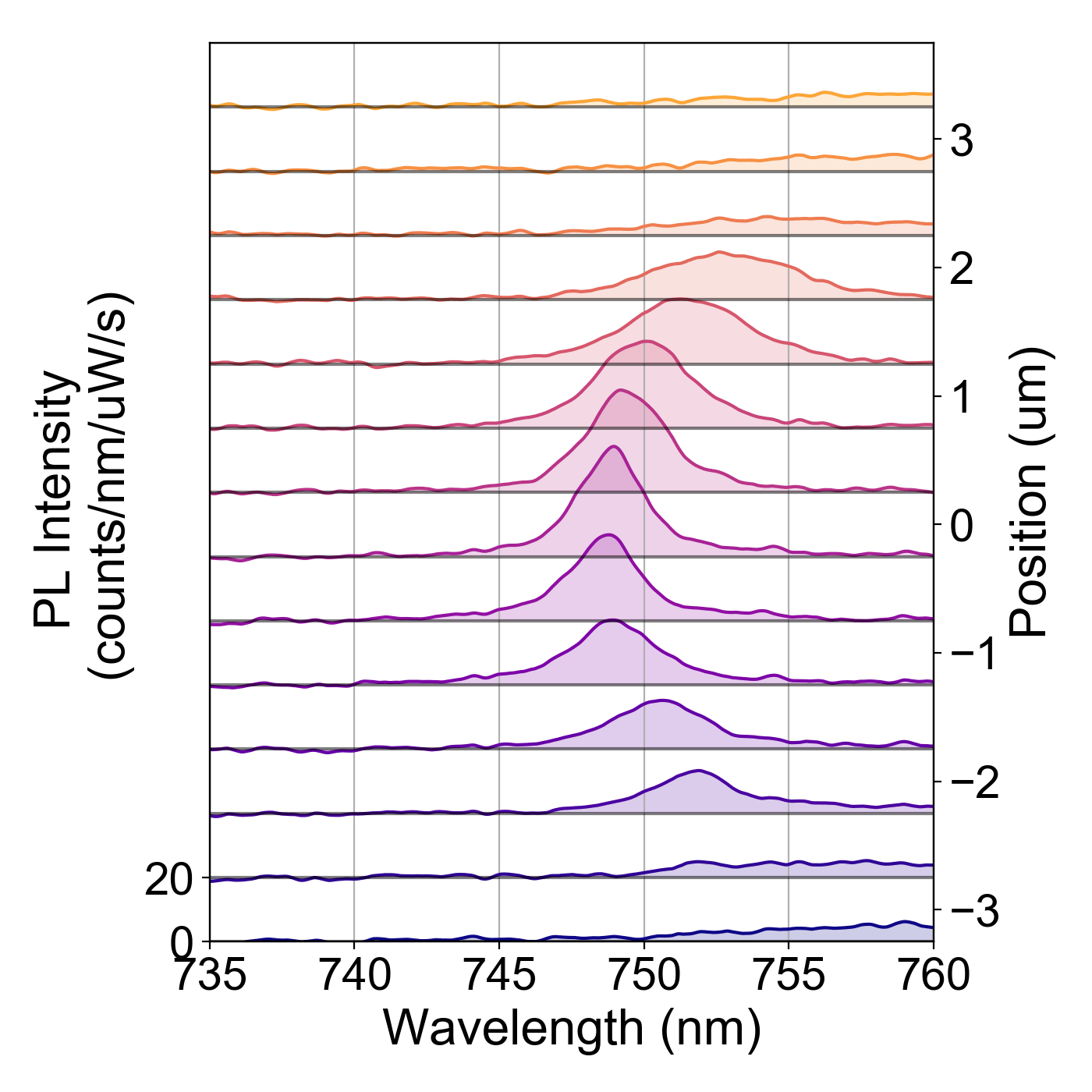}}
    \subfloat[ Center Anneal \label{sfig:RasterAnneal:ReCenterAnneal3}]
    {\includegraphics[width=0.3\textwidth]
    {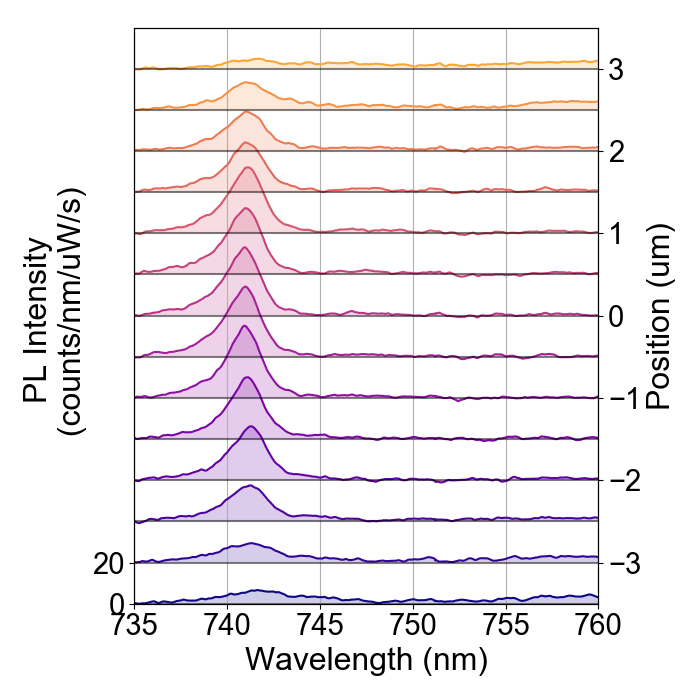}}

    \subfloat[Raster Scan Anneal \label{sfig:RasterAnneal:ReRasterAnneal4}]
    {\includegraphics[width=0.3\textwidth]
    {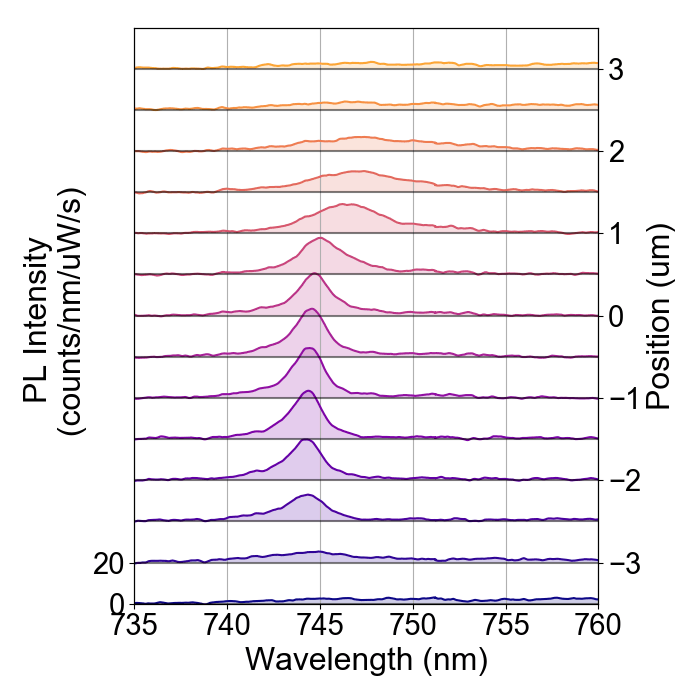}}
    \subfloat[ Center Anneal \label{sfig:RasterAnneal:ReCenterAnneal5}]
    {\includegraphics[width=0.3\textwidth]
    {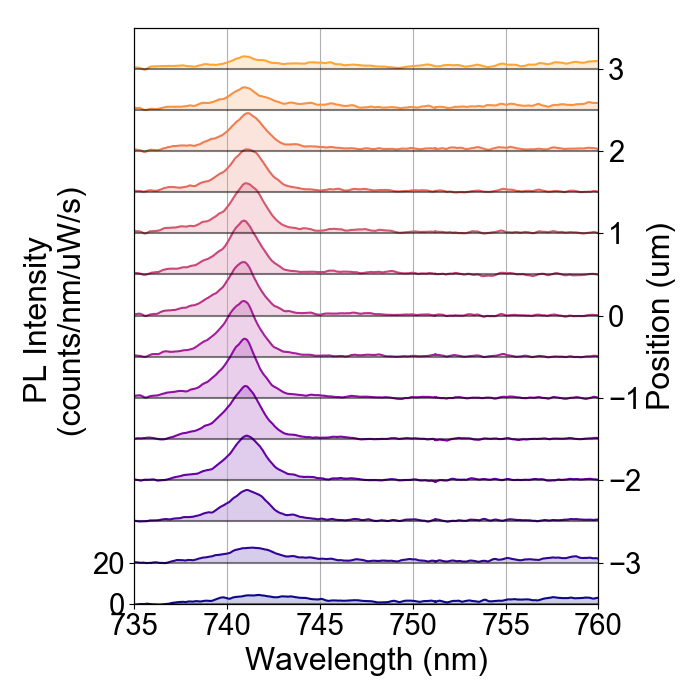}}
    \subfloat[Raster Scan Anneal \label{sfig:RasterAnneal:ReRasterAnneal6}]
    {\includegraphics[width=0.3\textwidth]
    {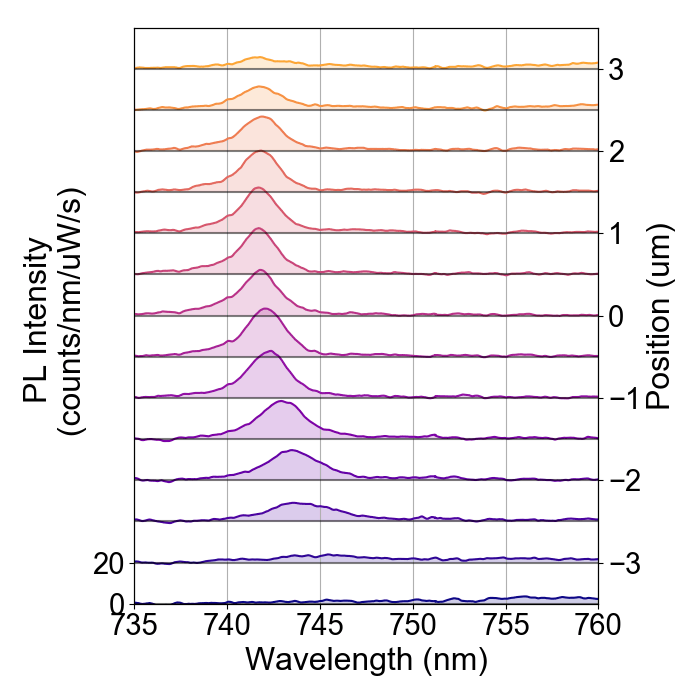}}

    \subfloat[ Center Anneal \label{sfig:RasterAnneal:ReCenterAnneal7}]
    {\includegraphics[width=0.3\textwidth]
    {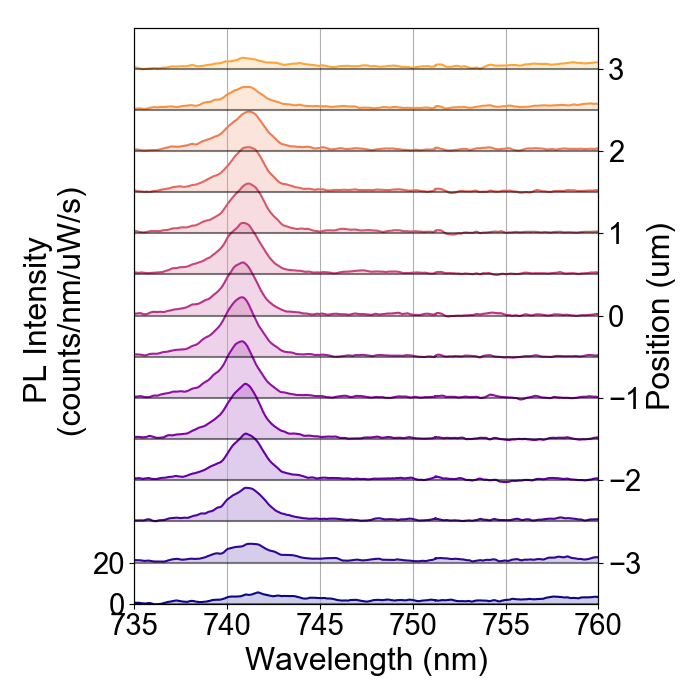}}
    \subfloat[Raster Scan Anneal \label{sfig:RasterAnneal:ReRasterAnneal8}]
    {\includegraphics[width=0.3\textwidth]
    {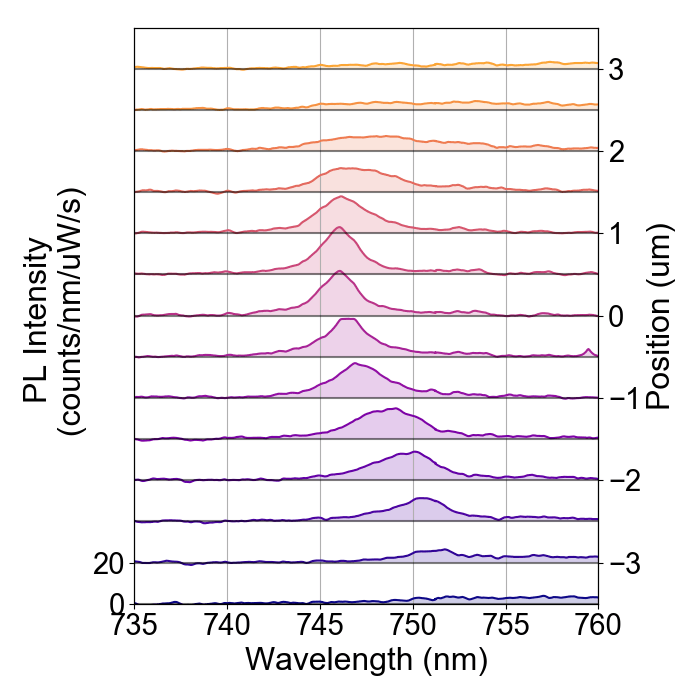}}
    \subfloat[ Center Anneal \label{sfig:RasterAnneal:ReCenterAnneal9}]
    {\includegraphics[width=0.3\textwidth]
    {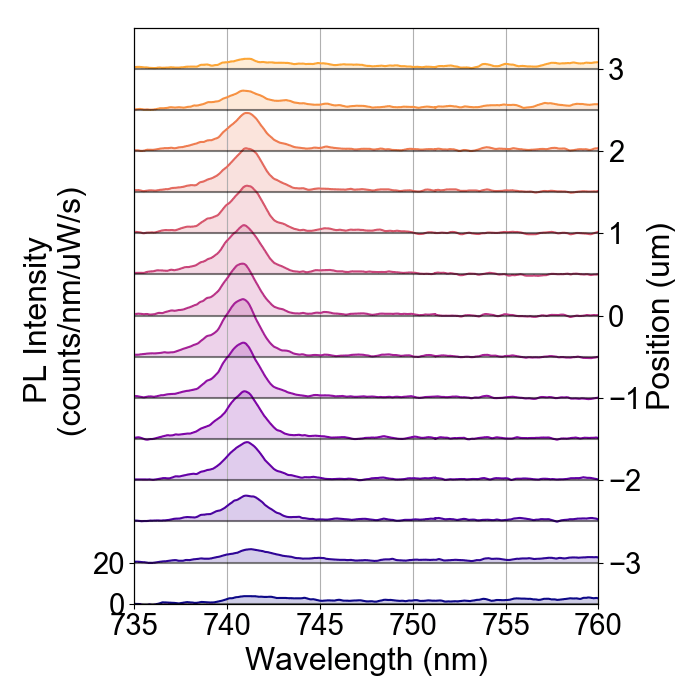}}
    \caption{PL homogeneity after a series of different anneals.
    Each line cut is along the same spatial line on the sample.  Raster scan anneals ending near the edge of the hole were alternated with annealing in the center of the hole.
    The anneals are shown in order from (a) to (i).
    PL in (a), (c), (e), (g) and (i) are after annealing in the center.
    PL in (b), (d), (f), and (h) are after a raster scanning the annealing beam.
    Note that the raster scan anneals ended with the beam near the edge of the hole.
    Each spot was annealed at \SI{400}{\micro\watt} with a 532 nm laser for 1s with a \SI{1}{\micro\meter} spot size.
    Note that (a) and (c) are also plotted in Fig. \ref{fig:RasterAnneal:CenterAnneal} and Fig. \ref{fig:RasterAnneal:ReCenterAnneal}, respectively.
    \label{sfig:RasterAnneal}}
\end{figure*}

\begin{figure*}
    \centering
    \subfloat[ \label{sfig:MultiHole:ReflectionBefore}]
    {\includegraphics[width=0.45\textwidth]
    {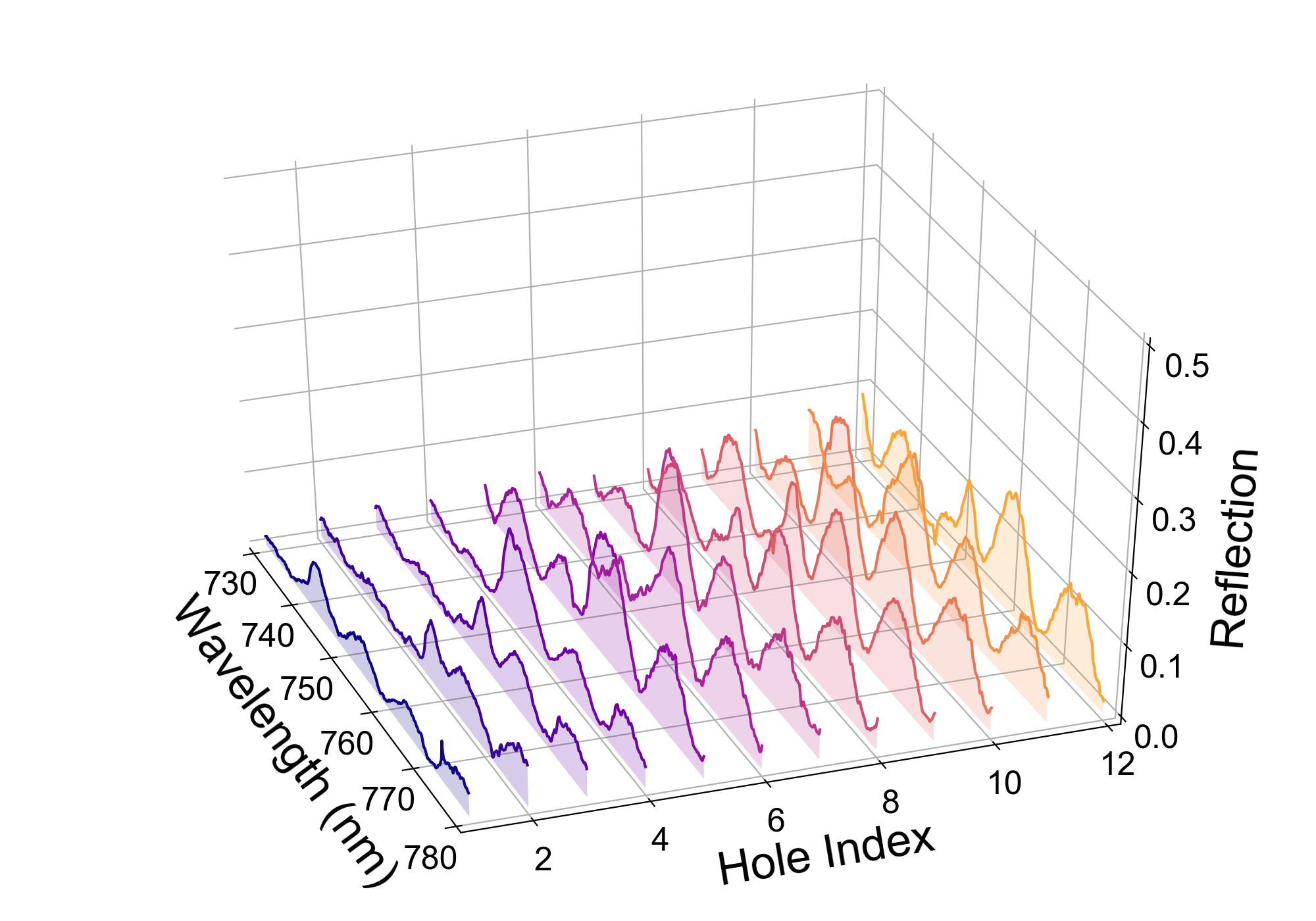}}
    \subfloat[ \label{sfig:MultiHole:ReflectionAfter}]
    {\includegraphics[width=0.45\textwidth]
    {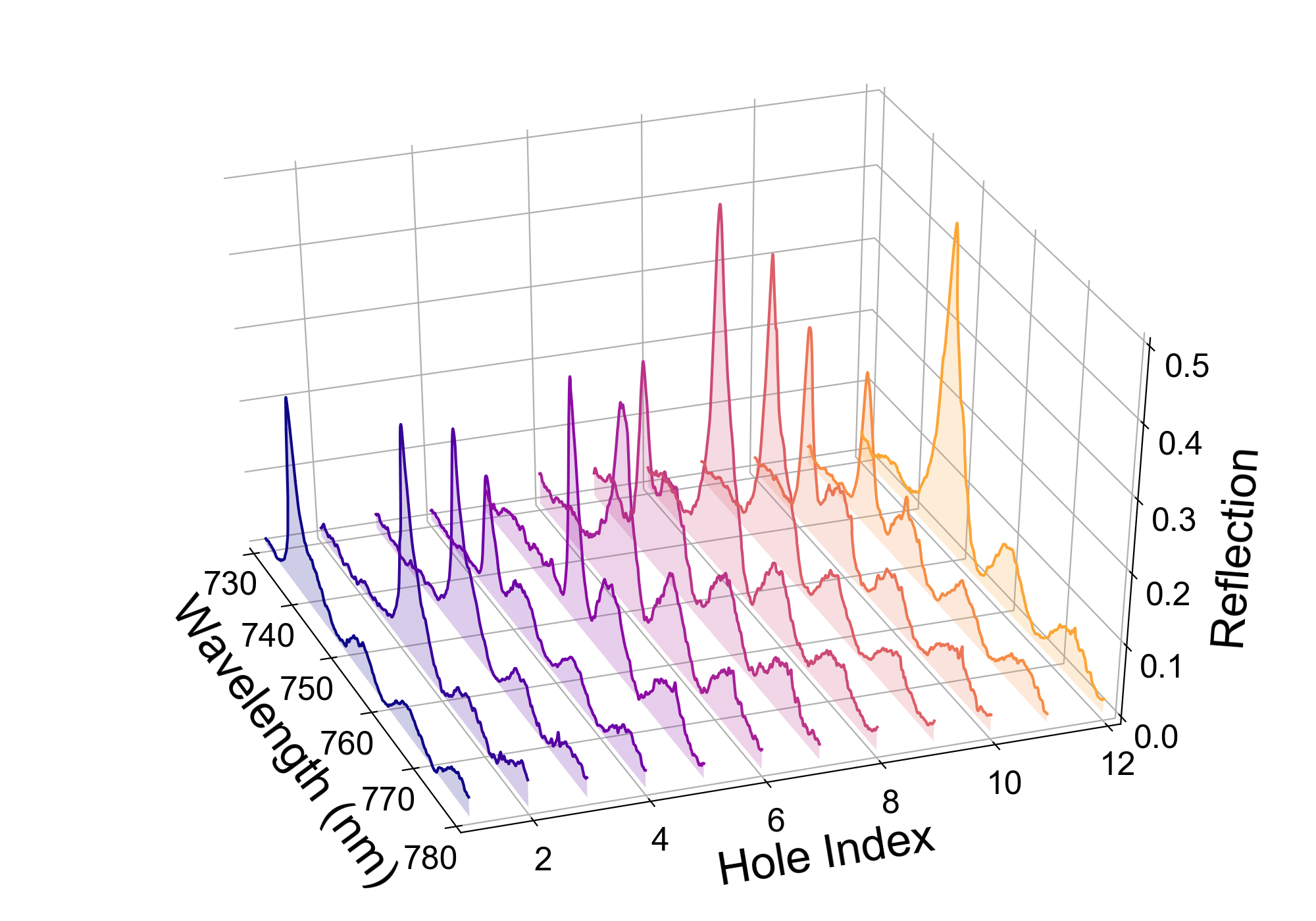}}
    \caption{Raw reflection spectra corresponding to Fig. \ref{fig:MultiHole}.
    Reflection before annealing, corresponding to Fig. \ref{fig:MultiHole:ReflectionBefore} (a) and reflection after annealing, corresponding to Fig. \ref{fig:MultiHole:ReflectionAfter} (b).
    \label{sfig:MultiHole}}
\end{figure*}

\begin{figure*}
    \centering
    \subfloat[ \label{sfig:XPS:Spectrum}]
    {\includegraphics[width=0.65\textwidth]
    {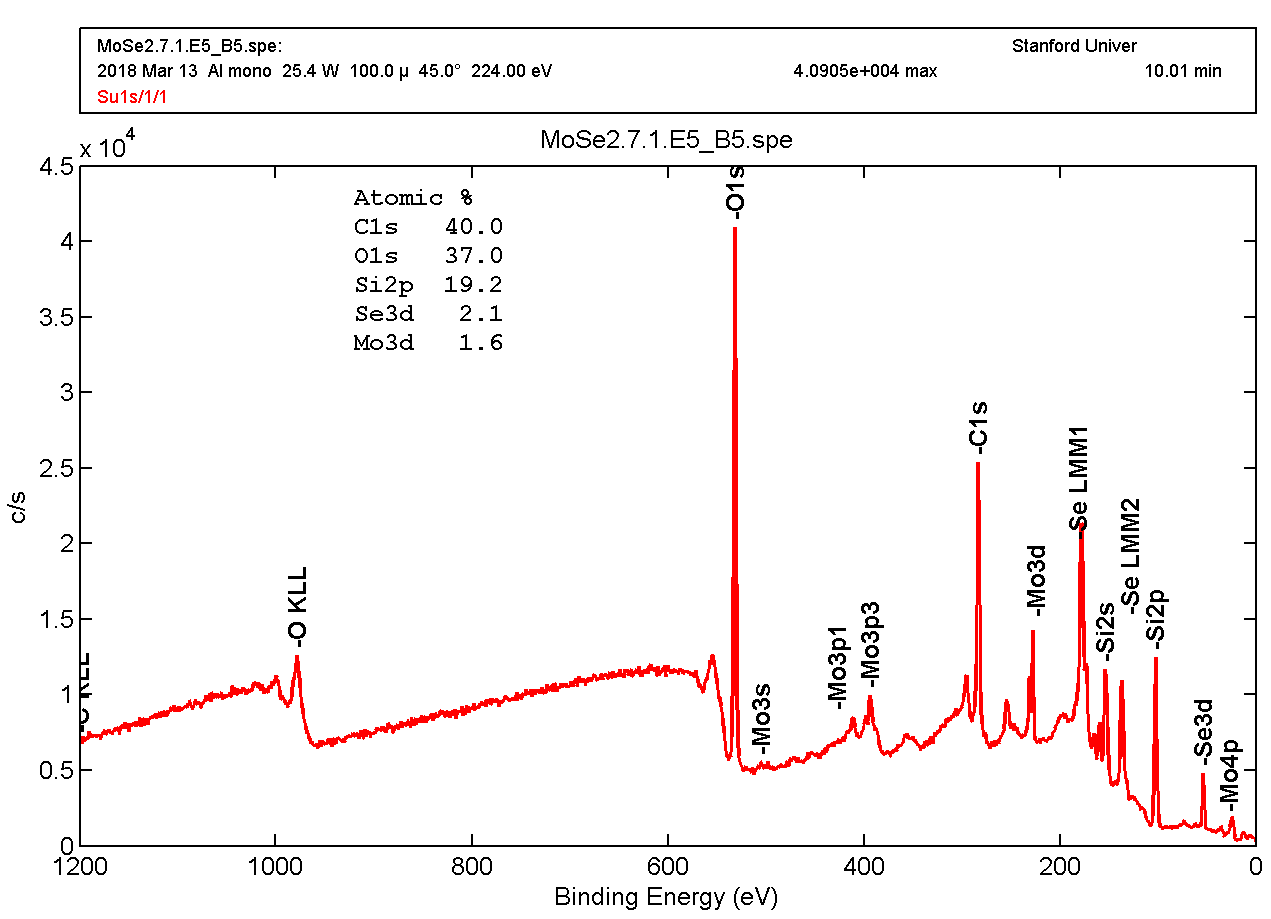}}

    \subfloat[ \label{sfig:XPS:MoSe2Reference}]
    {\includegraphics[width=0.55\textwidth]
    {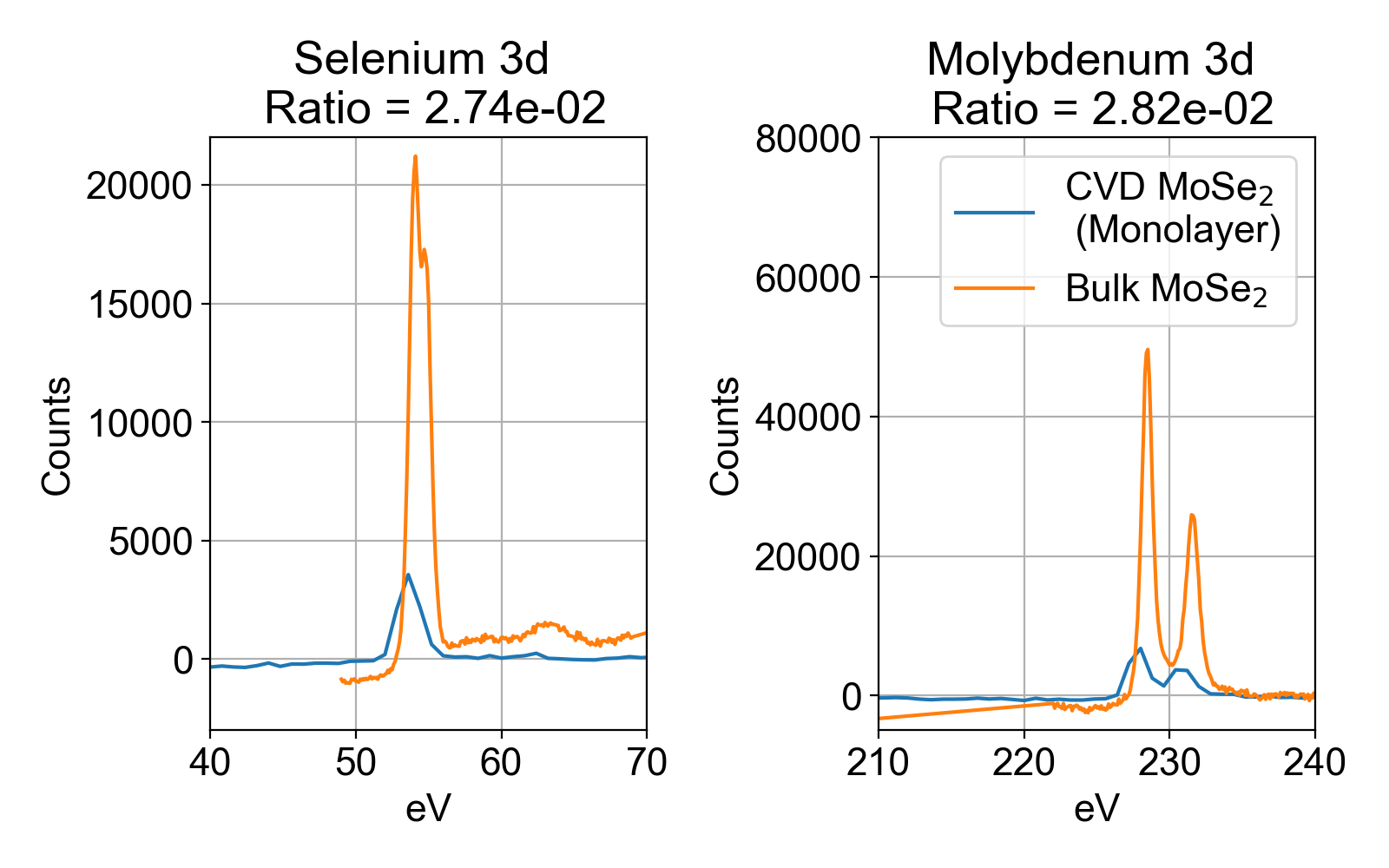}}
    \subfloat[ \label{sfig:XPS:CarbonZoom}]
    {\includegraphics[width=0.3\textwidth]
    {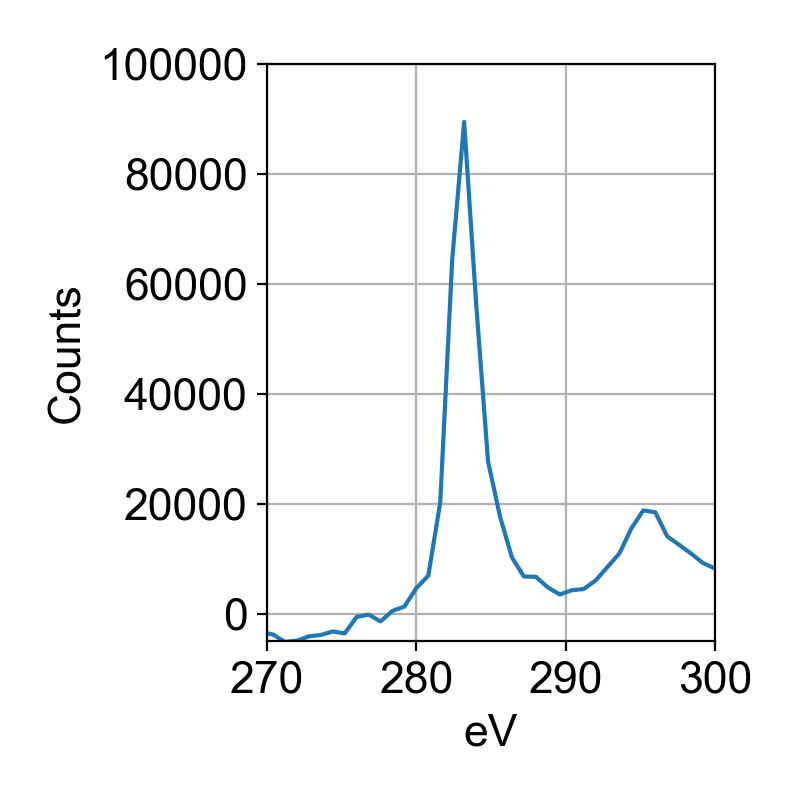}}
    \caption{XPS spectra from a supported area of monlayer $\mathrm{MoSe_2}$.
    The raw XPS spectra (a) along with extracted atomic percentages.
    The ratio of the integrated XPS signal from a bulk sample and the monolayer CVD sample (b) for both Mo and Se.
    The carbon 1s peak (c).
    \label{sfig:XPS}}
\end{figure*}

The raw XPS data for a supported flake is shown in Fig. \ref{sfig:XPS:Spectrum}.
We assume that the contaminants are qualitatively similar on suspended and supported areas.
We can see from this that by atomic percentage, the most abundant contaminant is carbon at 40\%.
While there is a significant amount of both oxygen and silicon these are present in a ratio of about 2:1, suggesting that these are primarily from the SiO$_2$ in the substrate.
Since the XPS is sensitive to atoms within the top 10-15nm of the sample and the SiO$_2$ thickness is 280 nm, we do not expect any signal from the silicon itself.
Note that the XPS is not sensitive to hydrogen.
This would indicate a carbon contamination thickness of about 8 nm, or $\sim$20 layers.

The C1s peak shown in \ref{sfig:XPS:CarbonZoom} shows little if any asymmetry, and no nearby peaks in the range 285-290 eV.
The C1s peak from C-C bonds is typically at 284 eV, while the C1s from C-O-C bonds is at slightly higher energies of 286 eV, and the C1s from C-O=C is at 288 eV.
Depending on the instrument resolution and the rest of the bonding structure, the combination of C-C bonds and C-O-C bonds from organic molecules can result in a asymmetric peak, or two separate peaks.
This is in addition to the C-O=C peak, which is usually well resolved if present.
However, the C1s peak here is symmetric and there are no other peaks within the expected C1s range, indicating that there is little C-O bonding in the contaminating carbon.
This is consistent with the observation that the oxygen to silicon ratio was close to that expected for SiO$_2$.
It is thus very likely that the carbon contaminants are primarily hydrocarbons.

A calibration of the XPS spectra for Mo and Se is shown in \ref{sfig:XPS:MoSe2Reference}.
The analysis using the most recent XPS machine calibration in \ref{sfig:XPS:Spectrum} gave a Se:Mo ratio of 1.3, which is substantially different from the expected ratio of 2.
However, we calibrated to a bulk sample of $\mathrm{MoSe_2}$ from HQ Graphene and found that the ratio of the integrated intensity between the monolayer and bulk was $2.74\cdot 10^{-2}$ for Se and $2.82\cdot 10^{-2}$ for Mo.
Assuming a ratio of 2:1 in the bulk sample, this indicates a Se:Mo ratio of 1.94 in the monolayer, which is quite close to the expected ratio.

\begin{figure}
    \centering
    \subfloat[ \label{sfig:BeforeAfterSiN:PL}]
    {\includegraphics[width=0.24\textwidth]
    {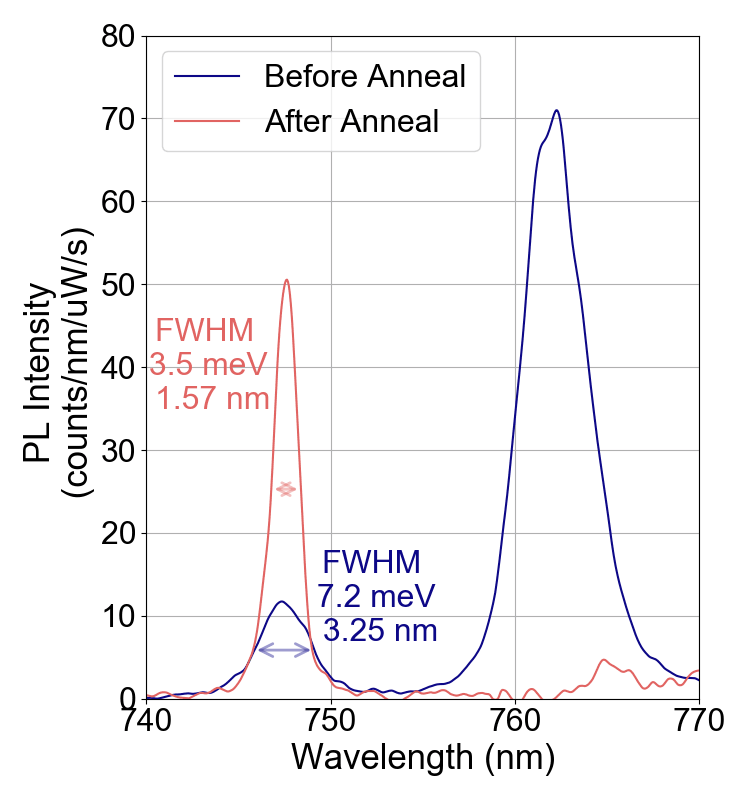}}
    \subfloat[ \label{sfig:BeforeAfterSiN:Reflection}]
    {\includegraphics[width=0.24\textwidth]
    {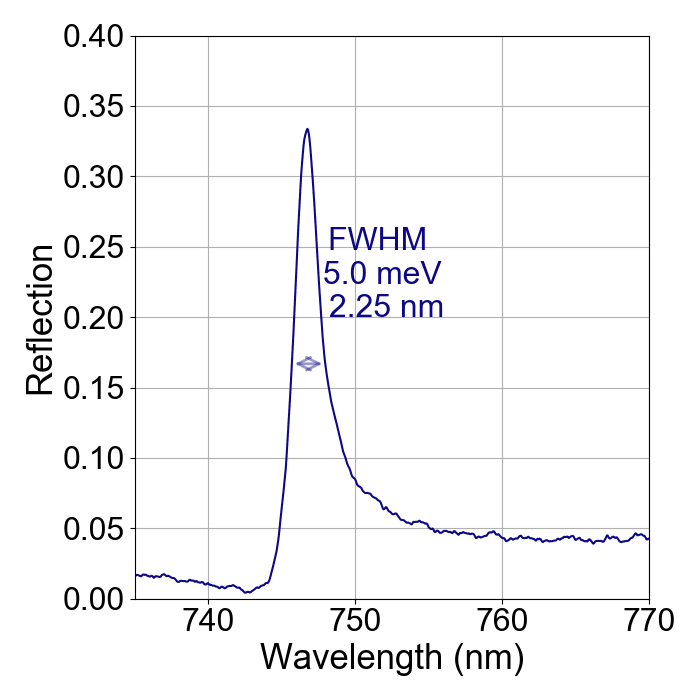}}
    \caption{Monolayer $\mathrm{MoSe_2}$ suspended on \SI{2}{\micro\meter} holes in a perforated silicon nitride membrane was annealed for 5 minutes with 2 mW laser power.
    The monolayer was exfoliated on to PDMS from bulk $\mathrm{MoSe_2}$, and then transferred to the perforated silicon nitride membrane.
    The PL (a) before and after annealing and reflection (b) after annealing.
    Note that (a) and (b) are from different holes.
    \label{sfig:BeforeAfterSiN}}
\end{figure}

Fig. \ref{sfig:BeforeAfterSiN} shows that the annealing process works for exfoliated $\mathrm{MoSe_2}$ as well.
Exciton PL is narrowed and trion PL is eliminated, in behavior consistent with that seen in the CVD samples.
The raw reflection spectrum shows no etalon fringes since the holes in the silicon nitride are through holes.
A significantly higher annealing power of 2 mW was used here, since the suspended holes are much smaller than for the CVD samples.

\begin{figure}
    {\includegraphics[width=0.37\textwidth]
    {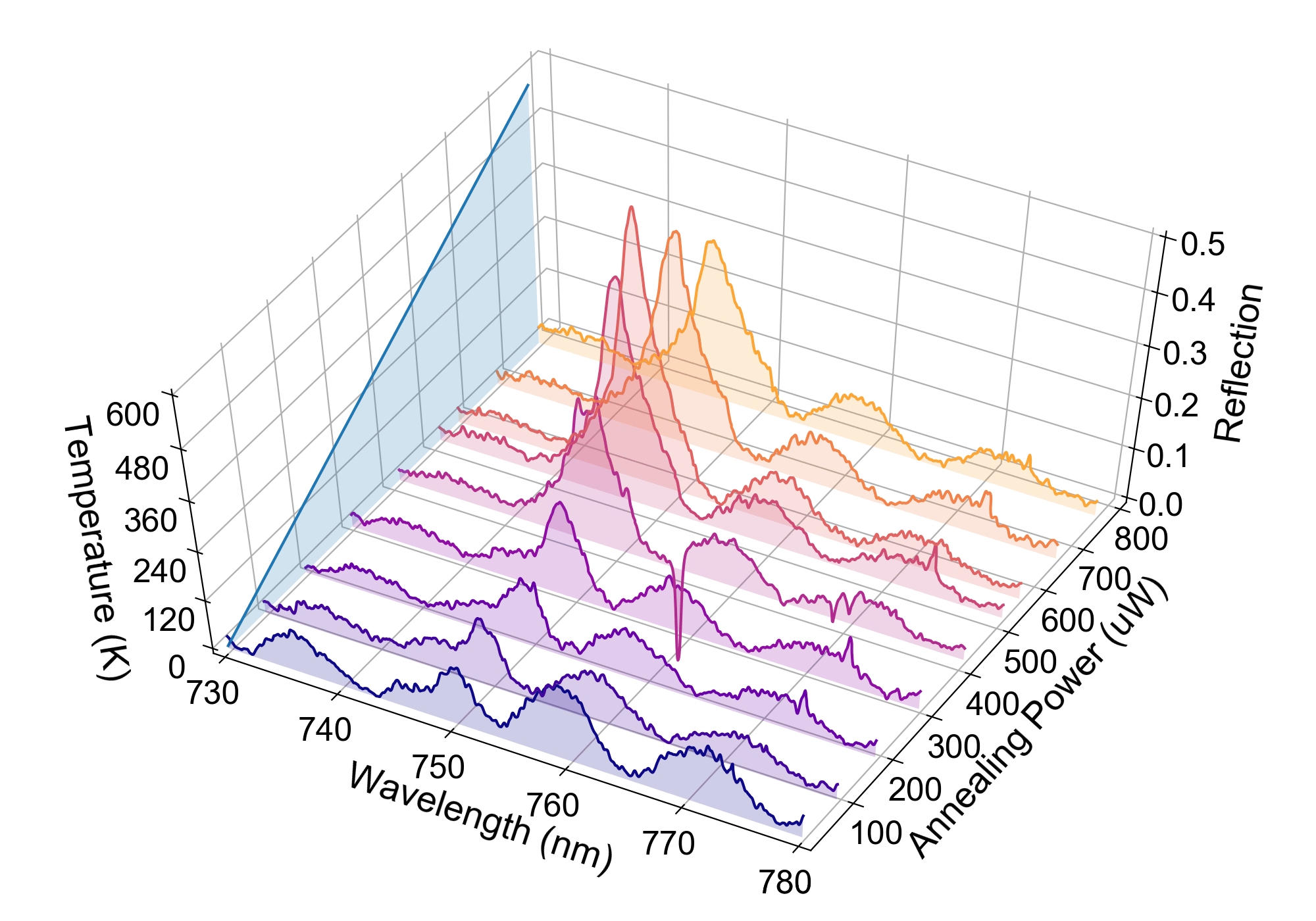}}
    \caption{The raw reflection after annealing a suspended $\mathrm{MoSe_2}$ monolayer at successively increasing laser annealing powers, corresponding to the fitted data in Fig. \ref{fig:AnnealPower:ReflectionAnneal}.
    \label{sfig:AnnealPower:ReflectionAnneal}}
\end{figure}

\begin{figure*}
    \centering
    \subfloat[ \label{sfig:AnnealTime:Reflection}]
    {\includegraphics[width=0.45\textwidth]
    {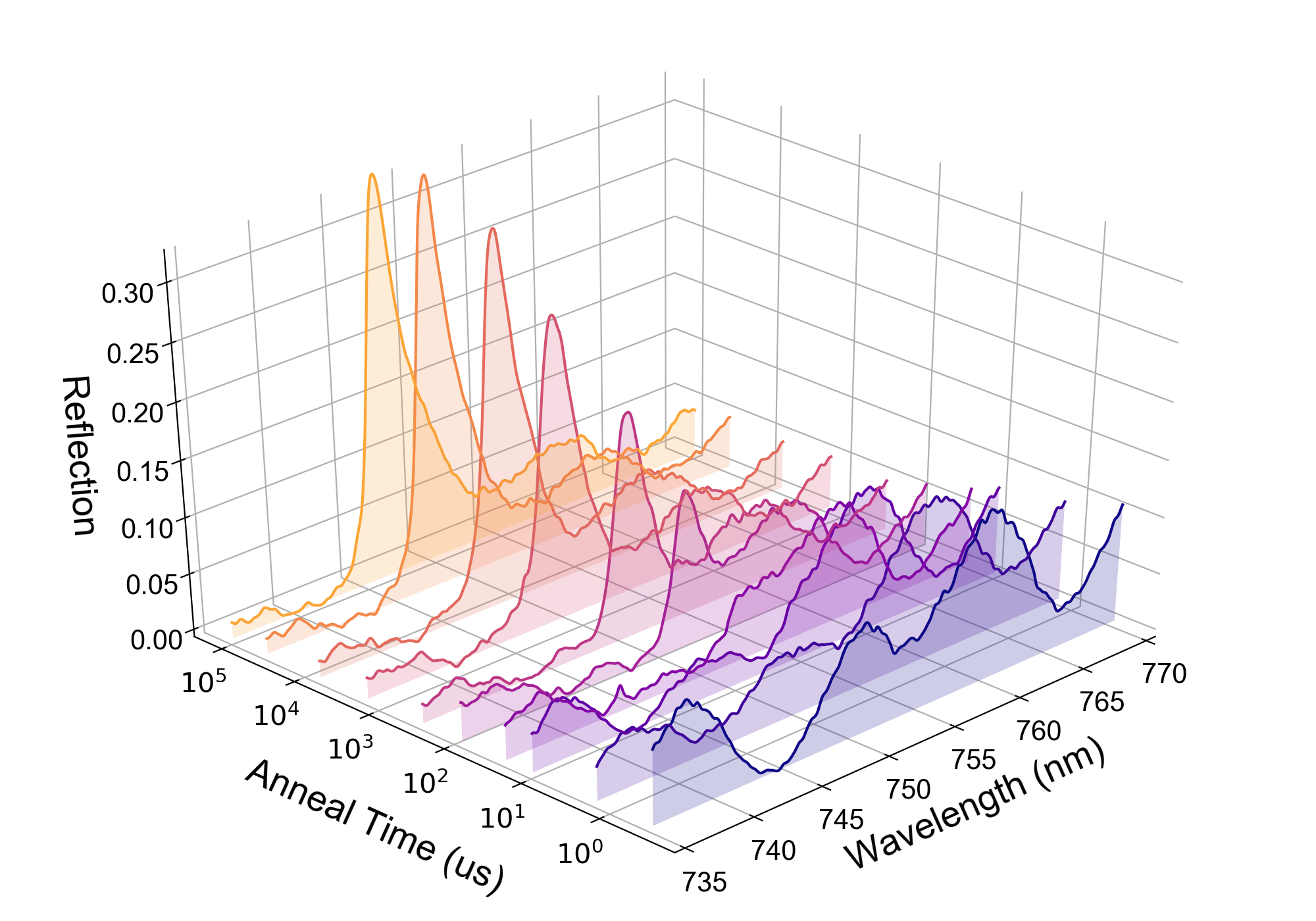}}
    \subfloat[ \label{sfig:AnnealTime:ReflectionFWHM}]
    {\includegraphics[width=0.35\textwidth]
    {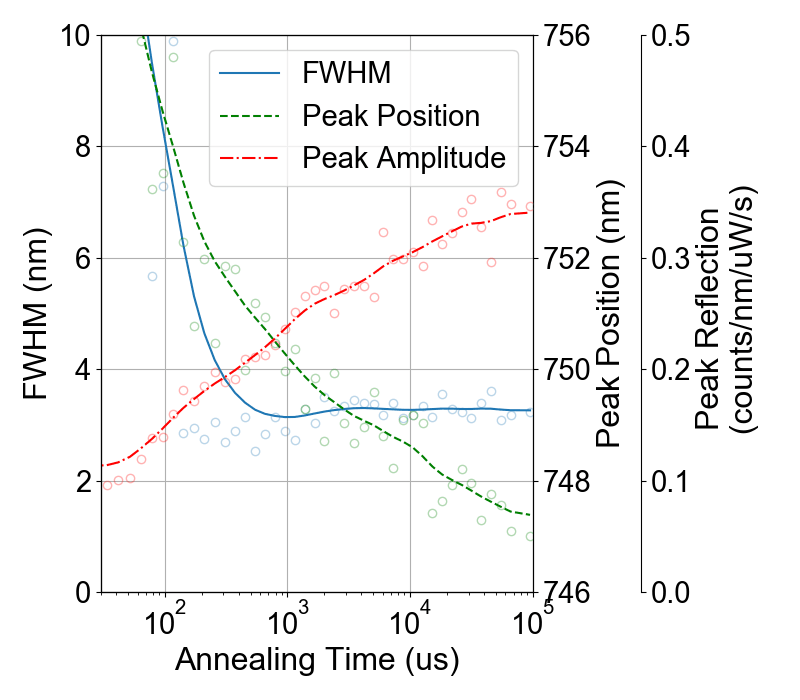}}
    \caption{The raw reflection spectra (a) corresponding to Fig. \ref{fig:AnnealTime:Reflection}.
    The extracted FWHM and amplitude (b) corresponding to Fig. \ref{fig:AnnealTime:ReflectionFWHM}.
    \label{sfig:AnnealTime}}
\end{figure*}

\begin{figure*}
    {\includegraphics[width=0.8\textwidth]
    {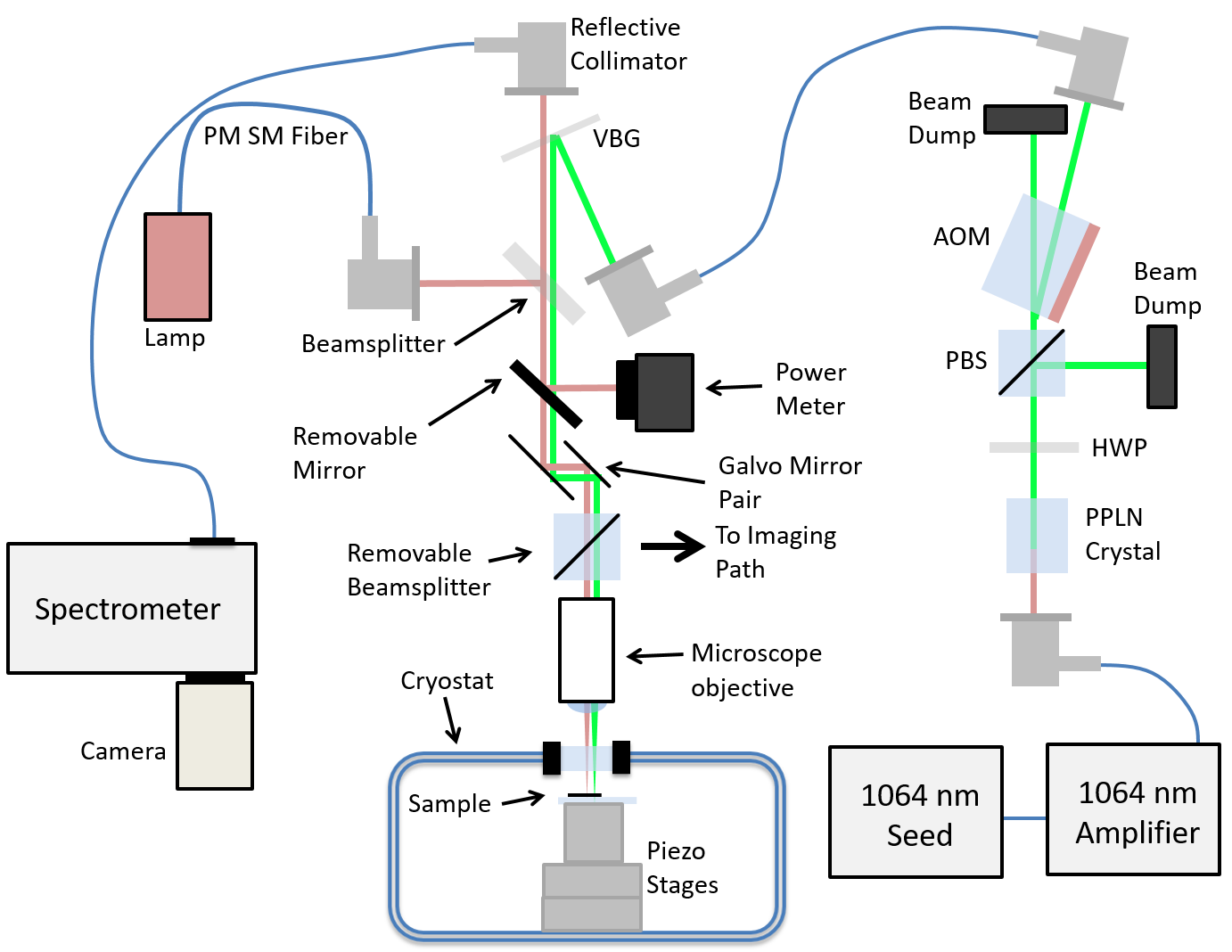}}
    \caption{Schematic showing the measurement setup for PL, reflection and Raman measurements.
    The volume Bragg grating (VBG) serves as a beamsplitter to couple the green excitation laser into the light path.
    Polarization maintaining single mode (PM SM) fiber is used to couple light onto the microscope and to the spectrometer.
    The half-wave-plate (HWP) and polarizing beamsplitter (PBS) serve as intensity control for the green laser generated from the periodically poled lithium niobate (PPLN) crystal.
    The acousto-optic modulator gives fast temporal control of the green intensity.
    \label{sfig:schematic}}
\end{figure*}

\begin{figure*}
    \centering
    \subfloat[ \label{sfig:FitExample:Annealed}]
    {\includegraphics[width=0.45\textwidth]
    {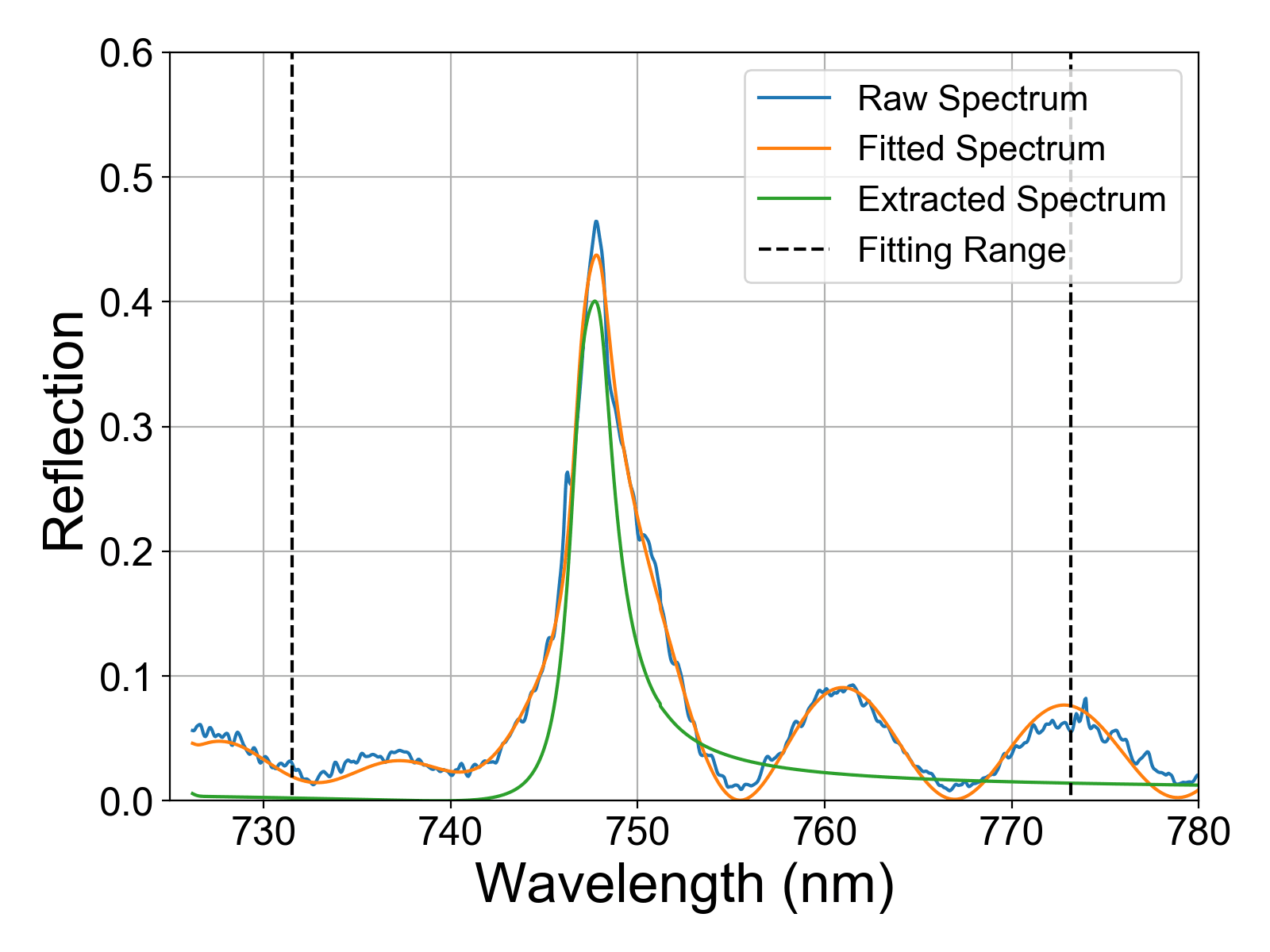}}
    \subfloat[ \label{sfig:FitExample:Unannealed}]
    {\includegraphics[width=0.45\textwidth]
    {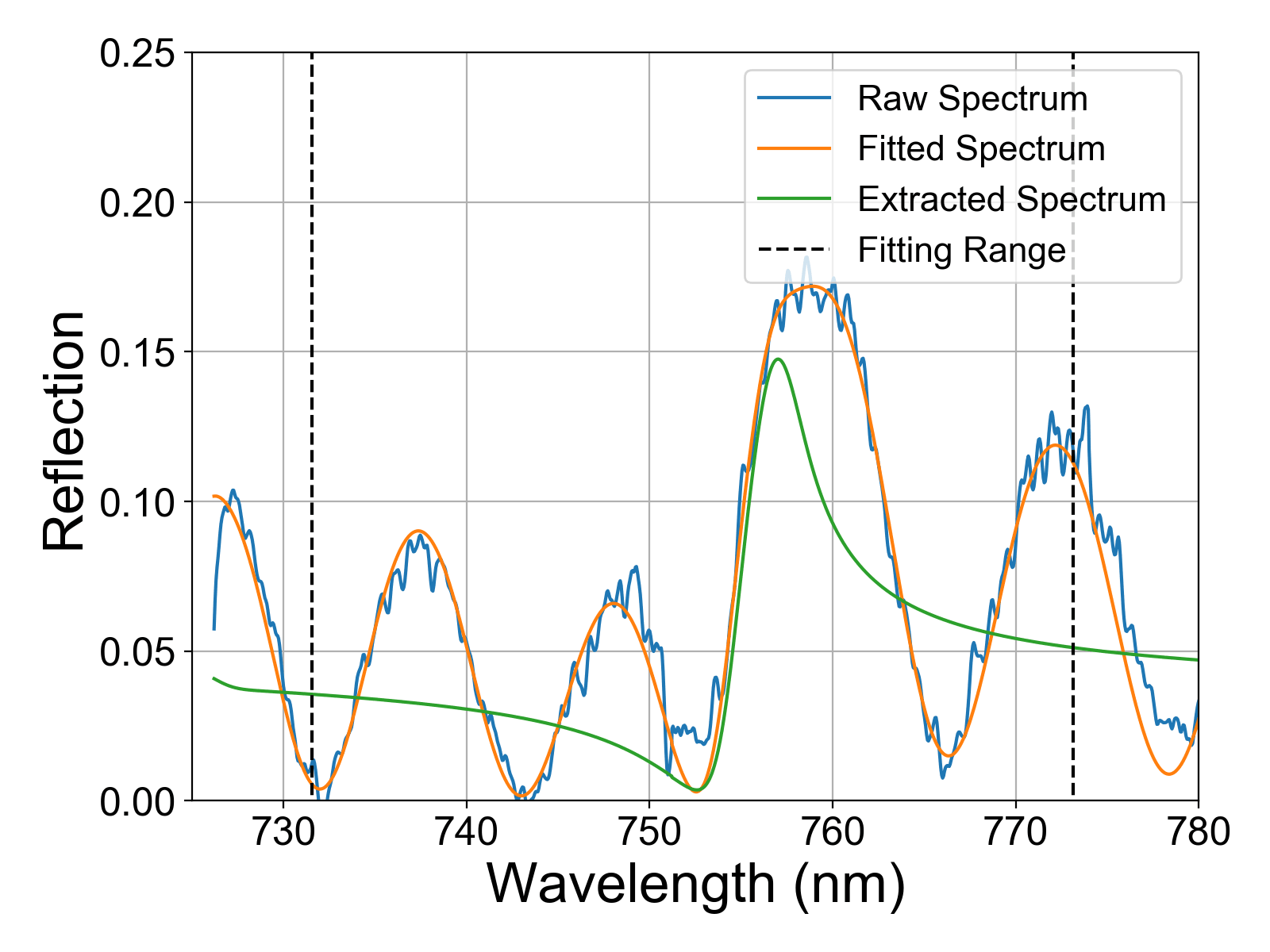}}
    \caption{Examples of fitting the reflection data and extracting the reflection of the suspended film by removing etalon effects.
    The reflection of an annealed sample (a) with a strong reflection peak.
    The reflection of an un-annealed sample (b) with a weak reflection peak.
    \label{sfig:FitExample}}
\end{figure*}

\begin{figure}
    {\includegraphics[width=0.45\textwidth]
    {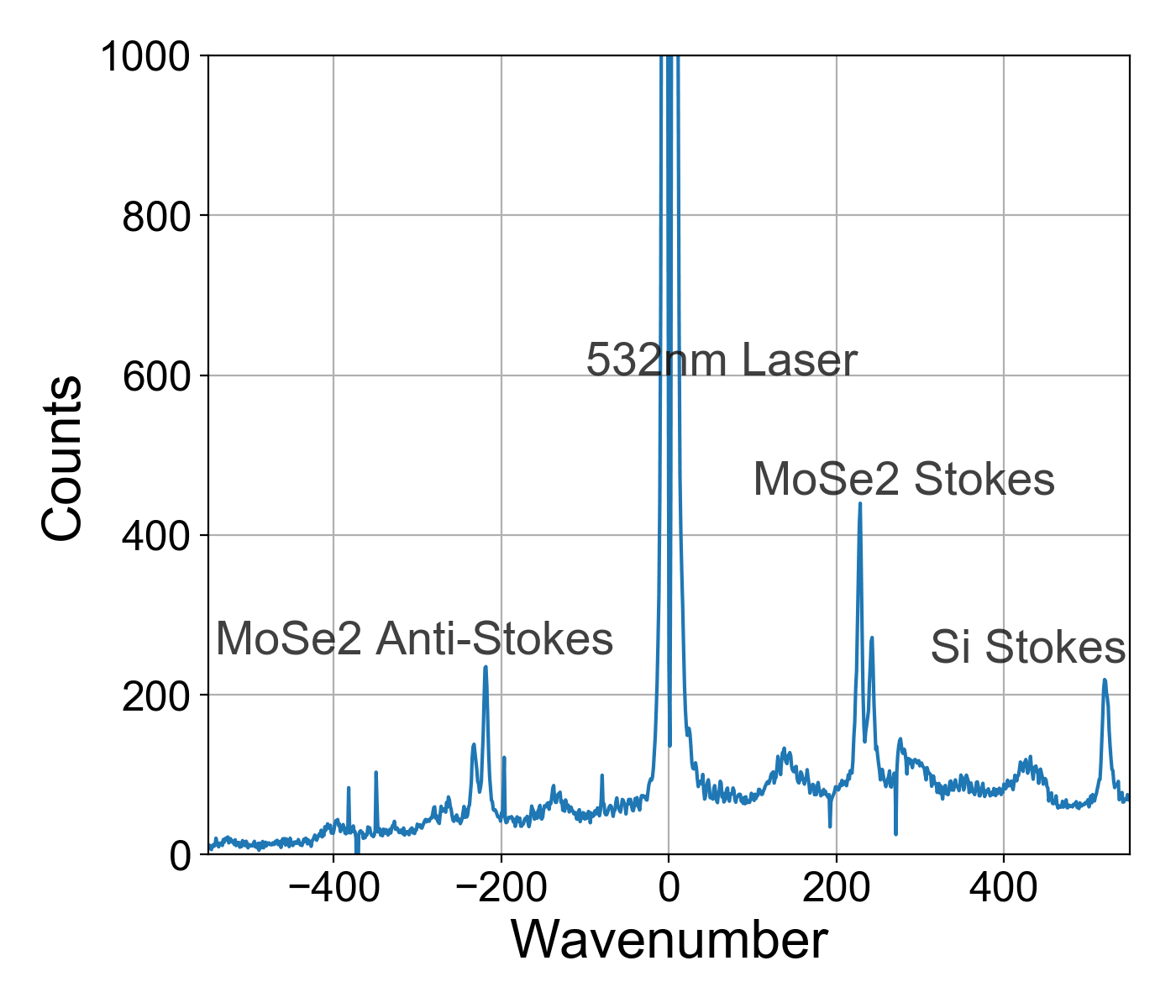}}
    \caption{An example Raman spectrum, showing both the Stokes and Anti-Stokes peaks of the 242 cm$^{-1}$ $\mathrm{MoSe_2}$ line from a suspended flake.
    The Si Stokes peak at 521 cm$^{-1}$ is also visible, because of the small back reflection  from the bottom of the hole.
    \label{sfig:raman}}
\end{figure}

An example of a Raman spectrum taken during annealing is shown in Fig. \ref{sfig:raman}.
The $\sim 242$ cm$^{-1}$ Stokes- and anti-Stokes peaks from $\mathrm{MoSe_2}$ are clearly visible, as well as a peak at slighly higher wavenumber.
Both peaks are observed in \cite{ResonanceEffectsInRamanScatteringOfMoSe2} for supported monolayer $\mathrm{MoSe_2}$.

%
%
%

\bibliography{LaserAnnealing}